\documentclass[rmp,aps,twocolumn,nofootinbib]{revtex4}

\usepackage{amsmath}
\usepackage{amssymb}

\pdfoutput=1

\usepackage{graphicx}
\usepackage{dcolumn}
\usepackage{bm}
\usepackage[usenames]{color}

\newcommand{\be}{\begin{equation}}
\newcommand{\ee}{\end{equation}}
\newcommand{\bea}{\begin{eqnarray}}
\newcommand{\eea}{\end{eqnarray}}

\begin{document}

\title{Vortices in quantum droplets: Analogies between boson and fermion systems}

\author{H. Saarikoski\footnote{
Present address: Institut f\"ur Theoretische Physik, Universit\"at Regensburg,
D-93040 Regensburg, Germany} and S.M. Reimann\footnote{Corrsponding
    Author, email: reimann@matfys.lth.se}}

\affiliation{Mathematical Physics, LTH, Lund University, 
SE-22100 Lund, Sweden
}

\author{A. Harju}

\affiliation{Department of Applied Physics and Helsinki Institute of Physics, 
Aalto University, FI-02150 Espoo, Finland}
\author{M. Manninen}

\affiliation{Nanoscience Center, Department of Physics, FIN-40014 University
  of 
Jyv\"askyl\"a, Finland}

\date{\today}
 
\begin{abstract}
The main theme of this review is the many-body physics of  vortices in 
quantum droplets of bosons or fermions, in the limit of small particle numbers.
Systems of interest include cold atoms in traps 
as well as electrons confined in quantum dots. 
When set to rotate, these in principle very different 
quantum systems show remarkable analogies. 
The topics reviewed include the structure of the finite rotating 
many-body state, universality of vortex formation and 
localization of vortices in both bosonic and fermionic 
systems, and the emergence of particle-vortex composites 
in the quantum Hall regime. 
An overview of the 
computational many-body techniques sets focus on 
the configuration interaction and density-functional 
methods. Studies of quantum droplets with one or
several particle components, where vortices as well 
as coreless vortices may occur, are reviewed, 
and theoretical as well as experimental challenges
are discussed. 
\end{abstract}

\maketitle
\setcounter{tocdepth}{4}
\tableofcontents

\section{Introduction}          

In recent years, advances in experimental methods in quantum optics 
as well as semiconductor physics have made it possible to create 
confined quantum droplets of particles, and to manipulate them with 
unprecedented control. 
Bose-Einstein condensates of ultra-cold atomic gases, for example, 
may be set rotating either by rotating the trap, or by ``stirring'' 
the cold atoms with lasers. These clouds of bosons are large in
present day experiments, but the regime of few-particle bosonic droplets 
ultimately may be reached. Confined electron droplets, on the other hand,
are nowadays routinely realized as low-dimensional nanostructured quantum
dots in semiconductors, where the droplet size and its angular momentum
can be accurately fixed by an external voltage bias and a magnetic field,
respectively. 
A bosonic atom cloud in a trap, and electrons confined
in quantum dots are very different systems by nature. However, when set
to rotate, their microscopic properties show remarkable analogies.
While quantum dots are usually quasi-two-dimensional due to the
semiconductor heterostructure, the dimensionality is reduced also
in a trapped rapidly rotating atom gas due to the centrifugal force, which
flattens the cloud of atoms.

The structure of a quantum state describing a rotating droplet fundamentally 
reflects how the system carries angular momentum. Intriguingly, some of the 
underlying mechanisms appear universal 
in two-dimensional systems regardless of the particle statistics, 
wave function symmetries, and the form of the interparticle interaction. 
For example, both bosonic and fermionic droplets show 
formation of vortices 
in the droplet with increasing angular momentum.
Eventually, in the regime of very rapid rotation, finite-size 
precursors of fractional quantum Hall states with 
particle-vortex composites are predicted to
emerge similarly in both bosonic and fermionic systems. 
Due to these universalities in the structure of the quasi-two-dimensional 
many-body state, rotating quantum droplets can often be described 
theoretically by similar concepts and analogous vocabulary. 
These analogies are the main theme of this review, where  boson and fermion 
systems are treated in parallel and similarities and 
differences between these systems are extensively discussed.

Despite the close connection between rotating cold atom gases
and electrons in nanostructured quantum systems in solids, 
research efforts in these fields have advanced mostly independently 
of each other. In this review we highlight the similarities between 
these fields, with the hope that it may serve  as
a source of inspiration for further studies on rotating 
quantum systems where complex and sometimes unexpected 
phenomena emerge.

\subsection{Finite quantum liquids in traps}

Confining elementary particles or indistinguishable composite 
particles, such as atoms, by cavities or external 
potentials at low temperatures, one may create finite-size quantum systems
with particle numbers ranging from just a few to millions. 
Cold atomic quantum gases in traps and lattices, 
photons in cavities and electrons confined in low-dimensional
semiconductor nanostructures are well-known examples.

\subsubsection{Atoms in traps}

Bose and Einstein predicted already in the 1920s the condensation 
of an ideal gas of bosonic particles into a single, coherent quantum 
state~\cite{bose1924,einstein1924,einstein1925}.
Apart from strongly interacting systems such as liquid helium,  
the experimental discovery of this phenomenon
had to wait many decades, until
advances in cooling and trapping techniques for dilute
atomic gases finally  made possible the observation of
Bose-Einstein condensation (BEC) in a cloud of cold bosonic alkali 
atoms~\cite{anderson1995,davis1995a,davis1995b,ensher1996,ketterle2002,
cornellwieman2002}.
These celebrated experiments clearly marked a new era in quantum physics
combining the fields of quantum optics, condensed matter physics and atomic
physics.  For the physics of BEC, see for example
the review article by~\textcite{leggett2001} as well 
as~\textcite{dalfovo1999}, the 
monographs by ~\textcite{pethicksmith2002,leggett2006,pitaevskiistringari2003},
and \textcite{inguscio1999}.

A BEC can be set rotating not only by rotating the trap, but also 
by stirring the bosonic droplet with 
lasers~\cite{madison2000,madison2001,chevy2000,aboshaer2001},
or by evaporating atoms~\cite{haljan2001,engels2002,engels2003}  
(see the discussion in the recent review by \textcite{fetter2008}). 
A weakly interacting dilute system becomes effectively two-dimensional when
set rotating, 
making a description in the lowest Landau level possible. 
We mainly restrict our analysis of BEC's in this review to this limit of 
quasi-two-dimensional droplets of atoms.

More recently, superfluid states have been realized also for trapped
fermionic atoms, where fermion pairing or molecule formation
can occur in two distinct regimes depending on the atomic
interaction strength. 
Pairing can take place in real space via molecule formation
and these composite bosons may then show Bose-Einstein
condensation~\cite{greiner2003,jochim2003,regal2004,zwierlein2004}.
Pairing can also occur in momentum space via
formation of correlated Cooper pairs and the superfluid state would
be analogous to the Bardeen-Cooper-Schrieffer (BCS)-type of a superconducting
state~\cite{zwierlein2005,chin2006}. This is a relatively novel field and not 
treated here; 
part of it has been reviewed by~\textcite{giorgini2008} 
and~\textcite{bloch2008}.

\subsubsection{Electrons in low-dimensional quantum dots}

Quantum dots are man-made nanoscale droplets of electrons
trapped in all spatial directions.
As they show typical properties of atomic systems, such as shell structure
and discrete energy levels, they are often referred to as 
artificial atoms~\cite{ashoori1996}. 
Electron numbers in quantum dots may
reach thousands. Quantum dots are often fabricated in semiconductor
materials, but the use of graphene has also been 
proposed~\cite{trauzettel2007,wunsch2008}.
These nanostructured finite fermion systems have been studied extensively
for (by now) two decades. 
Several review articles, 
discussing the quantum transport through
quantum dots~\cite{vanderwiel2003}, electronic structure~\cite{reimann2002}, 
the role of symmetry breaking and correlation~\cite{yannouleas2007} as
well as spin in connection with quantum
computing~\cite{coish2006,cerletti2005,hanson2007}, were published. 

The semiconductor quantum dots discussed here are 
of either lateral or vertical type. In a lateral device the electrons 
in a two-dimensional electron gas are trapped by external electrodes, 
while vertical dots are formed by, {\it e.g.}, etching out a pillar from a 
wafer containing a heterostructure. In both cases the motion of 
electrons is restricted into a thin disk, with a typical  
radius of few tens up to hundred nanometers, and a   
thickness that is often an order of magnitude smaller.
Electrons in quantum dots can be set rotating by external magnetic
fields perpendicular to the plane of motion.
Other stirring mechanisms have also been proposed, {\it e.g.} rotation in
the electric field of laser pulses~\cite{rasanen2007}.
Due to the band structure of the underlying semiconductor material,
magnetic field strengths giving rise to transitions
in the electronic structure of quantum dots
are orders of magnitude lower than in real atomic systems,
and attainable in laboratories.
Much of the information about the electronic structure 
must be extracted from electron transport
measurements~\cite{oosterkamp1999}. Direct imaging methods of electron 
densities in quantum dots have also been attempted, see for 
example~\cite{fallahi2005,dial2007}, but not yet proven equally useful 
in this context. 

Quantum dots in external magnetic fields
have a very close connection to quantum Hall systems,
the only difference being that the quantum Hall effect is measured
in a sample of the two-dimensional electron gas (2DEG),
which is often modeled as an infinite system. Quantum dots, 
however, are finite-size many-body systems.
At strong magnetic fields, where electrons occupy only the 
lowest Landau level, they are thus often 
referred to as ``quantum Hall droplets''~\cite{oaknin1995,yang2002}.
Many concepts familiar from the theory of the quantum Hall effect,
such as the Landau level filling factor, can be generalized
for these finite-size droplets~\cite{kinaret1992,reimann2002}.
However, due to the presence of the external
confining potential in quantum dots, the analogy to quantum Hall
states in the infinite 2DEG
is not exact and edge effects play an important 
role~\cite{viefers2008,cooper2008}.

\subsection{Vortex formation in rotating quantum liquids}

\begin{figure}[ptb]
\includegraphics[width=.45\textwidth]{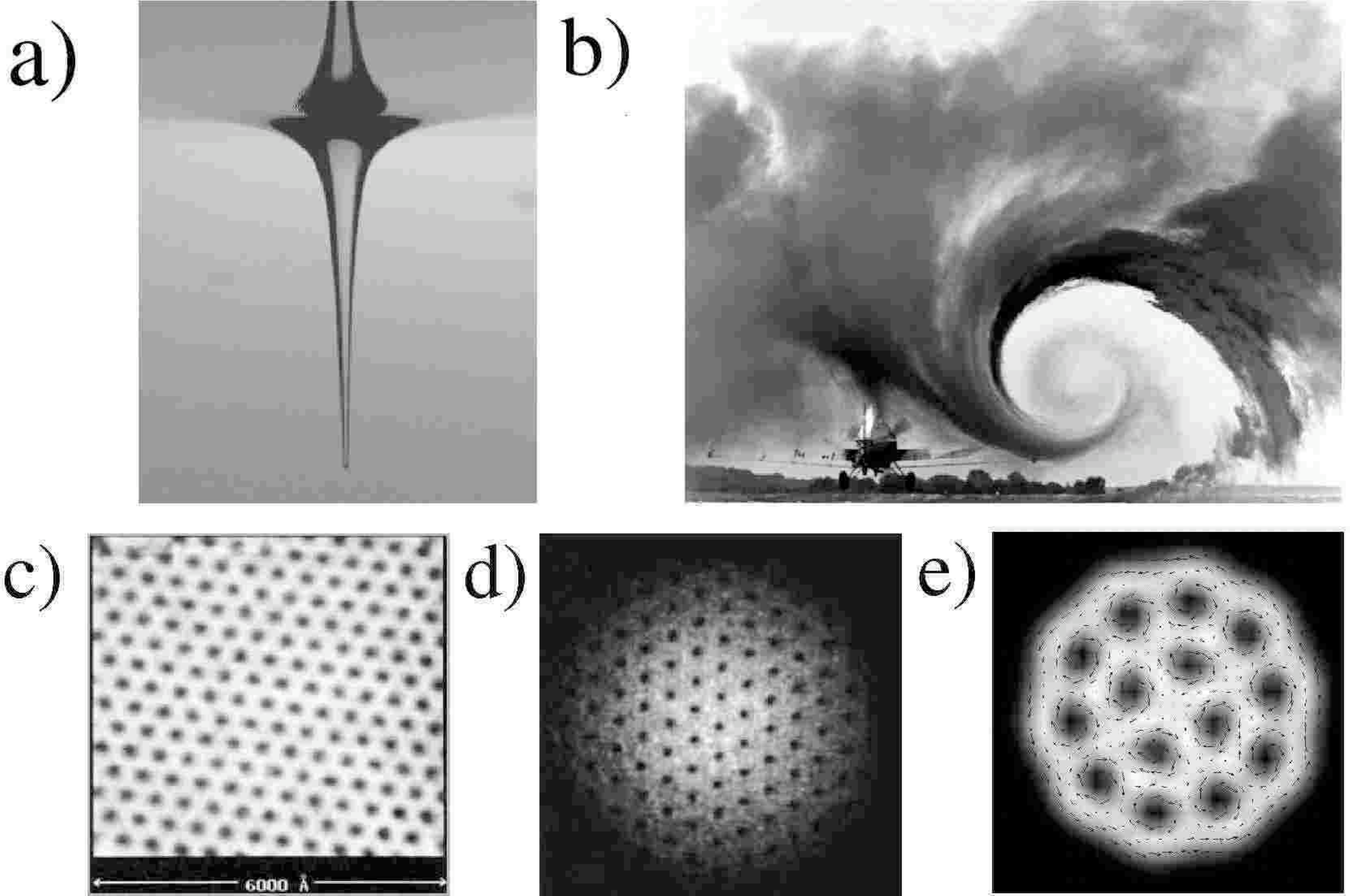}
\caption{Examples of vortices and vortex
lattices. Vortices are ubiquitous in both classical
and quantum systems: a) classical whirlpool vortex~\cite{andersen2003},
b) wake vortex of a passing airplane wing, revealed by colored smoke
(NASA Langley Research Center, Figure ID: EL-1996-00130) 
 c) STM-image of an Abrikosov vortex lattice~\cite{abrikosov57}
in a type-II superconductor~\cite{hess1989}, 
d) vortex lattice in a rotating Bose-Einstein condensate of $^{87}$Rb atoms
(adapted from \textcite{coddington2004}), 
e) cluster of vortices in the calculated electron density of a 24-electron
quantum dot, after~\textcite{saarikoski2004}.
In panels c)-e), the vortices appear as ``holes'' in the particle density.}
\label{overall}
\end{figure}

The formation of vortices in a liquid that is set to rotate is 
often a result of turbulent flow.
In the epic poem ``The Odyssey'', Homer describes
Ulysses' encounter with Charybdis, a monster-goddess who
sucked sea water and created a giant whirlpool~\cite{homer}.
This early account of vortex dynamics is strikingly accurate 
in identifying the characteristics of 
vortices, namely, the rotating current of the whirlpool and
the cavity at the center of the vortex which engulfed the ships
sailing nearby. 
Homer's description may well be illustrated by other examples of 
more harmless vortices, such as whirlpools in bathtubs
where water is draining out~\cite{andersen2003}. Other well-known examples
of vortices in air include tornadoes, or wake vortices created 
by an airplane wing (Figs.\ref{overall}(a) and (b)).

Vortices are ubiquitous also in quantum mechanical 
systems under rotation (see Figs. \ref{overall}(c)-(e)).
It is well known that the magnetic field 
in type-II superconductors penetrates through vortex lines~\cite{tinkham2004}
(see Fig.~\ref{overall}(c)). Superfluid  $^4$He is another 
example where vortices  may form in a strongly interacting bosonic 
quantum fluid~\cite{williams1974,yarmchuk1979,yarmchuk1982}.
(See also the early work 
by \textcite{onsager1949}, \textcite{london1954} and \textcite{feynman1955}, 
and for example the book by \textcite{donnelly1991}.) 
Vortices appear as a very general phenomenon 
in Bose as well as in Fermi systems with high as well as low particle density. 
They may emerge for short-range interactions between the particles,  
as in condensates of neutral atoms
(as shown in Fig.~\ref{overall}(d) for a rotating Bose-Einstein condensate 
of $^{87}$Rb atoms) 
or -- perhaps more surprisingly -- even in electron systems with 
long-range Coulomb repulsion, see Fig.~\ref{overall}(d) showing the vortices
in a quantum dot at a strong magnetic field. 

\subsubsection{Vortices in Bose-Einstein Condensates}

For vortices in rotating Bose-Einstein condensates, 
early theoretical descriptions have set focus on the Thomas-Fermi regime 
of strong interactions, see for 
example~\cite{rokhsar1997,ripoll1999,feder1999a,feder1999b,svidzinsky2000},
as well as weak interactions~\cite{mottelson1999,butts1999,kavoulakis2000}.
\textcite{baym1996} treated vortex lines in terms of the Gross-Pitaevskii
approach, and later on also discussed 
the transition to the lowest Landau level when the
rotation rate was increased~\cite{baympethick2004}. 
   
Intense experimental research efforts were made to 
observe vortices in rotating clouds of bosonic atoms, 
see {\it e.g.}, the early experimental work by~\textcite{matthews1999}, 
as well as~\textcite{madison2000}, \textcite{aboshaer2001}, 
\textcite{engels2002,engels2003},  
and \textcite{schweikhard2004}.
For recent reviews, we refer to the articles by~\textcite{fetter2008}, 
as well as ~\textcite{bloch2008}. 

In weakly interacting and dilute systems, an effective reduction of dimensionality
can for example be caused by rotation as a simple consequence of the increase
in angular momentum. 
Due to the reduction in dimensionality,
phase  singularities, {\it i.e.}, nodes in the wave functions, become
important. 

With increasing angular momentum, one finds successive transitions between 
patterns of singly-quantized vortices, arranged in regular arrays.
In finite-size systems, so-called ``vortex molecules'' 
are formed, in much analogy to finite-size 
superconductors~\cite{milosevic2003}.

There exist many analogies of a rotating cloud of bosonic atoms 
with (fractional) quantum Hall physics 
\cite{wilkin1998,cooper1999,viefers2000,ho2001}. This in fact  
may also give important theoretical insights into the regime of 
extreme rotation which has not yet been achieved
experimentally. (For related reviews, 
see~\textcite{cooper2008,viefers2008} and \textcite{fetter2008}).

\subsubsection{Vortices in quantum Hall droplets}

Vortices have been an integral part of the theory of quantum Hall states
in the 2D electron gas since the proposal of the Laughlin 
state~\cite{laughlin1983}. They  emerge also in 
quantum dots~\cite{saarikoski2004,toreblad2004} at strong magnetic fields, 
and close connections of these vortices to those
that can be found in rotating bosonic systems have been  
established~\cite{toreblad2004,toreblad2006,manninen2005,borgh2008}.
The vortex patterns in quantum dots depend on the 
strength of the external magnetic field, and on intricate details of particle
interactions~\cite{saarikoski2004,tavernier2004}.

In the regime of slow rotation, vortices 
(except those originating from the Pauli principle)
are not bound to particles 
and form charge deficiencies in the density distribution, 
which may localize to structures in the particle and 
current densities that resemble the aforementioned vortex molecules or 
regular vortex arrays in rotating Bose-Einstein 
condensates~\cite{saarikoski2004,saarikoski2005a,manninen2005}.
The emergence of vortices  that carry the angular momentum of the droplet 
is manifest in the structure of the many-body states. For fermions they  
may be described as hole-like quasiparticles~\cite{manninen2005}.
When the number of vortices increases with
the angular momentum, the electrons and vortices may form
composites well known from the theory of
the fractional quantum Hall effect, see for 
example~\textcite{jain1989} or \textcite{viefers2008}. 

\subsubsection{Quantum Hall regime in bosonic condensates}

In quantum dots, the fractional quantum Hall regime with a 
high vortex density  
can be readily attained at high magnetic fields.
For the case of rotating cold atom condensates,
despite extensive experimental studies~\cite{coddington2003,schweikhard2004},
this regime of extreme rotation is not yet within easy reach.
Very recently, however, it was suggested to  
exploit the equivalence of the Lorentz and the Coriolis force 
to realize ``synthetic'' magnetic fields in rotating neutral systems, 
which could be a very important step forward
in the efforts to realize BEC's at extreme rotation~\cite{lin2009}. 
To date, experiments with rotating BEC's
are only able to access states where the number of vortices is relatively 
small compared to the number of particles~\cite{matthews1999,madison2000,aboshaer2001,engels2002,engels2003,schweikhard2004,fetter2008}.
A high vortex density creates a highly correlated state.
Counterparts of typical quantum Hall states, such as the bosonic Laughlin
state and other incompressible states, as well as states having
non-Abelian particle excitations,
are predicted to emerge~\cite{wilkin1998,cooper1999,viefers2008,lin2009}.
Compared to the quantum Hall systems in
the 2D electron gas, rotating cold-atom 
condensates offer a high level of tunability
since particle interactions and trap geometries can be easily modified.
This makes bosonic quantum Hall states an extremely interesting
field of research~\cite{viefers2008,cooper2008}.

\subsubsection{Self-bound droplets}

A common feature of all the systems discussed above is that
the particles are bound 
by an external confinement, which often can be approximated to be 
harmonic. Nuclei, helium droplets and atomic clusters 
provide other interesting finite quantum systems where rotational states
have been studied. These systems are self-bound due to attractive
interactions between (at least some of) the components. 

Rotational states, shape deformations and fission of self-bound 
droplets are interesting topics in their own right.
However, while in a harmonic confinement the fast rotation 
causes the droplet to flatten into a quasi-two-dimensional 
circular disk, this is usually not the case in self-bound clusters,
where the rotation can be accompanied with a noncircular deformation,
often a two-lobed or even more complicated shape~\cite{hill2008}.
Eventually this can lead to a fission of the droplet to smaller fragments,
preventing the occurrence of very large angular momenta and vortex formation.
In the case of nuclei, the rotational spectrum is usually 
related to deformation~\cite{bohrmottelson}. 
Nevertheless, the possibility of vortex-like excitations has also been discussed, see \cite{fowler1985}, and nuclear matter is expected to carry vortices in 
neutron stars~\cite{baym1969,link2003}.

The only small self-bound system where vortices are likely to occur, is a
helium droplet. \textcite{grisenti2003} indicate that
anomalies in their cluster beam experiments could be caused by vortex
formation. However, no clear experimental evidence of vortex formation 
in small 
helium droplets has yet emerged, while theoretical 
studies suggest that vortices form in $^4$He 
nanodroplets~\cite{mayol2001,Lehmann2003,sola2007}. The properties of helium
nanodroplets have been recently reviewed by~\textcite{barranco2006}.

\subsection{About this review}

The main concern of this review are the structural properties
of the many-body states of small two-dimensional quantum droplets, 
where rotation induces strong correlations and vortex formation.
The direct connections between bosonic and fermionic systems,
as well as finite-size quantum droplets and infinite quantum Hall
systems are recurrent themes.
Other reviews complement our work by taking different approaches:
We refer to~\textcite{fetter2008} 
for a review of rotating BEC's especially in the regime which is
accessible with present day experimental setups, and
to~\textcite{viefers2008} for a review which focuses on the 
quantum Hall physics in rotating BEC's. Another 
recent review by ~\textcite{cooper2008}
describes rotating atomic gases 
in both the mean-field and the strongly correlated regimes.
A review on the many-body phenomena and correlations 
in dilute ultra-cold gases that also
discusses rotation, was recently published by~\textcite{bloch2008}. 

Quantum dot physics is a versatile field. We 
refer to~\textcite{reimann2002} and~\textcite{yannouleas2007}, as well 
as~\textcite{vanderwiel2003} and \textcite{hanson2007} 
for reviews on the electronic structure and spin-related phenomena. 
Vortices in superconducting quantum dots have also been much discussed in the
literature, but are not treated here. We instead refer the 
reader for example to the more recent 
articles by ~\textcite{baelus2001}, \textcite{baelus2002}, 
\textcite{baelus2004} and \textcite{grigorieva2006}.

We begin this review in Sec. \ref{sec:wave-function} by introducing
basic concepts to characterize the many-body states of 
rotating systems.
Section III discusses some of the computational
many-body methods used.
Section \ref{sec:single-component} discusses vortex formation in rotating
quantum liquids which are composed of one type
of particles (or one spin component), while 
Section \ref{sec:multi-component} is concerned with coreless vortices in
multi-component systems.
We conclude the review and discuss
possible future challenges in Sec. \ref{sec:summary}.

(Unless stated otherwise, equations are presented in SI units 
whereas most results of calculations are in atomic units.) 

%+++++++++++++++++++++++++++++++++++++++++++++++++++++++++++++++++++++

\section{Many-body wave function}

\label{sec:wave-function}

In the following, we briefly describe concepts and  methods  
to analyze the internal structure of the many-body states, 
such as pair-correlation functions and conditional probabilities.
We then proceed to show the connections between boson and fermion
states, and particle-hole duality that treats vortices as
hole-like quasi-particles. We finally give a brief overview of the  
connections to the quantum Hall physics in the (infinite) 
two-dimensional electron gas. 

\subsection{Model Hamiltonian}

\subsubsection{Rotating quantum droplets of bosons}

\label{sec:bosonicdroplets}
Clouds of bosonic condensates are usually confined by a harmonic trap that 
extends in all three spatial dimensions. 
An axisymmetric rotation with frequency $\Omega$
leads to centrifugal forces which
flatten the density by extending the radial size of the system, while
the cloud contracts in the axial direction.
The ratio between the axial thickness $R_z$
and radial thickness $R_{\perp }$
of the rotating cloud, {\it i.e.}, the aspect ratio, can be calculated within the 
Thomas-Fermi approximation \cite{fetter2008}
\begin{equation}
{R_z \over R _{\perp }}= {\sqrt{\omega_{\perp }^2 - \Omega^2} \over \omega_z},
\label{aspectratio}
\end{equation}
where $\omega_z$ and $\omega _{\perp }$ are the radial and axial trapping frequencies, respectively.
Imaging of the condensate~\cite{raman2001,schweikhard2004} confirms that the
rotation reduces the aspect ratio effectively.

With the trap rotating at an angular velocity 
$\Omega $, in the laboratory frame of reference the problem is 
time-dependent. One thus conveniently 
introduces a rotating frame at the angular 
velocity $\Omega $, 
%by the unitary transformation
%\begin{equation}
%{\cal U} (t) = \hbox{e} ^{i {\bf \Omega }\cdot {\bf l} t/\hbar } 
%\end{equation}
%where ${\bf l}$ is the angular momentum of the single atom. 
%Thus, in the rotating frame 
in which the (now time-independent) Hamiltonian contains an extra inertial 
term $-\Omega  {\bf L}$, where ${\bf L}$ is the total angular momentum
operator. 

In the case of circular symmetry of the 2D system, 
for its rotation around the $z$-axis, the angular momentum operator 
$L=L_z$ commutes with the Hamiltonian.
We may write
\begin{equation}
H_{\Omega}=H-\Omega  L_z,
\label{homega}
\end{equation}
where the many-body Hamiltonian in the rotating frame is
\begin{equation}
H=\sum^N_{i=1} \left ( {{\bf p}_i^2 \over 2m} +
{V}_{\rm ext}({\bf r}_i) \right ) + \sum_{i<j} V^{(2)}({\bf r}_i-{\bf r}_j)~.
\label{rotationhamiltonian}
\end{equation}
Here $V_{\rm ext} $ is the trapping potential that is usually harmonic with
oscillator frequency $\omega $, 
\begin{equation}
V_{\rm ext}={1\over 2} m\omega ^2{r}^2~,
\label{eq:harmonic}
\end{equation}
and $V^{(2)}$ is the two-body interaction between the trapped atoms. 

The ground states of Hamiltonian~Eq.~(\ref{homega})  
are then angular momentum eigenstates of 
Hamiltonian~Eq.~(\ref{rotationhamiltonian}) which have the lowest energy 
at some finite frequency of rotation $\Omega $. 

The effective interaction between the bosons is often assumed to be a contact 
interaction of zero range, 
\begin{equation}
V^{(2)}({\bf r}_i-{\bf r}_j) = {1\over 2} g\sum _{i\ne j} 
\delta ({\bf r}_i - {\bf r}_j)~, 
\end{equation}
where $g=4\pi \hbar ^2 a/M$, with atom mass $M$ and  
$a$ being the scattering length for elastic $s$-wave collisions between the
atoms.  
In the regime of weak 
interactions, $gn \ll \hbar \omega $, where $n$ is the particle
density and $\hbar \omega $ the quantum energy of the confining potential.
In a rotating system, the problem becomes effectively
two-dimensional when $gn$ is much smaller than the energy difference between 
the ground and first excited state for motion along the $z$-axis. 

The single-particle energies of the two-dimensional harmonic oscillator are 
$\epsilon = \hbar \omega (2n+|m|+1),$
where $n$ is the radial quantum number, and $m$ the single-particle angular 
momentum. In a non-interacting rotating many-particle system, consequently, 
the lowest-energy configuration is characterized by quantum numbers $n=0$, 
and $0\le m\le L$, where $m$ has the same sign as the angular momentum $L$. 
This single-particle basis is identical to the lowest 
Landau level (LLL) at strong magnetic fields. 
In this subspace, a configuration can be denoted by the Fock state
$\vert n_0 n_1 n_2 \cdots n_{m}\cdots n_L\rangle $, 
where $n_i$ is the (here bosonic) 
occupation number for the single-particle state with angular momentum
$m$, and $m=L$ is the largest single-particle angular momentum that can be 
included in the basis. 
As the angular momentum $L$ is a good quantum number, 
we have the restriction $\sum_m m n_m=L$. 

For a harmonic trap, 
there is a large degeneracy in the absence of interactions, 
which originates from the many different ways to distribute the $N$ bosons 
on the basis states with $0\le m \le L$~\cite{wilkin1998,mottelson1999}.  
Interactions break this degeneracy, and a particular state can be selected 
at a given $L$ that minimizes the interaction energy.
With reference back to the nuclear physics terminology, the highest angular 
momentum state at a given energy is called the {\it yrast} state~\cite{grover1967,bohrmottelson},  
the name originating from the Swedish word for ``{\it the most
dizzy}''. The line connecting the lowest energy
states in the energy-angular momentum diagram is consequently called 
the {\it yrast line}. 

For interacting particles, the yrast line is a non-monotonic function
of the angular momentum. At angular momenta corresponding to the 
ground states at a certain trap rotation frequency $\Omega $, 
it shows pronounced cusps reflecting the vortex 
structures of the system, as it will become clear later on.

\subsubsection{Electron droplet in a magnetic field}

We focus here on droplets of electrons trapped  in
a quasi-two-dimensional quantum dot~\cite{reimann2002}.
The spatial thickness of the confined electron droplet is of the order of 
nanometers for typical quantum dot samples. Electrons
in quantum dots are rotated, not by mechanical stirring, but
instead by applying an external magnetic field perpendicular to the dot
surface ({\it i.e.} along the $z$-axis) quite analogously to the
circular motion in a cyclotron.

A droplet of electrons in a quantum dot can be modeled using an 
effective-mass Hamiltonian in the $x$-$y$--plane, 
\begin{equation}
H=\left(\sum^N_{i=1} \frac{(-i\hbar \nabla_i+e {\bf A} )^2}{2 m^*}
+V_{{\rm ext}}(r_i)\right) + \frac{e^2}{4\pi \epsilon} \sum_{i<j}
\frac{1}{r_{ij}} \ ,
\label{hamiltonian}
\end{equation}
where $N$ is the number of electrons, 
$m^*$  and $\epsilon$ are the effective mass and dielectric 
constant of the semiconductor material, 
${\bf A}$ is the vector potential of the magnetic field,
${\bf B}=\nabla \times {\bf A}$, and the Zeeman term
has been omitted. 
The external confining potential $V_{\rm ext}$
is usually parabolic to a good accuracy~\cite{matagne2002}.
The single-particle states in 
the external harmonic potential Eq.~(\ref{eq:harmonic})
are known as Fock-Darwin states~\cite{fock1928,darwin1930}.
At strong magnetic fields the magnetic confinement
dominates over the electric confinement, 
and the Fock-Darwin states bunch to Landau levels, 
as described above for the case of rotation.
The LLL is then the most important subspace
for ground state properties of the system.

Using a symmetric gauge ${\bf A}=B(y{\hat e}_x-x{\hat e}_y)/2$
the first term in the Hamiltonian (\ref{hamiltonian})
can be expanded to give two terms that are proportional to the magnetic field.
The diamagnetic term  
is scalar, $e^2B^2/(8m^*)(x^2+y^2)$, and the other, the paramagnetic term, 
is proportional to the $z$-component of the angular momentum
$e\hbar/(2m^*i)B{\bf r}\times\nabla=e/(2m^*)BL_z$.
The scalar term depends on
the square radius from the center of the droplet and describes
the squeezing effect of the magnetic field. The latter term 
lowers the energy of the states that
circulate in the direction of the cyclotron motion, 
and favours alignment of the magnetic moment parallel to the external
magnetic field. 
By combining the diamagnetic term in the Hamiltonian 
Eq.~(\ref{hamiltonian}) with the external
confining potential and writing the paramagnetic term as
$e/(2m^*)BL_z=\Omega L_z$ we see directly that, except for the Zeeman
term and the type of interparticle interactions, the Hamiltonian
is exactly the same as that for a rotating bosonic system 
(\ref{rotationhamiltonian}).
The rotation corresponds
to a magnetic field strength of ${\bf B}=(2m^*\Omega /e){\hat e}_z$
in a weaker confinement
$V'_{\rm ext}={1\over 2} m^*(\omega_0^2-\Omega^2){\bf r}^2$.
This constitutes a close analogy between systems in mechanical rotation
and systems of charged particles in a perpendicular magnetic field.

\subsubsection{Role of symmetry breaking}
\label{sec:symmetry}
Even though the microscopic Hamiltonian 
often obeys certain symmetries, such as
rotation and translation, macroscopic systems may spontaneously
break these symmetries in order to attain lower energy and higher
order. In the thermodynamic limit, mean-field theories incorporating
order parameters can describe states with broken symmetries.
However, the exact wave function of the many-body system
must always preserve the underlying symmetry of the Hamiltonian.

Construction of a symmetry-broken state and a subsequent
restoration of symmetry has been proposed to construct 
wave functions in rotating,  
correlated many-particle systems~\cite{yannouleas2007}.
By construction, this approach focuses on the role of particle ordering
in the confining trap potential.
On the other hand, small perturbations in the symmetric potentials
can be used to probe the internal structure of the many-body states.
For vortices in small quantum droplets, 
this may be achieved effectively by 
using point perturbations,  or deforming
the external field slightly~\cite{saarikoski2005a,christensson2008b,parke2008,dagnino2009a,dagnino2009b}.

\subsection{Vortices in the exact many-body wave function}
\label{vortex-many-body}

Vortices in a complex-valued wave function are associated with 
phase singularities. They are manifested through a phase change 
of a multiple of $2\pi$ in every path encircling the singularity. 
The phase is not defined at the singularity, which means that the 
wave function must vanish at this point. The particle deficiency 
in the vicinity of the singularity gives rise to the vortex core. 
Different types of phase singularities can be
recognized: {\it (i)} those which are related to the antisymmetry of the fermion wave
function, {\it (ii)}
those which are largely independent of particle positions
and may be called isolated or free vortices (and occur for bosonic as well as
fermionic systems in a rather similar way),
and {\it (iii)} those which are attached 
to particles to form a bound system, {\it i.e.}, a  ``composite'' particle.

\subsubsection{Pauli vortices}

Exchange of two identical, indistinguishable bosons or fermions can change 
the wave function of the system at most by a factor $C=\pm 1$
so that $\Psi(\dots,{\bf r}_i,\dots,{\bf r}_j,\dots)
=C\Psi(\dots,{\bf r}_j,\dots,{\bf r}_i,\dots)$.
In the 2D plane, making two exchanges (with a total phase change of $2\pi$) 
is equivalent to rotating the particles in-plane with respect to each other.
In the LLL this phase change implies
that there is a vortex attached to the electron
(see Fig. \ref{sepitys}b below).
This vortex (related to the fermion antisymmetry) is called a 
``Pauli vortex'' (or as in quantum chemistry, also the ``exchange hole''). 
As a trivial consequence,  
a delta-function type interparticle interaction does not have any
effect on fermions with the same spin.

\subsubsection{Off-particle vortices}

Vortices that are not attached to any particles are called ``off-particle''
vortices. These elementary excitations may occur in boson as well as
in fermion systems. 

For the two-dimensional electron gas,  
off-particle vortices have been extensively
studied in connection with the quantum Hall effect, both for the bulk and  
in finite-size quantum dots.
The connection between the wave function phase and the vorticity
in such systems can most easily be seen by 
using the vector potential ${\bf A}({\bf r})$
of the magnetic field, that couples to the momentum operator in
the Hamiltonian, Eq.~(\ref{hamiltonian}).
A finite magnetic field leads to an extra phase change of
$\Delta\theta=e/\hbar \int_{\rm A}^{\rm B} {\bf A}({\bf r}) \cdot {\rm d\bf r}$
when the electron moves from A to B.
In a closed path in the 2D plane 
the phase shift must be $2\pi l$, where $l$ is an integer,
which causes the magnetic field to penetrate the 2D plane as vortices
carrying magnetic flux quanta $\Phi_0=h/e$. The integer $l$ is called
the winding number or vortex multiplicity ($l=0$ means no vortex).

\subsubsection{Particle-vortex composites}

When the total angular momentum (and thus 
also the number of vortices) increases, the correlations favour the 
attachment of additional vortices to the particles.
This is well established in the 2DEG, where it leads to  Laughlin type 
quantum Hall states at high magnetic fields. 
These states are discussed in Sec.~\ref{sec:qhstates} below.
Analogous Laughlin states are predicted to form also in
rotating bosonic 
systems~\cite{wilkin1998,cooper1999,wilkin2000,cooper2001}.
In general, the wave function antisymmetry requires that 
fermions must have an odd number of vortices attached to them, while
bosons have an even number of vortices.

In multi-component systems particle deficiency
associated with off-particle vortices in one component
may attract particles of other components. In finite-size 
quantum droplets this is usually energetically favourable. 
The structures that form are called ``coreless vortices'',
since vortex cores are filled by another particle component,
but the singularities in the phase structure remain.
Coreless vortices will be analyzed further in
Sec.~\ref{sec:multi-component}.

\subsection{Internal structure of the many-body states}

The {\it exact} many-particle wave-function
is in many cases known only as a numerical approximation, with the  
complexity growing exponentially with the particle number $N$.
Its dimensionality must be reduced to
allow visualization of the correlations and phase structures, since 
symmetries of the underlying Hamiltonian often hide the 
internal structures in the exact many-body state. 
Thus, pair-correlation functions and reduced wave functions
are often applied. The former
has been a standard tool in many-body physics for many years.
The latter, on the other hand, is more suitable to visualize 
the phase structure of the wave function and its singularities.

\subsubsection{Conditional probability densities}

The pair-correlation function is a conditional probability density
describing the probability of finding a particle at a position ${\bf r}$ 
when another
particle is at a position ${\bf r}'$. For systems with only one
kind of indistinguishable particles, one may write
\begin{eqnarray}\label{onecomp_paircorr}
P({\bf r},{\bf r}')&=&\langle \Psi \mid
\hat n ({\bf r}) \hat n  ({\bf r} ' )
\mid \Psi \rangle \\
  &=&\int\vert\psi({\bf r},{\bf r}',{\bf r}_3,\cdots,{\bf r}_N)\vert ^2
d{\bf r}_3\cdots d{\bf r}_N \nonumber
\end{eqnarray}
where $\mid \Psi \rangle $ is the many-body state, $\hat n$
the density operator and $\psi$ the many-body wave function.
For particles with spin (or another internal degree of freedom, as for example
in the case of different particle components), labeled by an index $\sigma $, 
the pair-correlation function is correspondingly 
defined as 
\begin{equation}
P_{\sigma , \sigma '  }({\bf r}, {\bf r}' ) = \langle \Psi \mid
\hat n _{\sigma } ({\bf r}) \hat n _{\sigma ' } ({\bf r} ' )
\mid \Psi \rangle,
\label{twocomp_paircorr}
\end{equation}
where
$\hat n _{\sigma } $ and $\hat n _{\sigma \prime } $ are the density
operators for the components.

In a homogeneous system $P$ depends only on the distance $\vert {\bf
r}-{\bf r}'\vert$
while in a finite system this is not the case. Instead, 
one has to choose a reference point ${\bf r}'$ around
which the pair-correlation function may then be 
plotted as a function of ${\bf r}$.
The details of the pair-correlation in finite systems are very sensitive to
the selection of this reference point. 
The inherent arbitrariness in choosing the off-centered
fixed point must be taken care of by sampling over a range
of values for ${\bf r} ' $ to allow any reasonable interpretation.
Usually, a position that does not coincide with any symmetry point  
and where the density of the system is at a maximum, 
gives the most informative plot.
Note, however, that in fermion systems the pair-correlations at short
distances are strongly dominated by the exchange-correlation hole of the 
probe particle, which may complicate the analysis. 

\subsubsection{One-body density matrix}

The one-body reduced density matrix is defined as
\begin{equation}
n^{(1)}({\bf r},{\bf r'})=
\langle \Psi\vert \hat\psi^\dagger({\bf r})\hat \psi ({\bf r'})\vert\Psi\rangle,
\end{equation}
where $\hat\psi^\dagger$ and $\hat\psi$ are field operators (with given
statistics),  creating and annihilating a particle.  
The eigenfunctions $\psi _i$ 
and eigenvalues $n_i$ of the density matrix are solutions 
of the equation 
\begin{equation}
\int d{\bf r'}n^{(1)}({\bf r},{\bf r'})\psi_i^*({\bf r'})=n_i\psi_i^*({\bf r}).
\end{equation}
For a noninteracting system, the eigenfunctions are simply the single-particle
wave functions, while the eigenvalues give the occupation numbers.
For interacting bosons, it is suggestive that 
the exact eigenstate corresponding to the highest 
eigenvalue ($n_1$) of the density matrix plays the role 
of a ``macroscopic wave function'' (order parameter) of the Bose 
condensate. This connection was established already many decades
ago in the context of off-diagonal long-range 
order~\cite{ginzburg1950,landaulifshitz1951,penrose1951,penrose1956,yang1962,pitaevskiistringari2003,pethicksmith2002}.
For a discussion of fragmentation~\cite{leggett2001} in this context, see for example
\textcite{baym2001}, \textcite{mueller2006} and \textcite{jackson2008}.  

Since the eigenstates of the density matrix can be complex, their phase can 
show singularities as they are characteristic for vortices. 
However,  the density matrix bears the same 
symmetry as the Hamiltonian and, consequently, so do its so-called 
``natural orbitals'' $\psi_i^*({\bf r})$. In a circular confinement, 
the eigenfunctions of the density matrix can thus only show an overall phase
singularity at the origin, but not at the off-centered vortex positions.

In a study of vortex formation in boson droplets this problem has been 
circumvented by adding a quadrupole perturbation to the confining
potential \cite{dagnino2007,dagnino2009a,dagnino2009b}. 
Indeed, then the positions of all vortices are seen as phase 
singularities of the complex ``order parameter'' 
$\psi_1^*({\bf r})$. With a related symmetry breaking of the 
external confinement, the vortices may also be seen as minima in the total
particle density~\cite{toreblad2004,saarikoski2005b,dagnino2007}, 
and as circulating currents as shown for example 
in Fig.~\ref{fig:edellipse} below.

\subsubsection{Reduced wave functions}

\label{sec:cond-wave}
Pair-correlation functions smoothen out the finer 
details of the many-particle wave function.
As real-valued functions, they are not suited to probe
the phase structure, and  
zeros (nodes) at the center of the vortex cores cannot be
directly identified either, since
integrations over particle coordinates blur their effect.
The concept of a reduced (or conditional) wave function has thus 
been introduced
to map out the nodal structure of the wave function as a ``snapshot''
around the most probable particle configuration.
For fermions, reduced wave functions were introduced 
in the context of two-electron
atoms~\cite{ezra1983} and coupled quantum dots~\cite{yannouleas2000},
and then generalized to many-particle
systems~\cite{harju2002b,saarikoski2004,tavernier2004}.
The basic idea is simple:
Instead of calculating average values, the wave function
is calculated in a subspace by fixing $N-1$ particles
to positions given
by their most probable configuration ${\bf r}^*_{2},\ldots, {\bf r}^*_{N}$.
The reduced wave function for the remaining (probing) particle
is then calculated at $\mathbf{r}$,
\begin{equation}
\psi_\mathrm{c}(\mathbf{r})=
\frac{\Psi(\mathbf{r},\mathbf{r}_2^*,\dots,\mathbf{r}_N^*)}
{\Psi(\mathbf{r}_1^*,\mathbf{r}_2^*,\dots,\mathbf{r}_N^*)} \,
\label{condwave}
\end{equation}
where ${\bf r}_1^*$ is the most probable position of the probe particle 
and the denominator is used to normalize the maximum value of
$\psi_\mathrm{c}$ to unity.
The most probable configuration for fixed particles 
$(\mathbf{r}_1^*,\mathbf{r}_2^*,\dots,\mathbf{r}_N^*)$
is obtained by maximizing the absolute square of $\psi _c$.

\begin{figure}
\includegraphics[width=.45\textwidth]{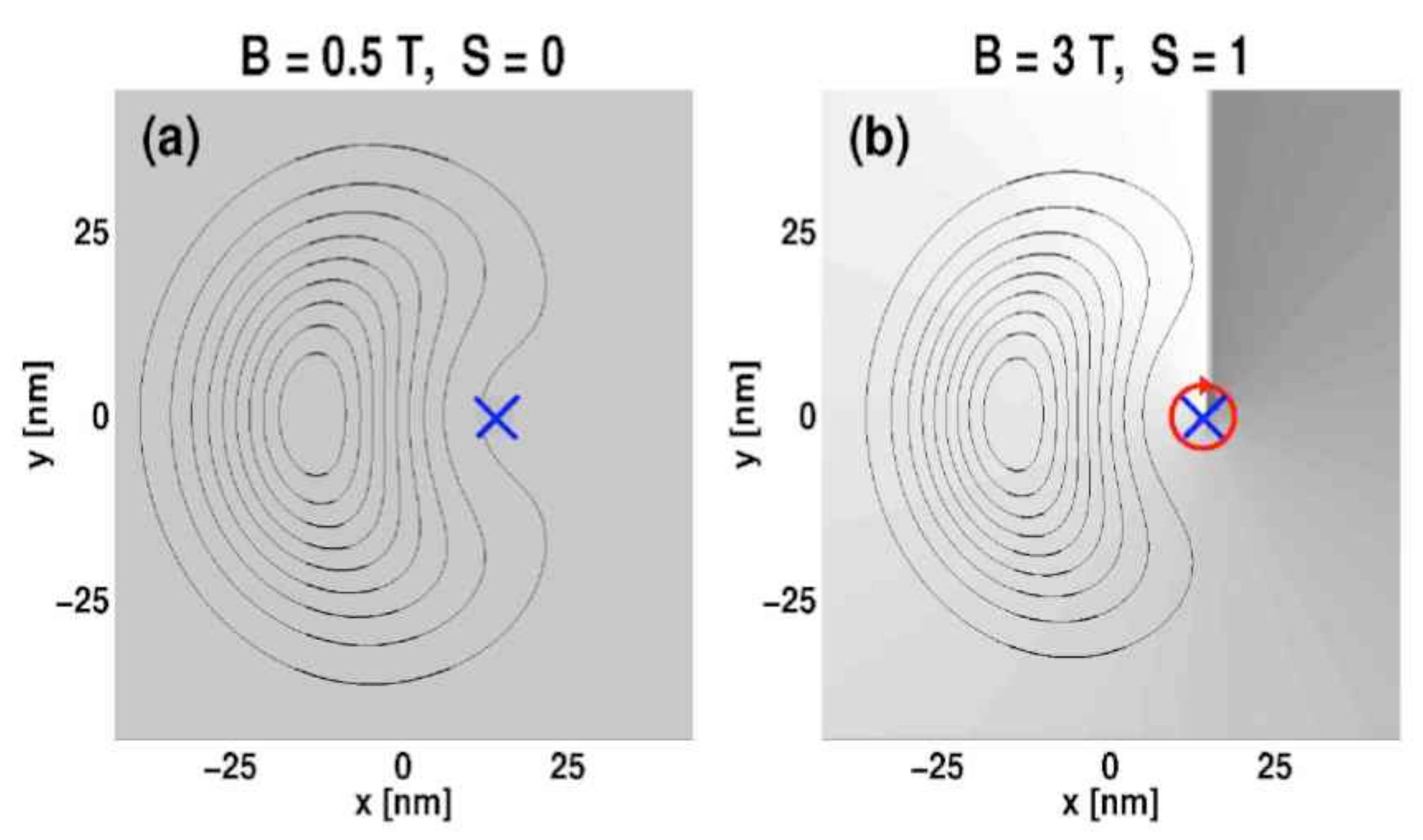}
\caption{Reduced wave function of a two-electron
quantum dot in (a) the singlet and (b) the triplet states.
The fixed electron is marked by the cross to the right.
The contours give the logarithmic electron density of the probing
electron and the gray scale illustrates the phase of the wave function.
The phase jumps from $0$ to $2\pi$ on the line where the scale
changes from white to darkest gray.
In the singlet state, the electrons have opposite spins and there is no
vortex. 
In the triplet state, the electrons have same spin and a vortex (circle
with an arrow in the direction of phase gradient)
is attached on top of the fixed electron in accordance with
the Pauli principle. Due to fermion antisymmetry
the phase changes by $2\pi$ if the probe electron is moved around the
fixed electron in this case. From~\cite{harju2005}.
}
\label{fig:condwave}
\end{figure}

It is often convenient to visualize $\psi_\mathrm{c}(\mathbf{r})$
by plotting its absolute value using
contours, usually in a logarithmic scale, together with its phase
as a density plot.
The resulting diagram represents a single-particle wave function
in a selected ``particle's-eye-view'' reference frame.
Nodes in the wave function can be identified
as zeros in $\psi_\mathrm{c}(\mathbf{r})$ associated with a phase change of
integer multiple of $2\pi$ for each path that encloses the zero.
Fig. \ref{fig:condwave} demonstrates the reduced wave function
in the simple case of a two-electron quantum dot in the spin singlet and
triplet state, respectively.  One electron position is fixed, as marked by the 
cross. 
In the singlet state, the electrons have opposite spins 
and there is no vortex. In the triplet, a vortex is attached to the fixed
electron in accordance with the Pauli principle.  

In the case of larger particle numbers, 
interpretation of the reduced wave function requires a 
careful analysis, since nodes for different reference frames of fixed particles
may not coincide~\cite{graham2003}.
However, localized nodes can be readily identified as
vortices. These include off-particle vortices, which
are associated with holes in the particle density.
Also particle-vortex composites can be identified
as nodes attached to the immediate vicinity of particles.

\begin{figure}
\includegraphics[width=0.45\textwidth]{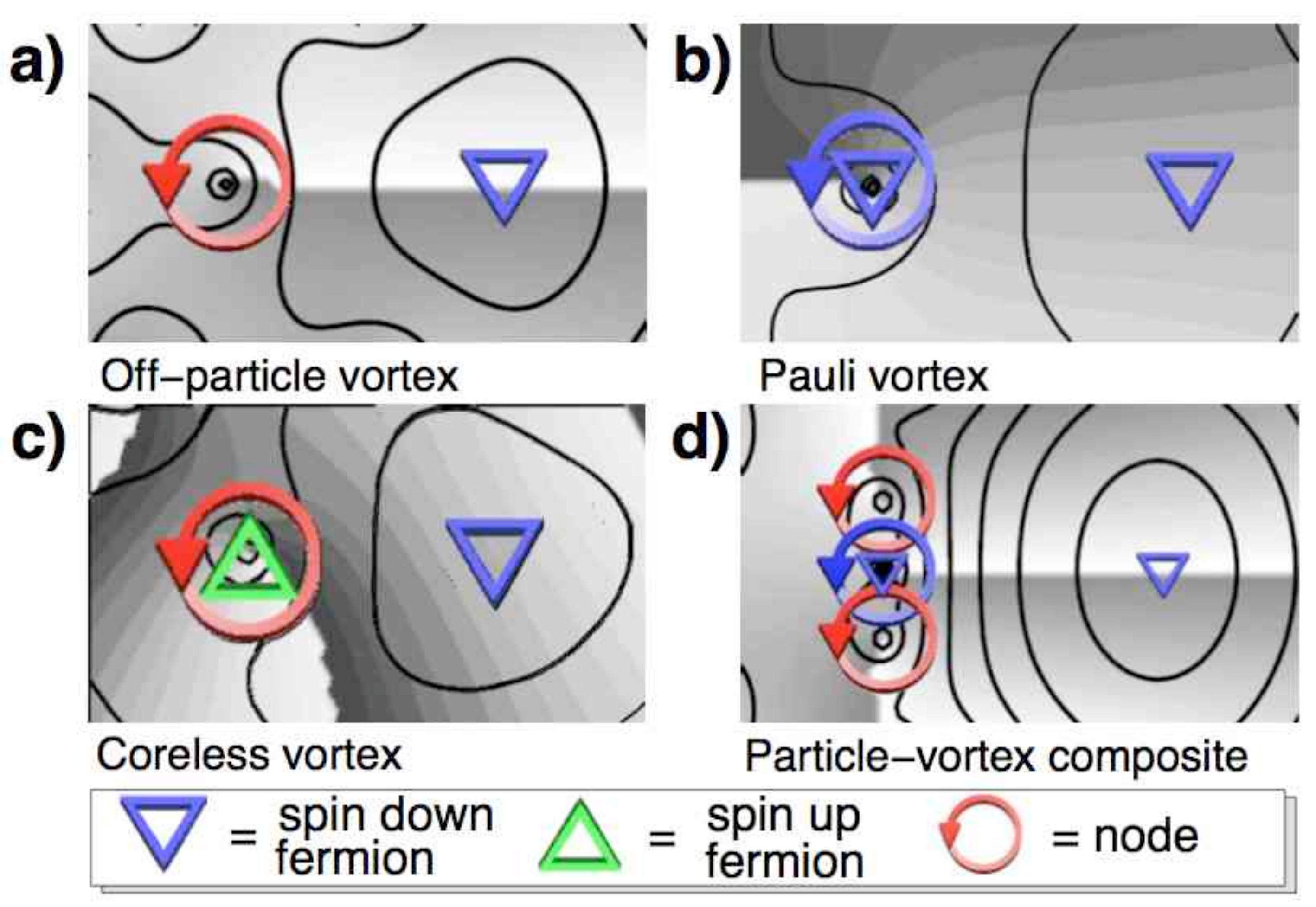}
\caption{({\it Color online}) Appearance of vortex structures in the reduced
wave function. The figures show details of reduced
wave functions for spin-$1/2$
fermions. The most probable position of the probing particle is to the right;
the contours show the magnitude (on a logarithmic scale), and the 
gray-scale shows the phase (darkest gray = $0$, lightest gray = $2\pi$).
a) An isolated, localized vortex which is
not attached to any particle. b) A Pauli vortex (exchange hole)
which is mandated by the wave function antisymmetry between interchange of
indistinguishable fermions. c) A coreless vortex where the vortex core
of spin-down component is filled by a spin-up fermion.
d) A composite of a fermion (with a Pauli vortex) and
two additional nodes which are bound to the particle, reminiscent
of the Laughlin $\nu={1\over 3}$ state. From~\cite{saarikoski2009}.
}
\label{sepitys}
\end{figure}

The reduced wave function as defined for single-component systems
in Eq.~(\ref{condwave}) can be readily generalized also for
multi-component systems with two or more particle species distinguishable
from each other. The wave function is then a direct product of the 
wave functions of different particle species. As a consequence, 
the reduced wave function can still be 
written as in Eq.~(\ref{condwave}), although different particle species
have to be distinguished.
The reduced
wave function depends on the species of the probe particle,
unless the number of particles of each species is equal.
The fact that  phase singularities of one
species coincide 
with particles of another species (see Fig.~\ref{sepitys}c)
indicates formation of coreless vortices. This is
discussed further in Sec.~\ref{sec:multi-component}.
As an example, 
Figure \ref{sepitys} exemplifies the appearance of the  
reduced wave functions for different nodal
structures, as here for fermions with spin-${1\over 2}$.
Correlations in the many-body state can be further studied 
by analysing the reduced wave function in the vicinity of the most probable
configuration(s).

\subsection{Particle-hole duality in electron systems}

\label{sec:duality}
In infinite quantum Hall liquids, particle-hole duality
can be used to study vortex formation by interpreting
holes as vortices~\cite{girvin1996,shahar1996,burgess2001}.
Similar arguments for the symmetry of
particle and hole states can be used in finite-size systems
to gain insight into issues like vortex localization and fluctuations.
We will here consider polarized electrons or, more generally, 
fermions of only one kind ({\it i.e.}, spinless fermions). 
However, much of the considerations 
can be generalized to systems with more degrees of freedom, such as for
example, spinor gases.

In the occupation number representation, 
the Hamiltonian for interacting electrons in the lowest
Landau level can be written as 
\be
H_p=\sum_i \epsilon_i c_i^\dagger c_i+\sum_{ijkl}
v_{ijkl}c_i^\dagger c_j^\dagger c_l c_k~,
\label{occuppres}
\ee
with annihilation and creation operators $c_i$ and $c_i^\dagger $
acting on determinants of states constructed from a given
single-particle basis.
Here we use the property that the occupation of each state for fermions can 
only be zero or one. We notice that the annihilation operator $c_i$
can be viewed as an operator creating a hole in the Fermi sea.
Formally we can define new operators $d_i=c_i^\dagger$ and 
$d_i^\dagger=c_i$ as creation and annihilation operators of the holes.
Equation (\ref{occuppres}) can then be written as a Hamiltonian of the holes.
For the lowest Landau level, considering only states with good total
angular momentum, it reduces to
\be
H_h=\sum_i \tilde\epsilon_i d_i^\dagger d_i+
\sum_{ijkl} v_{ijkl}d_k^\dagger d_l^\dagger d_j d_i + {\rm constant},
\ee
where
\be
\tilde\epsilon_i=
2\sum_{j}(v_{ijji}-v_{ijij})-\epsilon_i.
\ee
It is important to note that the interaction between the holes
is equal to the interaction between the particles
(assuming normal symmetry $v_{ijkl}=v_{klij}$), but the
single-particle energies of the holes are affected by the
interparticle interactions.
We can thus solve the many-particle problem either for the particles,
or for the holes.
The use of the holes, however, does not reduce the complexity of the problem: 
The same accuracy of the solution requires diagonalization of a matrix which
has the same size for particles or holes. 
However, considering holes instead of particles provides
an alternative way to understand the localization of vortices in 
fermion systems~\cite{jeon2005,manninen2005}.

Using the above particle-hole duality picture
we can treat the off-particle vortices as hole-like
quasi-particles~\cite{kinaret1992,ashoori1996,yang2002,saarikoski2004,manninen2005}.
In electron systems, these vortices carry a charge deficiency of 
an elementary charge $e$.
In the particle-hole duality picture the particles and holes (vortices)
can be treated on equal footing. They form a quantum liquid of
interacting electrons and vortices, where correlations play
an important role.

For a correct description of the internal structure of the many-body
system, we need to analyse all correlations between the constituents
of the system, {\it i.e.}, 
particle-particle, vortex-vortex, and particle-vortex
correlations. The relative strength of these correlations determines
the physics of the ground state. To give an example, 
clustering of electrons to a Wigner-crystal-like ``molecule'' of localized
electrons is a signature of 
particularly strong particle-particle correlations.
Analogously, the formation of a cluster or ``molecule'' 
of localized vortices  shows the
correlations between the vortex positions.
Since the vortex dynamics is not independent of the electron
dynamics, strong correlations between electrons and vortices may 
emerge, leading to the formation of 
particle-vortex composites.

\subsection{Quantum Hall states}

\label{sec:qhstates}
Vorticity increases with angular momentum, leading to the 
formation of particle-vortex composites at high magnetic fields.
In the theory of the quantum Hall effect they
were introduced to explain formation of incompressible electron
liquids at fractional filling~\cite{laughlin1983,jain1989}.
However, the phenomenon is more general,
and similar in both fermion and boson systems where vorticity
is sufficiently high~\cite{wilkin1998,cooper1999,viefers2008}.

It should be noted that the analogy between quantum Hall states in 
finite-size droplets and  corresponding states in the infinite
2D electron gas is only approximate, since the particle density
inside the trapping potentials is often inhomogeneous, and edge effects
play an important role.
Nevertheless, in order to (at least approximatively) 
relate the states in finite size electron droplets to
those in the infinite gas, the Landau level filling factor concept has
been generalized to finite size systems.
There is obviously no unique way to do such a generalization. However,
a definition
\begin{equation}
\nu=\hbar {N(N-1)\over 2L},
\end{equation}
which is based on the structure of Jastrow states,
has been used in the $\nu<1$ regime~\cite{laughlin1983,girvin1983}.
In large  fermion systems, the filling factor becomes 
equal to the particle-to-vortex ratio, being  a useful quantity also to
classify rapidly rotating bosonic systems.
Its relation to the fermion filling factor defined above 
is modified by the absence of Pauli vortices in
the bosonic wave function.

The quantum Hall liquid is theoretically described by the Laughlin wave
function~\cite{laughlin1983}
with its extensions, or by the related Jain construction~\cite{jain1989,jeon2004}.
These trial wave functions can be constructed just by using symmetry
arguments without any detailed knowledge of the interparticle interactions.
It has been shown that similar trial wave functions work for bosons and
fermions~\cite{regnault2003,regnault2004}.
Below we will discuss the vortex structures of these trial wave
functions and demonstrate that one can 
map the boson wave function onto the fermion wave function,
allowing a direct comparison of the vortex structures in these different
systems.

\subsubsection{Maximum density droplet state and its excitations}

\label{sec:mdd}
When an electron  droplet is placed in a sufficiently strong magnetic field, 
it may polarize and the single-particle orbitals in the lowest Landau
level become singly-occupied. 
(We remark that at some angular momenta,
the electrons may polarize even
if the Zeeman effect is 
ignored\footnote{Non-polarized states will be discussed 
in Sec. V.}~\cite{reimann2002,koskinen2007}).
The spin-polarized compact droplet of electrons in the LLL,
with total angular momentum $L=N(N-1)/2$,
is called the maximum density droplet (MDD) state~\cite{macdonald1993}.
The MDD has the lowest possible angular momentum 
which is compatible with the Pauli principle.
In the MDD, each electron carries a Pauli vortex
and the wave function can be written as
\begin{equation}
\Psi_{\rm{MDD}}=\prod_{i<j}^N (z_i-z_j) \exp\left[-\sum_{i=1}^N r_i^2/2\right] \ ,\label{eq:mdd}
\end{equation}
where $z_j=x_j+iy_j$, $r^2=x^2+y^2$ and $x$ and $y$ are coordinates
in the 2D plane.
The MDD can be written as a single-determinantal wave function;
for example, for seven particles it is $\vert 11111110000\cdots \rangle$,
where a ``1'' at position $i$ denotes an occupied state in the LLL
with single-particle angular momentum $i-1$.
Clearly, the MDD is the finite-size counterpart of the 
integer quantum Hall state with $\nu=1$.

Removing the Jastrow factor $\prod(z_i-z_j)$
({\it i.e.}, the Pauli vortices) from the MDD in Eq. (\ref{eq:mdd})
leaves just a product of Gaussians which form the non-rotating 
bosonic ground state.
The MDD state can therefore be interpreted as a fermionic ``condensate''-like 
state of particles that engulf the flux quanta and, in effect,
move in a zero magnetic field. 
In this way, the MDD state with $L_{\rm MDD}=N(N-1)/2$ is closely related
to the non-rotating $L=0$ state of a bosonic system.
We discuss this relation further in Sec.~\ref{sec:fermionboson},
where we show conceptually, that by removing the Pauli vortices from
each fermion, the wave function of a fermion system at $L$ is
often a good approximation
for a bosonic state with angular momentum $L'=L-L_{\rm MDD}$.

The first excitation of the MDD in the LLL can be approximated as
a single determinant where
one of the single-particle states is excited to a higher angular momentum.
This state can be understood in two different ways.
It is definitely a center-of-mass excitation, since
\be
\vert 11\cdots 110100\cdots\rangle=\sum_{i=1}^N 
z_i \vert \mathrm{MDD}\rangle ~.
\label{eqcomex}
\ee
On the other hand, this state is also a simple single-particle excitation
where a hole enters the droplet from the surface.
This hole is associated with a phase singularity in the reduced wave function.

\begin{figure}[h!!]
\includegraphics[width=.45\textwidth]{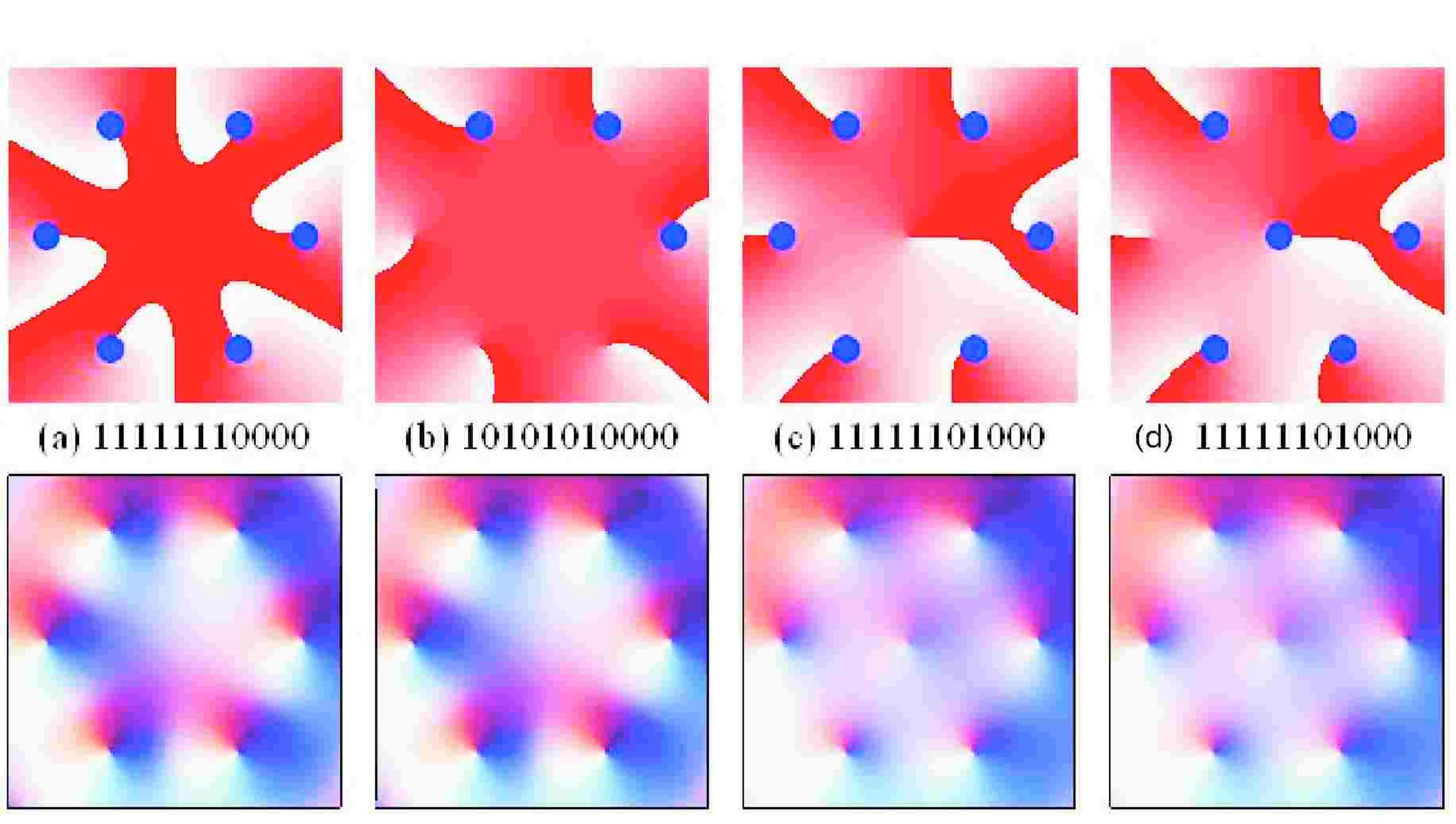}
\caption{({\it Color online}) (a) 
Reduced wave function of a 7-particle MDD-state, (b) 
the MDD state with three holes, and (c), (d) MDD state with a center-of-mass
excitation. The upper panels show the phase and the lower
panels the magnitude of the reduced wave function.
The bullets mark the fixed particle positions.}
\label{mattif1}
\end{figure}

To illustrate the nodal structure of a MDD, we show in Fig. \ref{mattif1} (a),
with seven particles as an example,  
the reduced wave function for this state.
Figure \ref{mattif1} (b) shows the reduced wave function
of the four-particle state
$\vert 1010101000\rangle$ with three holes in the MDD,
demonstrating that the holes  
localize on the sites of the ``missing'' electrons, each of 
them carrying a vortex that is characterized 
by zero density at the core, and the corresponding  phase change. 
It is important to note that in the reduced wave function,  
only the positions of the particles are
fixed, while the vortices are free to choose their optimal positions.
This is illustrated in Fig. \ref{mattif1} (c) and (d) for a center 
of mass excitation: When
one of the atoms (here fixed at the vertices of a hexagon) is moved 
to the center, the free vortex correspondingly 
moves from the center to the hexagon.

\subsubsection{Laughlin wave function}

The angular momentum of a quantum Hall state increases with
the formation of additional vortices. 
When there are three
times more vortices than electrons ($\nu=1/3$),
fermion antisymmetry is preserved if two additional vortices 
(on top of Pauli vortices) are attached to each fermion.
The corresponding wave function is the celebrated Laughlin state
\begin{equation}
\Psi_{\rm{m}}=\prod_{i<j}^N (z_i-z_j)^m \exp\left[-\sum_{i=1}^N
r_i^2/2\right] \ ,
\end{equation}
where the antisymmetry of fermion wave functions
requires that the exponent $m$ is an odd integer~\cite{laughlin1983}.
The analogous wave function for a boson system in a trap is
given by even values of $m$.
The exponent $m$ is related to the filling factor, $\nu=1/m$,
and to the angular momentum $L=mN(N-1)/2$.
According to computational studies that apply diagonalization schemes to the 
many-body Hamiltonian (see Sec.~\ref{sec:many-body} below), 
the Laughlin wave function with $\nu=1/3$ gives a good approximation of
the ground state at the corresponding filling factors in
finite-size quantum Hall droplets.
We discuss this regime of strong correlations in the context
of rapidly rotating quantum droplets in Sec.~\ref{sec:rapid}.

\subsubsection{Jain construction and composite particles}

The composite-fermion (CF) theory \cite{jain1989,jain2007} generalizes the
Laughlin wave function to a larger set of possible filling fractions.
The basic idea is that when an even number of vortices, or flux
quanta, is bound to electrons, these interact less as the
vortices keep them apart, {\it i.e.}, the exchange hole is widened by the cores 
of bound vortices. In addition, the composites move 
in an effective 
magnetic field that is weaker than the original one.  

Formally, the composite fermion wave function can be written as~\cite{jain2007}
\begin{equation}
\Psi_{\rm CF}={\cal P}_{\rm LLL}\psi_{\rm S} \prod_{i>j}(z_i-z_j)^m,
\label{jainconst}
\end{equation}
where $\psi_{\rm S}$ is a single Slater determinant of single-particle
states, the product $\prod(z_i-z_j)^m$ adds $m$ vortices at each electron
and the operator ${\cal P}_{\rm LLL}$ projects the wave function to the
lowest Landau level. If $\psi_{\rm S}$ is taken to be the MDD, 
Eq.~(\ref{eq:mdd}) and  
Eq.~(\ref{jainconst}) lead to the Laughlin wave function for
the fractional Hall effect with filling factor $\nu=1/(m+1)$
and no projection to the LLL is needed.
However, $\psi_{\rm S}$ is not restricted to the LLL, which allows 
constructing the states along the whole yrast line. 
For example, in order to get the MDD of composite particles,
we have to take for $\psi_{\rm S}$ a MDD of states with negative
angular momenta, which means replacing 
$z_i$ and $z_j$ with their complex conjugates $z_i^*$ and $z_j^*$ 
in Eq. (\ref{eq:mdd}). Note that the states with negative
angular momenta are at higher Landau levels. Multiplying this
by $\prod(z_i-z_j)^2$ and projecting to the LLL gives the normal
MDD wave function of Eq. (\ref{eq:mdd}).
Wave functions between $\nu=1$ and $\nu=1/3$ can be obtained
by starting from properly chosen Slater determinants for 
$\psi_{\rm S}$~\cite{jain2007}. The projection to the LLL, however, is
the most difficult part of the Jain construction. In practice,
it can be done by replacing $z_i^*$'s by partial 
derivatives~\cite{Girvin1984}.

The composite fermion picture accurately describes states
at high angular momentum ($L\gg L_{\rm MDD}$) where
two  vortices (in addition to the Pauli vortex) 
are attached to each electron. However, for the states 
immediately above the MDD ($L\approx L_{\rm MDD}$) 
the CF theory still requires 
the attachment of two vortices  to each electron. This means that the
composite particle (electron and two vortices) has to move 
in an effective magnetic field which 
is {\it opposite} to the true magnetic field. In this case
the projection operator ${\cal P}_{\rm LLL}$ will
remove the two vortices (attached by the product $\prod(z_i-z_j)^2$)
and leads to the physically correct result that only one (Pauli) vortex is
attached to each electron. The true number of vortices
attached to each electron can thus be determined only after the 
projection to the lowest Landau level.

Comparison with exact numerical calculations have shown that
the CF theory in the mean-field approximation does not predict
all ground states correctly~\cite{harju1999, yannouleas2007}.
It is possible to go beyond
mean-field theory, but the price to pay is that the beauty
of not having variational parameters in the wave function is
lost~\cite{jeon2007}.

The CF theory has also been used for
bosons~\cite{cooper1999,viefers2000,cooper2008,viefers2008}. 
In this case an
odd number of vortices are attached to each particle, {\it i.e.},
the exponent $m$ in Eq. (\ref{jainconst}) is odd.
Interestingly, the boson wave function is constructed as
a product of two antisymmetric fermionic wave functions. 
The composite fermion picture naturally predicts a close
relation between the bosonic and fermionic states along 
the yrast line, discussed in the next section.

\subsection{Mapping between fermions and bosons}

\label{sec:fermionboson}
In the Laughlin state, the difference in angular momentum between the 
boson and fermion states equals 
that of the maximum density droplet, since trivially, 
\be
\prod_{i<j}^N(z_i-z_j)^m=\prod_{i<j}^N(z_i-z_j)\prod_{i<j}^N(z_i-z_j)^{m-1}.
\ee
As long as the single-particle basis is restricted to the lowest Landau level,
a similar transformation can be used to add a Pauli vortex to each bosonic
particle, {\it i.e.}, by multiplying the boson wave function with
the determinant of the MDD,
\be
\Psi_{\rm fermion}=\prod_{i<j}^N(z_i-z_j) \Psi_{\rm boson}.
\label{mapping}
\ee
This transformation is valid, in addition to the Laughlin states, also for the 
Jain construction. It is expected that the same mapping is a good 
approximation for any many-particle state in the lowest
Landau level~\cite{ruuska2005}. 

The accuracy of the boson-fermion mapping has been studied in detail
by computing the overlaps between the exact fermion wave
function, and the wave function obtained by 
transforming the exact boson state to a fermion state
using Eq. (\ref{mapping})~\cite{borgh2008}.
At high angular momenta where the particles localize,
the mapping becomes exact, while at small angular momenta
the mapping is justified by the small number of possible 
configurations in the LLL. It is important to note that
the free vortices of the bosonic system stay as free vortices also
in the fermionic state. Only the Pauli vortices which localize at 
the particle positions are added. After transforming the bosons to fermions,
particle-hole duality allows a detailed study of the vortex structure
of the bosonic many-body wave function.

\begin{figure}[h!!]
\includegraphics[width=.45\textwidth]{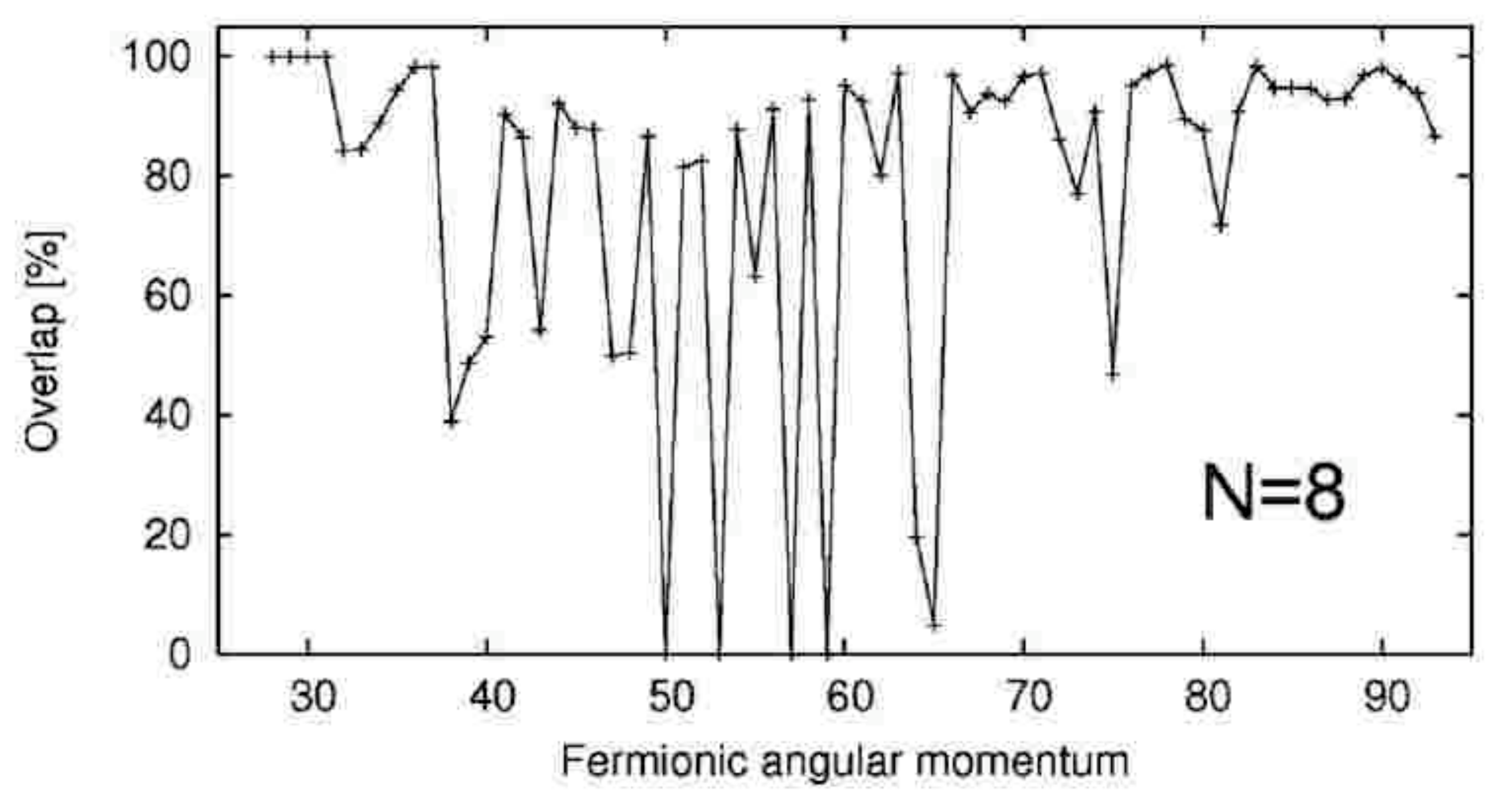}
\caption{Overlap between the fermion ground state and the
transformed boson ground state as a function of the total angular momentum,
for eight particles. From~\cite{borgh2008}.}
\label{bostofer2}
\end{figure}

Figure \ref{bostofer2} shows the calculated 
overlap between the transformed boson state and the exact fermion
state as a function of the total angular momentum for eight particles.
The transformation described by Eq. (\ref{mapping}) does
not always result in the ground state of the fermion system at given
angular momentum. Instead, it may be one of the low-lying
excitations and, consequently, the overlap drops to zero in these cases,
as shown in Fig.~\ref{bostofer2}.
Moreover, the complexity of the wave function increases, while
overlaps of the transformed wave function
with the true fermion ground states tend to decrease with 
the number of particles $N$.

\begin{figure}[h!!]
\includegraphics[width=.35\textwidth]{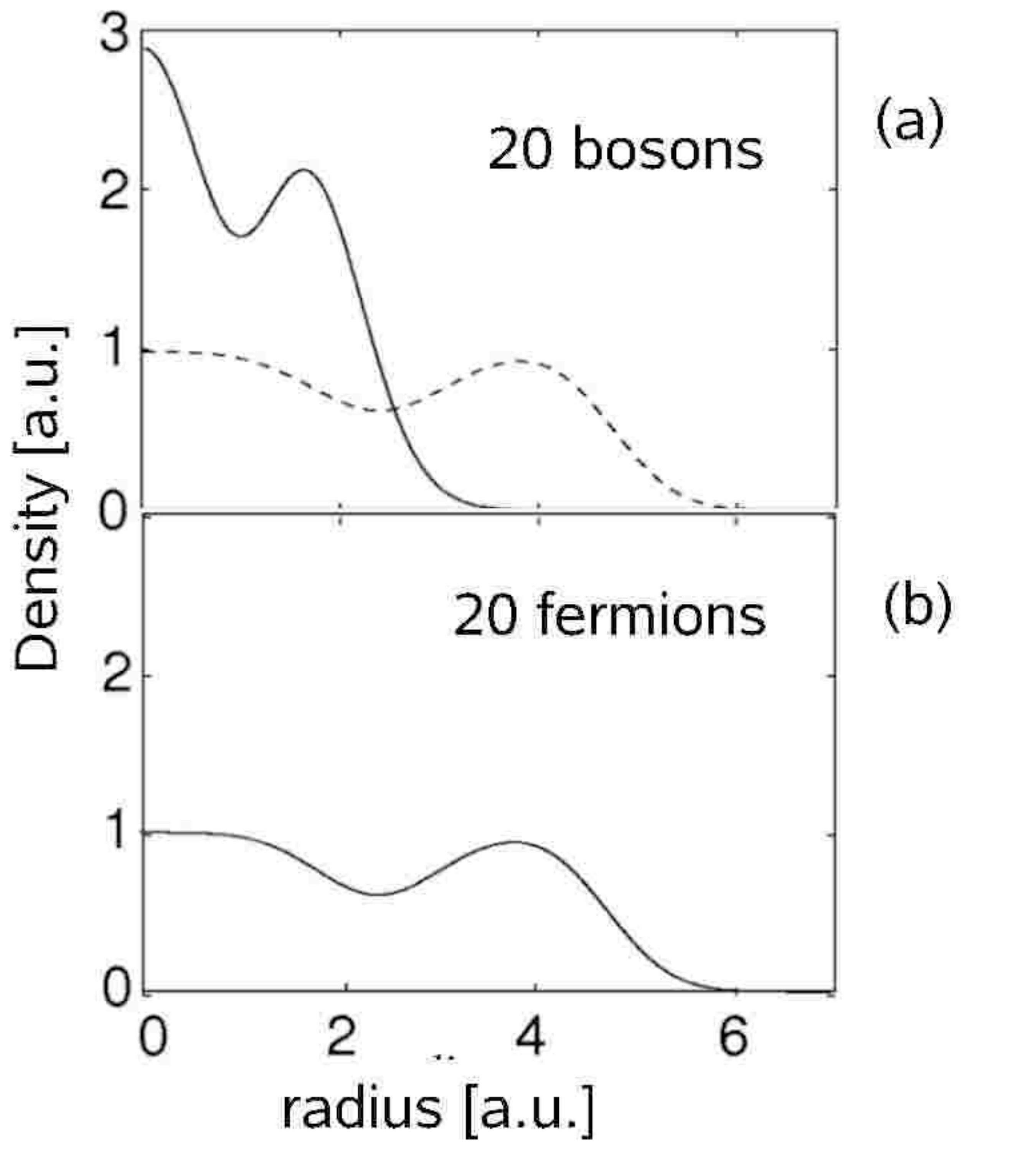}
\includegraphics[width=.35\textwidth]{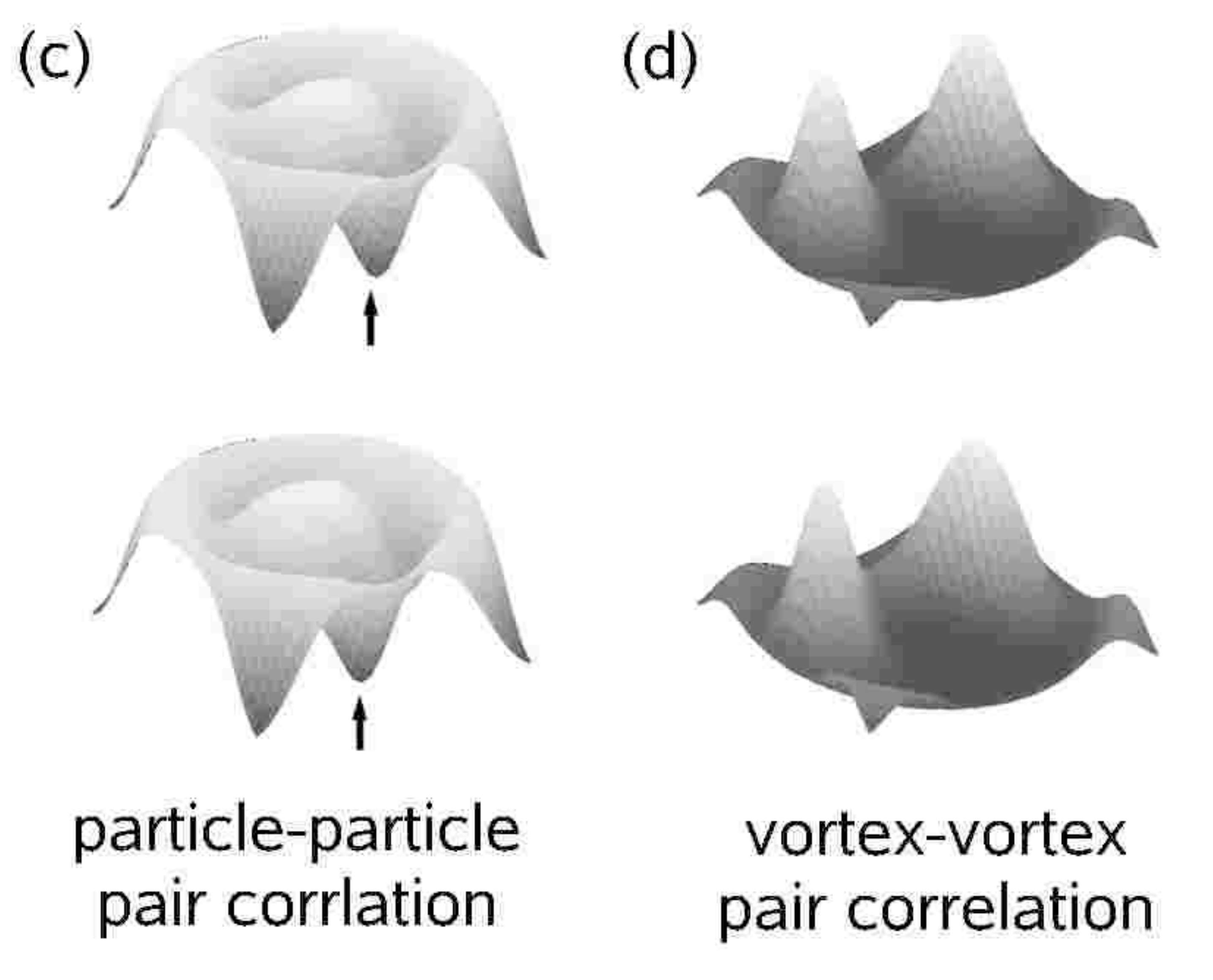}
\caption{Mapping between boson and fermion states. 
The upper panels show the particle density of 
20 bosons (a) and fermions (b) with Coulomb interactions, 
in the region of three vortices as a function of the
radial distance of the droplet center. For bosons, the density
of the mapped fermion system is shown as a dashed line. 
The lower panels show in coulmn (c) 
the particle-particle pair-correlations determined from the fermion wave
functions.  The position of the reference point is marked by the arrow
at the bottom of the exchange-correlation hole. In column (d) 
the corresponding vortex-vortex pair-correlations are shown. }
\label{bostofer}
\end{figure}

Figure \ref{bostofer} illustrates the effect of the mapping for a 
droplet with $N=20$ particles in a harmonic trap at angular momenta where 
three free vortices form. The radial density profile of the
bosonic state shows a minimum at the expected radial distance. 
When the bosonic state is transformed to a fermionic one, 
its radial density expands and becomes 
nearly identical to the exact density of the corresponding fermion system.
The mapping allows to study the internal structure of the 
vortex lattice in the particle-hole duality picture: Figure \ref{bostofer}
also shows the  particle-particle and vortex-vortex correlation functions,
indicating similar localization of three vortices in both cases.

The simple mapping of Eq. (\ref{mapping}) is computationally demanding when
the particle number increases. This is due to the fact that 
every configuration of
the boson wave function fragments to numerous fermion configurations.
A simpler mapping was suggested by~\textcite{toreblad2004}
with a one-to-one correspondence between each boson and fermion configuration
in the few-body limit. 
This mapping captures the most important configurations, but could
not give as good overlaps.

The above transformation, Eq. (\ref{mapping}), can be 
generalized to two-component quantum droplets. The transformation 
$L_{\rm boson}=L_{\rm fermion} - L_{\rm MDD}$ would attach a Pauli 
vortex to each boson. It is apparent
that fermion states with $L_{\rm fermion} < L_{\rm MDD}$ cannot have bosonic
counterparts in the LLL. Nevertheless, suggestive analogies in the (coreless)
vortex structures between bosonic and fermionic states have been obtained
in the few-particle limit~\cite{koskinen2007,saarikoski2009}.

\section{Computational many-body methods}

\label{sec:many-body}

The complexity of the many-body wave function grows exponentially with 
the particle
number $N$, which makes computational studies 
indispensable. We here give a brief overview of the central methods
used in the computational approaches to physics of rotation in both bosonic
and fermionic systems, and their applicability to small droplets.
As it is often the case for approximate approaches, 
the methods presented here have their limits of usability -- no ``universal''
method exists which is superior to the others in capturing the essential
physics in all parameter regimes. 

The exact diagonalization or so-called configuration interaction (CI) method 
does not introduce any approximations
to the solution of the Schr\"odinger equation apart from a 
cut-off in the used basis set. Therefore it is ideally suited to
analyse correlations in the system. This method is, however, limited to
relatively small particle numbers. Mean-field and density-functional
methods are often needed to complement data for larger systems. 
In the density-functional approach, correlation effects are usually
incorporated using local functionals of the spin densities.
The method is able to reveal some of the underlying
correlations in the system, but local approximations 
may fail to describe properly
the complex particle-vortex correlations and formation of particle-vortex
composites~\cite{saarikoski2005a}.
In the following, we draw upon the analogies between 
(a conventionally fermionic) density-functional 
theory and the Gross-Pitaevskii approach for bosons. We finally
summarize the configuration interaction method 
for the direct numerical diagonalization of the many-body
Hamiltonian. 

Rather generally, the 
ground-state energy of an interacting many-body system trapped 
by an external potential $V_{\rm ext} ({\bf r})$ can be written as a
functional of the particle density $n({\bf r})$, summing up the kinetic, 
potential and interaction energy contributions,
\begin{eqnarray} \label{energyfunctional}
E[n({\bf r}) ] &=&   
T_0[n({\bf r})] 
+ \int \hbox{d} {\bf r} ~n({\bf r}) V_{\rm ext} ({\bf r})+ \\\nonumber
&& {1\over 2} \int \hbox{d} {\bf r}  \int \hbox{d} {\bf r'} 
~n({\bf r}) n({\bf r'}) V^{(2)}({\bf r},{\bf r} ')  
+ E_{\rm xc}~,
\end{eqnarray}
where $T_0[n({\bf r})]$ is assumed to be the 
{\it non-interacting} kinetic energy functional,
the second term accounts for
the trap potential, the third term is the 
Hartree term for a  two-particle potential $V^{(2)}$,
and the exchange-correlation energy 
$E_{\rm xc} $ is defined to include 
all other many-body effects.

Introducing a set of 
single-particle orbitals $\psi _i ({\bf r})$, the density may be  
expressed as
\begin{equation}
n({\bf r}) = \sum _{i=0}^{\infty } 
f_i \mid \psi _i ({\bf r}) \mid ^2~,  
\label{orbitals}
\end{equation}
with occupancies $\sum _i f_i =N$, following either bosonic or fermionic 
statistics.  
One can then write the non-interacting 
kinetic energy functional for the orbitals $\psi _i$ in the form
\begin{equation}
T_0[n({\bf r})] = 
\sum _i f_i \int \hbox{d}{\bf r} ~ \psi _i^* ({\bf r}) 
\left (-{\hbar ^2 \nabla ^2\over 2m}\right ) \psi _i ({\bf r})~.  
\label{kinenergyf}
\end{equation}
The crux of the matter is that 
Eq.~(\ref{kinenergyf}) not necessarily 
holds for the {\it exact} kinetic energy 
functional $T[n({\bf r})]$. In many cases there will be a substantial 
correlation part in the kinetic energy functional that is not accounted 
for by the expressions above. 
In the spirit of  
density-functional theory~\cite{dreizlergross1990}, the last term 
in Eq.~(\ref{energyfunctional}), $E_{\rm xc}$, thus has the task to  
collect what was neglected by this assumption, together with the 
effects of exchange and correlation that originate from the 
difference between the 
true interaction energy, and the simple Hartree term. 
It is important to note that the Hohenberg-Kohn theorem guarantees that this 
quantity is a functional of only the density, $E_{\rm xc}=E_{\rm xc}[n({\bf r})]$.

\medskip
\subsection{The Gross-Pitaevskii approach for trapped bosons}
\label{sec:gp}

\subsubsection{Gross-Pitaevskii equation for simple condensates}

In the case of bosons, for a simple condensate 
all bosons are in the lowest state $\psi _0({\bf r})$ 
and the particle density is 
\begin{equation} 
n({\bf r}) =  |\psi _0 ({\bf r})| ^2 
 = N \mid \phi _0({\bf r}) \mid ^2
\label{eq:gp1}
\end{equation}
where the condensate wave function $\psi _0({\bf r})$ is normalized to $N$, 
and the corresponding 
``order parameter'' $\phi _0({\bf r})$ to unity.

By using contact interactions and ignoring the correlations in 
Eq.~(\ref{energyfunctional}) one obtains the well-known
Gross-Pitaevskii energy functional, 
\begin{eqnarray}
E[n({\bf r})] &=& \int \hbox{d} {\bf r} \left[ -{\hbar ^2 \over 2m} \mid 
\nabla \psi _0({\bf r})\mid ^2 \right.\\\nonumber
&&\left.+ V_{\rm ext} ({\bf r} ) \mid \psi _0({\bf r}) \mid ^2
+ {1\over 2} g \mid \psi _0({\bf r} )\mid ^4 \right] ~.
\label{grosspitaevskiifunctional}
\end{eqnarray}

Finding the ground state usually amounts to a variational procedure, 
{\it i.e.}, 
independent variations of $\psi $ and $\psi ^*$ under the condition that 
the total number of particles in the trap is constant. 
For the variation with respect to $\psi _0^*$,
\begin{equation}
{\delta \over \delta \psi _0^* ({\bf r}) } 
 \left[ E[ \psi _0, \psi_0^*] -  \mu \int \hbox{d} {\bf r} \mid \psi _0
({\bf r})\mid ^2 \right] = 0~, 
\end{equation}
where the chemical potential $\mu $ plays the role of a Lagrange multiplier
to fulfill the constraint.
We then arrive at the time-independent Gross-Pitaevskii equation,
\begin{equation}
\left [ -{\hbar ^2 \over 2m} \nabla ^2  + V_{\rm ext} ({\bf r}) 
 + g \mid \psi _0({\bf r}) \mid ^2 \right ] \psi _0({\bf r}) 
= \mu \psi _0({\bf r})~  
\label{grosspitaevskiiequation}
\end{equation}
having the typical form of a self-consistent mean-field equation.
The corresponding $N$-particle bosonic wave function is 
\begin{equation}
\psi ({\bf r}_1,{\bf r}_2,...,{\bf r}_N)=\prod _{i=1}^N \psi _0({\bf r}_i)~.
\end{equation}
The Gross-Pitaevskii approach, derived already in the 60's 
independently by ~\textcite{gross1961} and \textcite{pitaevskii1961}, has  
been applied extensively for the theoretical description of inhomogeneous and
dilute Bose gases at low 
temperatures\footnote{For a more detailed discussion, see for example the textbooks by
~\textcite{pitaevskiistringari2003} and~\textcite{pethicksmith2002}.}.
It is often convenient to 
solve the Gross-Pitaevskii equations
in the imaginary-time evolution method, using a fourth-order
split-step scheme~\cite{chin2005}.

\subsubsection{Gross-Pitaevskii approach for multi-component systems}

The above Gross-Pitaevskii equation 
for a simple single-component Bose 
condensate~Eq.~(\ref{grosspitaevskiiequation})  
can be straightforwardly generalized also to multiple components 
of distinguishable species of particles.
Let us consider as an example a two-component gas of atoms 
of kinds $A$ and $B$, that are interacting through the usual $s$-wave
scattering with equal interaction strengths $g= g_{AA}=g_{BB}=g_{AB}$.
The order parameters  $\psi _A$ and $\psi _B$  of the 
two components then play an analogous role than the spin ``up'' and ``down'' 
orbitals in the spin-dependent Kohn-Sham formalism 
(see Sec.~\ref{sec:dft}). 
The corresponding Gross-Pitaevskii energy functional in the rest frame 
is simply
\begin{eqnarray}\nonumber
&& E = \sum _{\sigma =A,B} \int \mathrm{d} {\bf r}
\psi _{\sigma } ^* \biggl (- {\hbar ^2 \nabla ^2\over 2M} + V_{\rm ext} ({\bf r})
\biggr ) \psi _{\sigma } +  \\
&&   {g\over 2} \int \mathrm{d} {\bf r} \bigl (\mid \psi _A \mid ^4 +
\mid \psi _B \mid  ^4  + 2 \mid \psi _A \mid ^2 \mid \psi _B \mid ^2\bigr ) ~~
\end{eqnarray}
where $\sigma = \{A,B\}$ plays the role of a pseudospin $1/2$. 
In analogy to the single-component case described above,
we minimize the energy functional with respect to $\psi _A^*$ and $\psi _B^*$, 
which results in two coupled Gross-Pitaevskii equations: 
\begin{eqnarray} \label{eq:MULTIGPeq}\nonumber
\biggl( \frac{{\bf p}^2} {2M} + {1\over 2} M \omega ^2 
{\bf r} ^2 + g(|\psi_A|^2 + |\psi_B|^2)\biggr) \psi_A 
 &=& \mu_A\psi_A \\ \nonumber
 \biggl(\frac{{\bf p}^2}{2M} + {1\over 2} M \omega ^2 {\bf r} ^2
 + g(|\psi_B|^2 + |\psi_A|^2)\biggr) \psi_B
 &=& \mu_B\psi_B~.~~
\end{eqnarray}
Naturally, it is required that  
$N_{A}=\int \mathrm d {\bf r} |\psi_A|^2$
and $N_{B}=\int \mathrm d {\bf r} |\psi_B|^2$, which determines the 
chemical potentials $\mu_A$ and $\mu_B$. 
One may choose to normalize the order parameter of one of the components,
say B, to unity. Then, $N_A$ is determined by the ratio $N_A/N_B$.
For the total angular momentum,
$L= \int \mathrm{d}{\bf r} (\psi _A^* \hat L_z \psi _A + 
\psi _B^* \hat L_z \psi _B )= L_A + L_B$.
The above mentioned imaginary-time evolution method is
also in the multi-component case the method of choice to numerically solve the 
Gross-Pitaevskii equations.

\subsection{Density-functional approach}
\label{sec:dft}

The density-functional theory  for the solution of many-body problems
in physics and chemistry was proposed by Hohenberg, Kohn
and Sham in the 1960's~\cite{hohenberg1964,kohn1965}. It is 
a correlated many-body theory where all the ground-state properties can
in principle be calculated from the particle
density~\cite{hohenberg1964,kohnrmp,dreizlergross1990,parryang}.
The original density-functional theory did not take into account
the effects of a non-zero spin polarization and currents induced by 
an external magnetic field. Since these effects have marked consequences on
the ground-state properties of the rotating many-body systems,
for a description of quantum dots in strong magnetic fields, 
extensions such as the spin-density-functional 
method~\cite{gunnarsson1976,vonbarth1979}
and the current-spin-density-functional 
method~\cite{vignale1987,vignale1988,rasolt1992,capelle1997} 
were applied.
For a very pedagogic review on density-functional theory, we 
refer to \textcite{capelle2006}.

\subsubsection{Spin-density-functional theory for electrons}

In the spin-density-functional formalism one can derive self-consistent
Kohn-Sham equations for the Hamiltonian Eq.~(\ref{hamiltonian}) that describes
$N$ interacting electrons in an external magnetic field:
\begin{eqnarray}
& \nabla^2 V_{\rm H}=-{n / \epsilon} &
\label{hartpot} \\
& n_\sigma({\bf r})=\sum^{N_\sigma}_i |\psi_{i,\sigma}({\bf r})|^2 &
\label{density}  \\
& \left\lbrace {1 \over 2 m^*} \left[ {\bf p}+
{e}{\bf A}({\bf r})\right]^2+
V_{\rm eff,\sigma}({\bf r})\right\rbrace \psi_{i,\sigma} =
\epsilon_{i,\sigma} \psi_{i,\sigma} ~.&
\label{sch}
\end{eqnarray}
Eq.~(\ref{hartpot})
is the Poisson equation for the solution of the Hartree potential
$V_{\rm H}$, {\em i.e.} the Coulomb potential for
the electronic charge density $n$,
where $\epsilon$ is the dielectric constant of the medium.
Eq.~(\ref{density}) determines the spin densities,
where $\sigma=\{\uparrow,\downarrow \}$
is the spin index, $N_\sigma$ is the number of electrons
with spin $\sigma$, the 
$\psi_{i,\sigma}$'s are the one-particle wave functions, and
the summation is over the $N_\sigma$ lowest states
(which here have fermionic occupancy).
In Eq.~(\ref{sch}), the {\em effective} scalar potential for electrons
\begin{equation}
V_{\rm eff,\sigma}({\bf r}) = V_{\rm ext}({\bf r}) + V_{\rm H}({\bf r})
+ V_{\rm xc,\sigma}({\bf r}) + V_{\rm Z} 
\label{effpot}
\end{equation}
consists of the external scalar potential $V_{\rm ext}$, the Hartree potential
$V_{\rm H}$, the exchange-\-cor\-re\-la\-tion potential
$V_{\rm xc}$ and the Zeeman term $V_Z=g^* \mu_B B s_\sigma$,
where $\mu_B$ is the Bohr magneton, $s_\sigma=\pm 1/2$, $B$ is the magnetic
field and $g^*$ is the gyromagnetic ratio.
All the interaction effects beyond the Hartee potential $V_{\rm H}$
are incorporated in the exchange-correlation potential $V_{\rm xc}$.
A more fundamental generalization of the density-functional method for systems
in external magnetic fields is the current-density-functional
method~\cite{vignale1987,vignale1988}, where the vector potential $\hbox{\bf
  A}$ is replaced by an effective vector potential $\hbox{\bf
  A}_{\hbox{eff}} = \hbox{\bf A} + \hbox{\bf A}_{\hbox {xc}}$ accounting for
many-particle effects on the current densities. 
In the above equations, only the conduction electrons of the 
semiconductor are explicitly included in the theory, 
while effects of the lattice are incorporated via material 
parameters such as effective mass, dielectric constant and 
effective $g$-factor.

Density-functional
 approaches are often combined with local approximations
for the exchange-correlation potential where 
$V_{\rm xc}$ in actual calculations is usually taken as the
exchange-correlation potential of the uniform electron gas.
In 2D electron systems, approximate parametrizations 
have been calculated~\cite{tanatar1989,attaccalite2002} and
the approach leads to a set of mean-field-type equations. 
It should be emphasized that density-functional theory {\it a priori} 
is not a mean-field
method but a true many-particle theory. Its strength is
that it very often may provide accurate approximations to the ground
state properties such as the total energy
with the computational effort of a mean-field method.
It is important to note that 
single-particle states (Kohn-Sham orbitals) and their eigenenergies
are auxiliary parameters in the Kohn-Sham equations.  
However, as an approximation, the Kohn-Sham orbitals may still be used to 
construct a single Slater determinant to account for the nodal structure.    

The density-functional approach in the 
local density approximation, 
as well as the unrestricted Hartree-Fock method, may show
broken symmetries in particle and current densities.
This is often interpreted as reflections of the internal structure
of the exact many-body wave 
function\footnote{For a comprehensive discussion of this issue in the context of quantum dots,
see \textcite{reimann2002}.}. 
However, a caveat is that
implications of symmetry-breaking patterns may in some cases 
yield wrong implications
of the actual many-body structure of the exact wave function.
This problem is well-known in quantum chemistry as ``spin contamination'', 
and we refer to~\textcite{szabo1996} as well as the more recent articles 
by~\textcite{schmidt2008}, as well as 
\textcite{harju2004} and \textcite{borgh2005} for a thorough discussion. 
This conceptual problem of spin-density-functional theory 
often calls for an analysis by more exact
computational methods.

\subsubsection{Density-functional theory for bosons}
\label{gpeq}

The Gross-Pitaevskii mean-field approach discussed above certainly is the 
most widely used theoretical tool to describe Bose-Einstein condensates,
and has been extensively applied to investigate vortex structures in 
rotating systems. 
Clearly, it is a density-functional method based on the 
functional Eq.~(\ref{grosspitaevskiifunctional}) where the density
is a square of a single one-particle state, Eq. (\ref{eq:gp1}).
However, there are many situations where correlations determine 
the many-body states, that cannot be captured by the standard 
Gross-Pitaevskii approach~\cite{bloch2008}.

On the other hand, the exact diagonalization method, which captures
all correlation effects, cannot be used for systems
which consist of more than just a few particles.
A bosonic density-functional theory has been introduced as one 
possible way to go beyond  the mean-field approximation
\cite{hunter2004,braaten1997,zubarev2003,griffin1995,nunes1999,rajagopal2007,capelle_notes}.
For ground states this approach is not very efficient 
due to a lack of nodal structure in the wave function. 
This, however, is different in the case of 
fragmented or depleted condensates~\cite{mueller2006,capelle_notes}.

Following the well-known Hohenberg-Kohn theorem, 
the energy functional $E[n({\bf r})] $ is minimized by the ground-state 
density. This in fact is independent of whether the particles 
are bosons or fermions, and a corresponding density-functional approach to
bosonic systems was more recently formulated by~\textcite{capelle_notes}.
Taking the $E_{\rm xc}$ contributions into account, the variation of 
Eq.~(\ref{energyfunctional}) adds the potential \cite{nunes1999}
\begin{equation}
V_{\rm xc} = {1\over \psi ({\bf r})} {\delta E_{\rm xc}\over \delta \psi ({\bf r})}~. 
\end{equation}
However, $\psi ({\bf r}) $ cannot describe
correctly  the many-body state, if
the ground state contains ``uncondensed'' bosons, or requires a macroscopic
occupation of more than one single-particle state.
\textcite{capelle_notes} showed that 
since the Hohenberg-Kohn  theorem still holds in these cases, the
Gross-Pitaevskii equation, Eq.~(\ref{grosspitaevskiiequation}), can be more
generally expressed as 
\begin{eqnarray}
\left[ -{\hbar ^2\over 2m} \nabla ^2 \right. &+& V_{\rm ext} ({\bf r}) + \int 
\hbox{d} {\bf r} ~n({\bf r}) n({\bf r}')V^{(2)}({\bf r}-{\bf r}') \nonumber \\
&+& \left. {\delta E_{\rm xc}[n]\over \delta n({\bf r})} \right] \psi _i ({\bf r}) 
= \epsilon _i ({\bf r}) \psi _i ({\bf r})~,  
\end{eqnarray}
with the label $i$ now running over all solutions of the equation. 
The orbitals $\psi _i$ do not have a simple relation to the 
Gross-Pitaevskii order parameter, but they 
do provide the correct density via Eq.~(\ref{orbitals})
with (bosonic) occupancies $f_i$ of the states $\psi _i$~. 
These equations took 
a form that is indeed very familiar from the usual Kohn-Sham equations 
for fermions discussed above ~\cite{capelle_notes}. 
For an account of viable approximations to $E_{\rm xc}$, we refer to 
\textcite{capelle_notes}, as well as ~\textcite{nunes1999} and ~\textcite{zubarev2003}.
%In a sense, the above 
%approach is related to the multi-orbital mean field approach
%by ~\cite{alon2005a,alon2005b}, who evaluated the energy functional on the 
%Hartree-Fock level. 

\subsection{Exact diagonalization method}

\label{sec:exact}
The configuration interaction (CI) method, also called ``exact
diagonalization'', 
is a systematic scheme to expand the many-particle
wave function. It traces back to the early days of quantum mechanics, 
to the work of~\textcite{Hylleraas1928} on the Helium atom. 
It has been extensively used in quantum chemistry, but nowadays found 
its use also for quantum nanostructures as well as cold atom systems. 
In the basic formulation of this approach, one takes the eigenstates
of the non-interacting many-body problem (called configuration) as
a basis and evaluates the interaction matrix elements between these
states.  The resulting Hamiltonian matrix is then diagonalized. 

Rules to calculate the matrix elements were originally derived 
by~\textcite{Slater1929,Slater1931} and 
\textcite{Condon1930},
and developed further by~\textcite{Lowdin1955}.
We note that the use of the term
``exact diagonalization'' that has been widely adopted by the 
community, often  replacing the quantum-chemistry terminology of  
``configuration interaction'',  
might in some cases
be misleading, as truly exact results are obtained only in the limit
of an infinite basis.

Consider a Hamiltonian split into two parts
$\mathcal{H} = \mathcal{H}_0 + \mathcal{H}_I$, where the Schr\"odinger
equation of the first part is solvable,
\begin{equation}
\mathcal{H}_0 | \phi_i \rangle = \varepsilon_i | \phi_i \rangle ~,
\label{ED_basis}
\end{equation}
and the states $|\phi_i\rangle$ form an orthonormal basis. The solution
for the full Schr\"odinger equation can be expanded in this basis as
$| \psi \rangle = \sum_i \alpha_i | \phi_i \rangle$.  Inserting
this into the Schr\"odinger equation
\begin{equation}
\mathcal{H} | \psi \rangle = E | \psi \rangle 
\end{equation}
results in
\begin{equation}
(\mathcal{H}_0 + \mathcal{H}_I) \sum_i \alpha_i | \phi_i \rangle = E \sum_i \alpha_i | \phi_i \rangle \ ,
\end{equation}
or a matrix equation
\begin{equation}
(H_0+H_I) \mbox{\boldmath $\alpha$} = E \mbox{\boldmath $\alpha$} \ ,
\end{equation}
where $H_0$ is a diagonal matrix with 
$\langle \phi _i \mid H_0 \mid \phi _i \rangle =\varepsilon _i $
and the elements of $H_I$ are 
$ \langle \phi_j | \mathcal{H}_I | \phi_i \rangle $.  
The vector $\mbox{\boldmath $\alpha$}$ contains the values 
$\alpha_i$. In
principle, the basis $\{|\phi_i\rangle \}$ is infinite, but the actual
numerical calculations must be done in a finite basis. The main
computational task is to calculate the matrix elements of $H_I$ and to
diagonalize the corresponding matrix. The convergence 
as a function of the size of the basis set depends on 
the problem at hand, and is of course fastest for the cases where
$\mathcal{H}_I$ is only a small perturbation to $\mathcal{H}_0$.

The basic procedure is straightforward text-book knowledge of 
quantum mechanics. However, one should bear in mind that much of the 
state-of-the-art computational knowledge must be employed
when it comes to numerical implementations, in order to model large and 
highly-correlated systems. 

The usual starting point for the exact diagonalization method 
is the non-interacting problem. 
In 2D harmonic potentials, harmonic oscillator states - 
or Fock-Darwin states of non-interacting particles in a 
magnetic field - can be used to construct a suitable basis, 
but it can also be optimized by using states from, {\it e.g.}, 
Hartree-Fock or density-functional methods 
(for a recent example, see the work by~\textcite{emperador2005}). 
For fermions, the solution is a Slater determinant formed
from the eigenstates of the single-particle Hamiltonian. The
corresponding symmetric $N$-boson state is a permanent. 
In the non-interacting ground state, all the
bosons occupy the same state. On the other hand, fermions occupy the
$N$ lowest states due to the Pauli principle. Due to
interactions, other configurations than the one of the non-interacting
ground state have a finite weight in the expansion of the
many-particle wave function.
Often, the increasing complexity of the quantum states with large 
interaction strengths and large system sizes causes severe convergence 
problems, where the number of basis states needed for an accurate description  
of the many-body system increases far beyond computational reach. 

In rotating weakly-interacting systems confined by harmonic potentials,
a natural restriction of the 
single-particle basis is the LLL. It provides a 
well-defined truncation of the Hilbert space for 
the given value of the angular momentum $L$ and particle number $N$.
The LLL approximation in the harmonic confinement
implies that the diagonal part of the Hamiltonian is independent of the 
configuration, and solving the Hamiltonian reduces to the diagonalization 
of the potential energy of the interparticle interactions.
This truncation eliminates also the usual issue of regularization
that emerges with the use of contact forces in exact diagonalization
schemes, see for example, ~\textcite{huang1963}:
The direct diagonalization of the Hamiltonian with contact interactions 
on a {\it complete} space yields unphysical solutions 
unless the class of allowed basis functions obeys special and often 
impractical boundary conditions~\cite{esry1999}. 
The Hamiltonian matrix in the LLL is often sparse, and in the limit of 
large $N$ and $L$ it is usually diagonalized in a 
Lanczos scheme~\cite{arpack}. 

\section{Single-component quantum droplets}
 \label{sec:single-component}

In the following, we describe the structure of single-vortex states and 
the formation of vortex ``clusters'' or  vortex ``molecules'', as they are
also often called, in single-component systems. 
In the strongly-correlated regime of rapid rotation, the increased vortex
density leads to finite-size counterparts of fractional quantum Hall states,
both with bosons and fermions.
The existence of giant or multiple-quantized vortices in anharmonic traps is 
also discussed. 

\subsection{Vortex formation at moderate angular momenta} 
 
\label{sec:vortexformation}

\subsubsection{Vortex formation in trapped bosonic systems} 
\label{sec:onevortexboson} 
 
\begin{figure}[ptb]  
\includegraphics[width=0.45\textwidth]{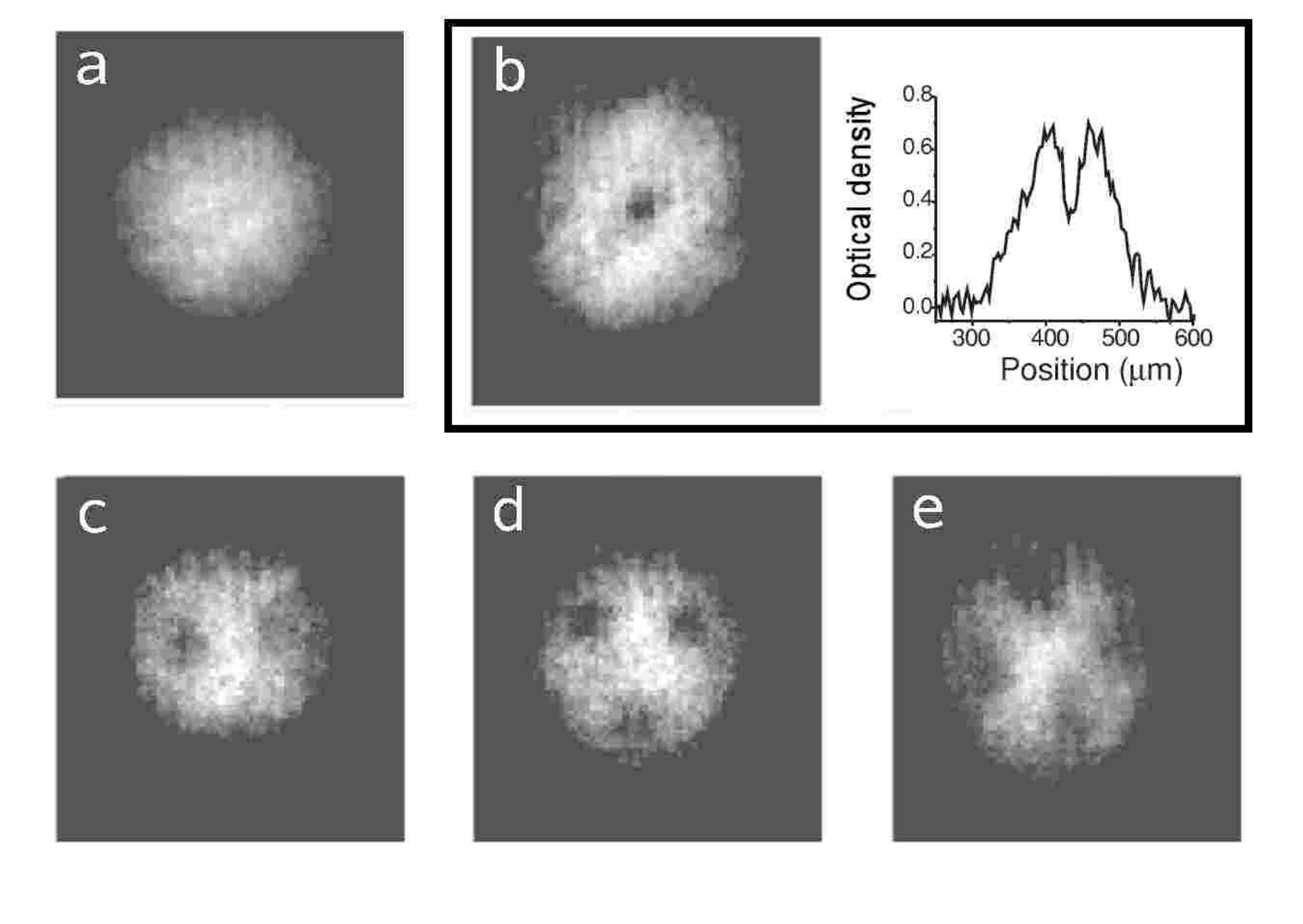} 
  \caption[]{Transverse absorption images of a Bose-Einstein condensate 
 of $^{87}$Rb, stirred with a laser beam.  
As the rotation of the trap increases from a) to e), a clear vortex pattern evolves. 
The inset to the right of panel b) shows the cross section of the optical 
density which shows a pronounced minimum at the center. 
After ~\textcite{madison2000}.} 
\label{fig:madisonvortices}  
\end{figure}

Following the achievement of Bose-Einstein condensation in trapped 
cold atom gases, experimental setups were devised to study their rotational  
behavior. The first observation of vortex patterns in these systems was made  
for a two-component Bose condensate consisting of two internal spin states  
of $^{87}$Rb, where the formation of a single vortex was  
detected~\cite{matthews1999}. 
Soon after this seminal experiment, evidence 
for the occurrence of vortices was found by 
literally ``stirring'' a one-component 
gaseous condensate of rubidium by a laser beam 
~\cite{madison2000}. 
While the vortex cores are too small to be directly observed 
optically (the core 
radius is typically from 200 to 400 nm), vortex imaging is 
possible if the atomic cloud first is allowed to expand by  
turning off the trap potential~\cite{madison2000}. 
In this way, regular patterns of 
vortices were observed in the transverse absorption images of the  
rubidium condensate (see Fig.~\ref{fig:madisonvortices}).  
At moderate rotation, above a certain critical frequency  
$\Omega _c$,  first a central ``hole'' occurred,  
clearly identified as a pronounced minimum in 
the cross-section of the density distribution, shown to the right  
in Fig.~\ref{fig:madisonvortices}b). 
As the rotation of the trap increases, a 2nd, 3rd and 4th  
vortex penetrates the bosonic cloud. The vortices then    
arrange in regular geometric patterns.  
Intriguingly, these patterns coincide with the geometries of Wigner crystals   
of repulsive {\it particles}, as they have been found for example in  
quantum dots at low electron densities, or strong magnetic  
fields~\cite{reimann2002}. 
Vortices with the same sign of the vorticity 
effectively repel each other~(see for example, \textcite{castin1999}). 
This supports the view of Wigner-crystal-like 
arrangement of vortices, throwing an interesting 
light on the much debated melting of the vortex lattice at extreme rotation
(see also Sec.~\ref{Sec:Melting} below).  
The interplay between vortex- and particle localization  
in a rotating harmonic trap is further discussed in 
Section~\ref{sec:lattices} below.  

The theory of vortices in rotating BEC's has attracted a lot of 
attention in the recent years, and much work has been published 
for the Thomas-Fermi regime of strong interactions, see for example,  
\cite{rokhsar1997,ripoll1999,feder1999a,feder1999b,svidzinsky2000}.  
In this limit, which applies to most experiments on rotating BEC's,  
the coherence length $\xi = (8\pi n a )^{1/2} $, where $n$ is 
the density and $a$ the scattering length, is much smaller than the  
extension of the bosonic cloud, and some properties of the system resemble 
those of a bulk superfluid~\cite{baym1996}.  
In the case of weakly  interacting bosons in a harmonic trap, 
however, the coherence length becomes 
larger than the size of the cloud, and the interaction energy plays the 
dominant role: the mesoscopic limit is reached, where the system becomes like 
a quantum-mechanical Knudsen gas~\cite{mottelson2001}.
In this mesoscopic limit, the analogies  
between trapped bosons and quantum dots at strong magnetic fields become  
apparent. This regime of weak interactions is our primary concern in the  
following.   
 
\subsubsection{Weakly interacting bosons under rotation} 

Let us now consider a dilute system of 
$N$ spinless bosons in a harmonic trap,  
weakly interacting by the usual contact force $g\delta ({\bf r}-{\bf r}_0)$, 
where $g=4\pi \hbar ^2 a/M$ is the strength of the effective two-body 
interaction with scattering length $a$ and atom mass $M$. 
The condition for weak interactions is that the interaction energy  
is much smaller than the quantum energy of the confining  
potential, {\it i.e.},  
\begin{equation} 
ng \ll \hbar\omega\;, 
\end{equation} 
where $n$ is the particle density.
As explained in Sec.~\ref{sec:bosonicdroplets} above,  
requiring maximum alignment of the total angular momentum,  
the relevant single-particle states of the oscillator are those 
of the LLL. 
This approach, which has earlier proven very successful for the description
of the fractional quantum Hall regime for the electron gas, 
has been introduced for bosonic systems by~\textcite{wilkin1998}.  

As mentioned in ~\ref{sec:bosonicdroplets}, the large degeneracy 
originating from the many different ways to distribute the $N$ bosons 
on the single-particle states of the LLL, is lifted by the interactions. 
 
Identifying the elementary modes of excitation,  
\textcite{mottelson1999} showed that besides the usual condensation 
into the lowest state of the oscillator, the yrast states ({\it i.e.} 
the states maximizing $L$ at a given energy, 
see Sec.~\ref{sec:bosonicdroplets})
involve additional kinds of condensations that are associated with 
the many different possibilities for distributing the angular momentum on the  
degenerate set of basis states in the LLL.  
For $1 \ll L \ll N$, 
the yrast states and low-energy excitations as a function of 
$L$ can be constructed by a collective operator  
\begin{equation} 
Q_{\lambda } = { 1\over \sqrt{2N\lambda !} } 
\sum _{p=1} ^N z _p ^{\lambda } ~, 
\end{equation} 
with coordinates of the $p$th particle $z_p = x_p + iy_p$. 

In the case of attractive interactions, 
the lowest-energy state for fixed angular momentum is 
the one involving excitations of the center-of-mass of the cloud.  
The yrast state is then described by  
$\mid\Psi _L \rangle \sim (Q_1)^L \mid\Psi _{L=0}\rangle $ 
~\cite{wilkin1998,mottelson1999}. 

In the case of repulsive interactions,
for bosons in the LLL at $L=0$ the only possible state is the pure 
condensate in the $m=0$ single-particle orbital, thus maximizing the 
interaction energy.  Increasing the angular momentum by one 
is possible via a center-of-mass excitation of the non-rotating state. 
For $L>1$ and $L\ll N$, the excitation energies of the 
modes $Q_{\lambda \ne 1}$ show that 
the yrast states are predominantly obtained by a condensation
into the quadrupole ($\lambda =2$) and octupole ($\lambda =3$) 
modes, as shown by \textcite{mottelson1999}. 

\textcite{bertsch1999} compared these results to a numerical computation of 
the yrast line. For the harmonic trap in the lowest Landau 
level, the problem can be solved straightforwardly by numerical  
diagonalization of the Hamiltonian Eq.~(\ref{rotationhamiltonian}). 
(See the discussion in Sec.~\ref{sec:exact} above).  

\begin{figure}[ptb]  
\includegraphics[width=0.45\textwidth]{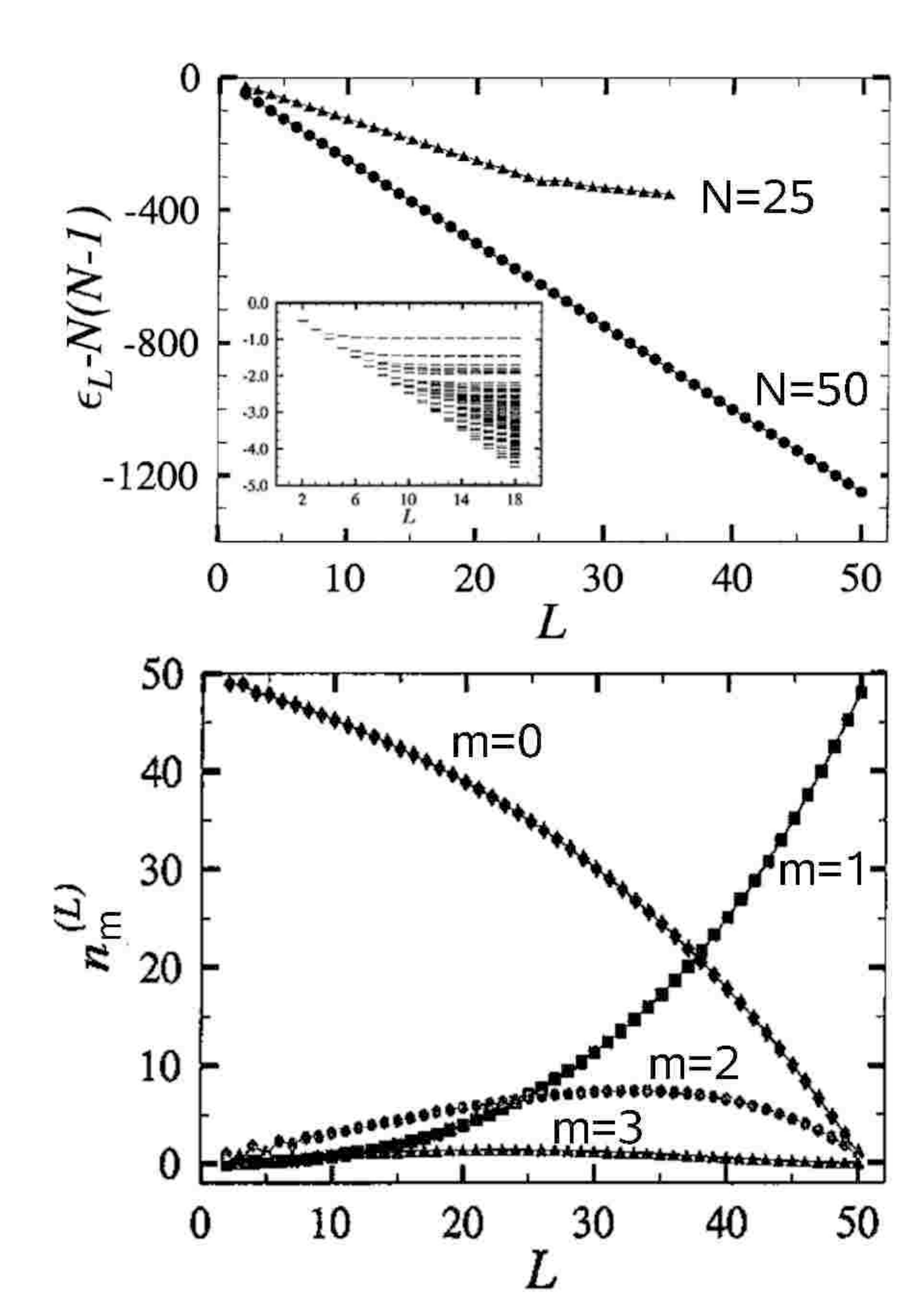} 
  \caption[]{{\it Upper panel:} Many-body yrast lines for $N=25$ and   
$N=50$ spinless bosons in a harmonic confinement for angular  
momenta $2\le L\le 50$. The inset shows the excitation spectrum for $N=50$ and 
$L\le 18$, excluding the spurious center-of-mass excitations,  
see text. From~\textcite{papenbrock2001}.  
{\it Lower panel:} Occupancies of the lowest single-particle states of the  
harmonic oscillator in the lowest Landau level, for $m=0$ {\it (diamonds)},  
$m=1$ {\it (squares)}, $m=2$ {\it (circles)} and $m=3$ {\it (triangles)}.  
From ~\textcite{bertsch1999}. Calculations are within the lowest Landau level.}  
\label{fig:bosonspec} 
\end{figure}

The resulting yrast line decreases with increasing $L$ for repulsive 
interactions, since centrifugal forces tend to keep the particles 
further apart  when rotation increases 
(see Fig.~\ref{fig:bosonspec} for the example of $N=25$ and $N=50$ bosons).
It shows a linear decrease in energy with $L$, that  
extends up to $L=N$.  
This linearity was also 
found in a study within the Gross-Pitaevskii  
approach by~\textcite{kavoulakis2000} (see below). 
The inset to Fig.~\ref{fig:bosonspec} shows the excitation spectra  
for  $N=50$ bosons at angular momenta $L\le 18$.  
``Spurious'' eigenstates occur that originate from a $SO(2,1)$ symmetry 
~\cite{pitaevskii1998}  only exciting the center-of-mass, {\it i.e.},  
the yrast spectrum at $L+1$ includes the full set of states at angular  
momentum $L$. (These center-of-mass excitations were  
excluded in the spectra shown in Fig.~\ref{fig:bosonspec}.) 
In a harmonic confinement the center-of-mass 
excitations are exactly separated
from the internal excitations and they are known to exist also in 
Fermi systems \cite{trugman1985,reimann2002}.

The lower panel in Fig.~\ref{fig:bosonspec} shows the 
occupancies of the lowest single-particle states for a $N=50$ bosonic 
state with angular momentum up to $L=N$.
In agreement with the aforementioned 
results of ~\textcite{mottelson1999}, at small $L/N$  
the yrast states are mainly built from single-particle states  
with $m=0$, $m=2$ and $m=3$, 
respectively, where $m$ is the angular momentum of the 
single-particle state~\cite{bertsch1999}.  
Approaching $L/N=1$, the yrast state takes a much 
simpler structure, with a dominant occupancy of the $m=1$ single-particle 
orbital. At $L/N=1$, a single vortex locates at the center of the cloud. 

An analytic expression for the exact energies for $2 \le N \le N$ was 
conjectured by  \textcite{bertsch1999} and subsequently derived by
\textcite{jackson2000}; in atomic units it reads 
\begin{equation} 
\epsilon _L = {1\over 2} N (2N-L-2)~.  
\end{equation} 
~\textcite{smith2000} derived analytically the 
exact eigenstate as an elementary symmetric polynomial 
of coordinates relative to the center-of-mass. 
Later, exact yrast energies for a  universality class of interactions
were derived~\cite{hussein2002a,vorov2003}.
Generalizing a conjecture by~\textcite{wilkin1998}  
for the structure of the unit vortex at $L=N$,  
\begin{equation} 
\mid L=N\rangle = \Pi _{p=1}^N (z_p-z_c)\mid 0 \rangle  
\end{equation} 
where  
$z_c=(z_1+z_2+\cdots +z_N)/N$ is the center-of-mass coordinate,   
~\textcite{bertsch1999} could demonstrate that the exact wave function  
in the {\it whole} region $2\le L\le N$ is given by  
\begin{eqnarray}\nonumber 
\mid L \rangle &=& {\cal N} \sum _{p_1<p_2<...<p_L} (z_{p_1} - z_c )\\ 
&&\times  (z_{p_2} - z_c) ... (z_{p_L} - z_c)\mid 0 \rangle, 
\label{bertschpapenbrock} 
\end{eqnarray} 
where ${\cal N}$ is a normalization constant, and the indices run over all 
particle coordinates, up to the total particle number $N$.
 
\begin{figure}[ptb]  
\includegraphics[width=0.45\textwidth]{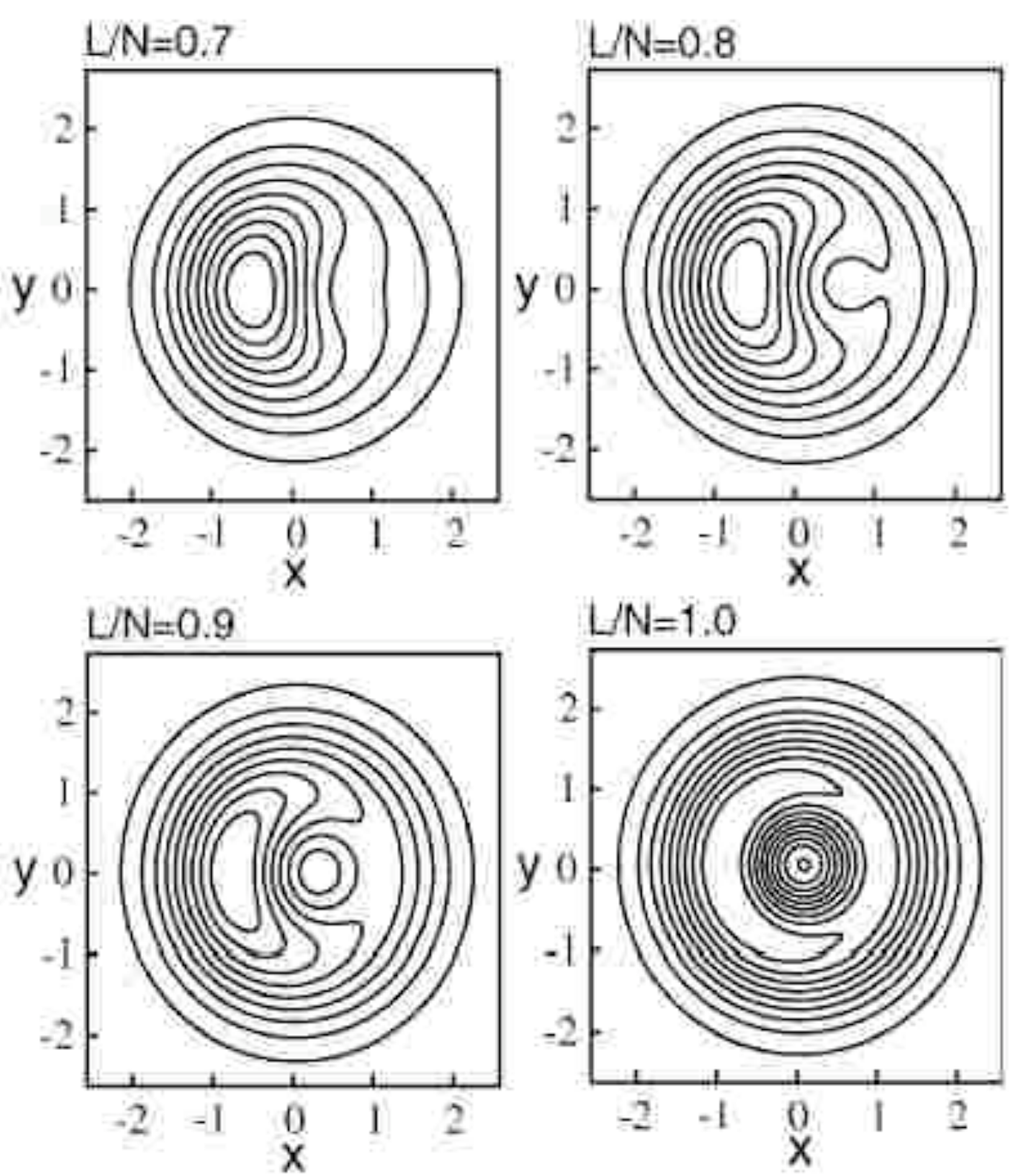} 
  \caption[]{Equidensity lines of the pair-correlation function $P({\bf r}, 
  {\bf r}_A)$ for $N=40$ spinless bosons at $L=28, 32, 36$ and $40$. For 
  clarity, the reference point was located outside the cloud at ${\bf r} 
  _A=(3,0)$. The vortex, which approaches the center from the right 
with increasing $L$, 
gives rise to a pronounced minimum in the pair-correlation plots. 
From ~\textcite{kavoulakis2002}.}  
\label{fig:vortexentryci} 
\end{figure}

Let us now investigate the evolution of the pair-correlated densities,
defined in section ~\ref{sec:cond-wave} above.  
Fig.~\ref{fig:vortexentryci} shows their contours, 
for $N=40$ bosons with the reference  
point located at a distance $r_A=3 \ell _0$ (chosen  
outside the bosonic cloud for clarity; $\ell _0 $ is the oscillator
length). 
Starting from a homogeneous Gaussian density distribution at $L=0$, as $L/N$ 
increases, clearly the first vortex enters the cloud from its outer parts. 
At $L=N$, the (azimuthally symmetric) particle density has developed  
a pronounced central hole, that is also apparent from the  
correlation function shown in the lower right panel of Fig.~\ref{fig:vortexentryci}. 
The nodal pattern of this state, as probed by conditional wave functions, 
clearly confirms the simple structure of the unit vortex 
(see {\it e.g.} the $L=N=5$ state in Fig.~\ref{fig:bosoniclaughlin}). 
For a discussion of the low-energy excitations at and around the unit vortex,
we refer to~\textcite{ueda2006}. 

Recently, \textcite{dagnino2009a,dagnino2009b} 
studied the vortex nucleation process by calculating 
the density matrix obtained from the CI eigenstates 
for a trap with a small quadrupole deformation. 
A related early study was presented by \textcite{linn2001}, who applied
a variational method to investigate the ground state phase diagram
in an axially asymmetric BEC. 
The analysis by ~\textcite{dagnino2009a,dagnino2009b}
indicated that when the rotation frequency of the axially deformed trap
is increased and the system passes through the first vortex transition, 
two of the ``natural orbitals'' of the density matrix have equal
weight. \textcite{nunnenkamp2009} 
also studied the noise correlations at criticality
for the elliptic trap, while \textcite{parke2008} relate the transition to 
vortex tunneling in the process of nucleation.  

In the light of the above-mentioned findings, however, it is worth  noting 
that the overall picture strongly depends on the symmetry of the chosen 
trap deformation, and is further complicated 
by finite-size effects -- the latter being an inevitable 
restriction in the CI method that becomes more severe, when the angular
momentum no longer commutes with the Hamiltonian. 

\begin{figure}[ptb]  
\includegraphics[width=0.45\textwidth]{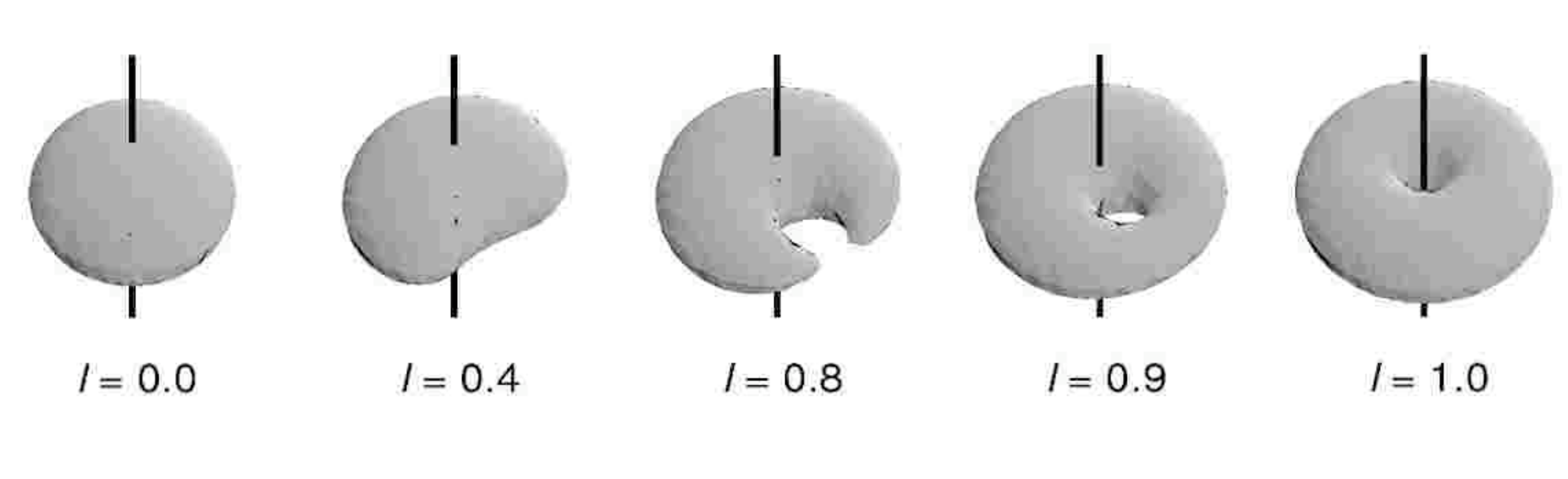} 
  \caption[]{Vortex entry for a spherical bosonic cloud at angular momenta  
$l=L/N$. Shown are the surfaces of constant density obtained by the 
Gross-Pitaevskii  method.  The cloud flattens with increasing 
angular momentum. From ~\textcite{butts1999}.}
\label{fig:buttssinglevortex} 
\end{figure} 

In the Gross-Pitaevskii approach, the vortices are directly visible 
in the density as well as the phase of the order parameter, 
which breaks the rotational symmetry.
\textcite{butts1999} and \textcite{kavoulakis2000} were among the first to
apply this method to a weakly interacting, dilute condensate of bosonic atoms 
in a rotating harmonic trap. 
Fig. \ref{fig:buttssinglevortex} shows the equidensity surfaces for 
the Gross-Pitaevskii order parameter $\Psi ({\bf r})$ for 
the states along the yrast line between $L=0$ and $L=N$
\cite{butts1999}, demonstrating how the first vortex enters the cloud. 
In the non-rotating case, the condensate forms a lump 
with zero angular momentum at the center of the trap. 
Beyond a certain critical rotation, however,  
the ground state becomes a vortex state with one single-quantized vortex
that manifests itself as a central hole in the 
density (see $l=1.0$ in Fig.~\ref{fig:buttssinglevortex}).
The phase of the order  
parameter changes by $2\pi $ when encircling this hole  
(see Fig.~\ref{fig:vorticesgrosspit}, upper panel, left).  
The value of the critical rotation frequency depends on the system parameters, 
but the angular momentum per particle $l=L/N$ 
equals unity 
when the vortex reaches the center. This result is also confirmed by 
the exact diagonalization calculations in the few-particle 
regime (see Fig.~\ref{fig:angularmomentum}a). 
The same mechanism of vortex entry was also found in the  
Gross-Pitaevskii study by~\textcite{kavoulakis2000}.  
In the limit of large $N$, \textcite{jackson2001} compared the energies
obtained in the Gross-Pitaevskii approach to those obtained by the CI method, 
and found that the mean-field results provide the
correct leading-order approximation to the exact energies within the same
subspace. 
For a more complete discussion of the mean-field theory 
of single-vortex formation in bosonic condensates, we refer 
to~\textcite{fetter2008}. 
 
\subsubsection{Single-vortex states in electron droplets} 
 
\label{sec:singlevortex} 
Two-dimensional electron droplets in quantum dots 
can be rotated by applying a perpendicular magnetic 
field. The number of confined electrons, 
as well as the rotation frequency can be controlled
by an external gate voltage and the field strength, respectively. 

In symmetric quantum dot devices the confining potential can often
be modeled accurately by a 2D 
harmonic potential~\cite{matagne2002,bruce2000,nishi2006}. 
These systems would therefore 
be ideal testbeds for analysis of vorticity in rotating fermionic systems 
with repulsive interactions. 
However, direct experimental 
detection of signatures of vortex formation in 
the electron density is very difficult due to small charge densities 
inside the electron droplet, that is often buried in a semiconductor  
heterostructure. Attempts to extract any signatures of vortex 
formation have usually focused on the analysis of 
quantum transport measurements~\cite{saarikoski2005c,guclu2005a}.  
 
In weak magnetic fields, electron droplets in quantum dots  
are composed of electrons which have their spin either 
parallel or antiparallel to the magnetic field. 
As the strength of the magnetic field increases,  
the system gradually spin-polarizes. 
For details on this process and electronic structure of quantum dots 
in this regime we refer to the reviews by~\textcite{kouwenhoven2001} 
and~\textcite{reimann2002}. 
The first totally spin-polarized state in the LLL 
is the maximum density droplet (MDD) state~\cite{macdonald1993} 
discussed in Sec.~\ref{sec:mdd}. 
The existence of this state was firmly established  
experimentally~\cite{oosterkamp1999} using quantum transport measurements.  
When the angular momentum is further increased with 
the magnetic field, the MDD state
reconstructs, and a vortex may form inside the electron droplet. 
 
The breakdown mechanism of the MDD and its interpretation has been 
one of the most discussed subjects in the early theoretical studies 
of quantum dots. Many of these works were inspired by 
the theory of excitations of the quantum Hall states. 
\textcite{macdonald1993} as well as \textcite{chamon1994}
discussed the possibility of edge excitations 
in large quantum Hall systems. Their 
studies suggested that the MDD would break up via reconstruction 
of the MDD edge. This possibility was 
examined further by \textcite{goldmann1999}  
using a set of trial wave functions 
which described a MDD state surrounded by a ring of 
localized electrons.
In large quantum dots, density-functional studies indicated 
a charge-density wave (CDW) solution 
along the edge of the dot~\cite{reimann1999} around a rigid MDD-like 
dot center. These studies showed that for larger dot sizes,  
a rotating single-component fermion liquid 
would not develop vortex states but instead the edge of the 
system would be excited around a rigid MDD-like center. 
However, Hartree-Fock calculations for small electron droplets 
predicted that holes are created inside 
the droplet that would bunch to minimize 
the exchange energy~\cite{ashoori1996}. 
\textcite{yang2002} used the exact diagonalization  
approach and also found the MDD state unstable towards creation of 
internal holes in high magnetic fields.
A skyrmion type of excitation above the MDD state was considered 
by~\textcite{oaknin1996}. This study generalized the theory of skyrmion 
type of excitations in the 2D electron gas (2DEG) 
\cite{ezawa2000} to finite-size quantum Hall droplets, which was motivated 
by localization of skyrmions in a Zeeman field. 
They proposed a wave function whose form for large particle numbers 
is that of a mean-field type of skyrmion excitation. 
\cite{heinonen1999} found also edge spin textures in 
an ensemble density-functional approach.
A skyrmion-type spin texture can be treated as another 
manifestation of vorticity, as pointed out in the context of 
two-component bosonic condensates, see Sec.~\ref{sec:multi-component}.
For quantum dots with four and six electrons, a recent study within the CI
method showed that meron excitations are dominant for the lowest-lying
states in very small quantum dots at strong magnetic fields 
(in the limit of vanishing Zeeman coupling), see \cite{petkovic2007}.   

\begin{figure} 
\includegraphics[width=.45\textwidth]{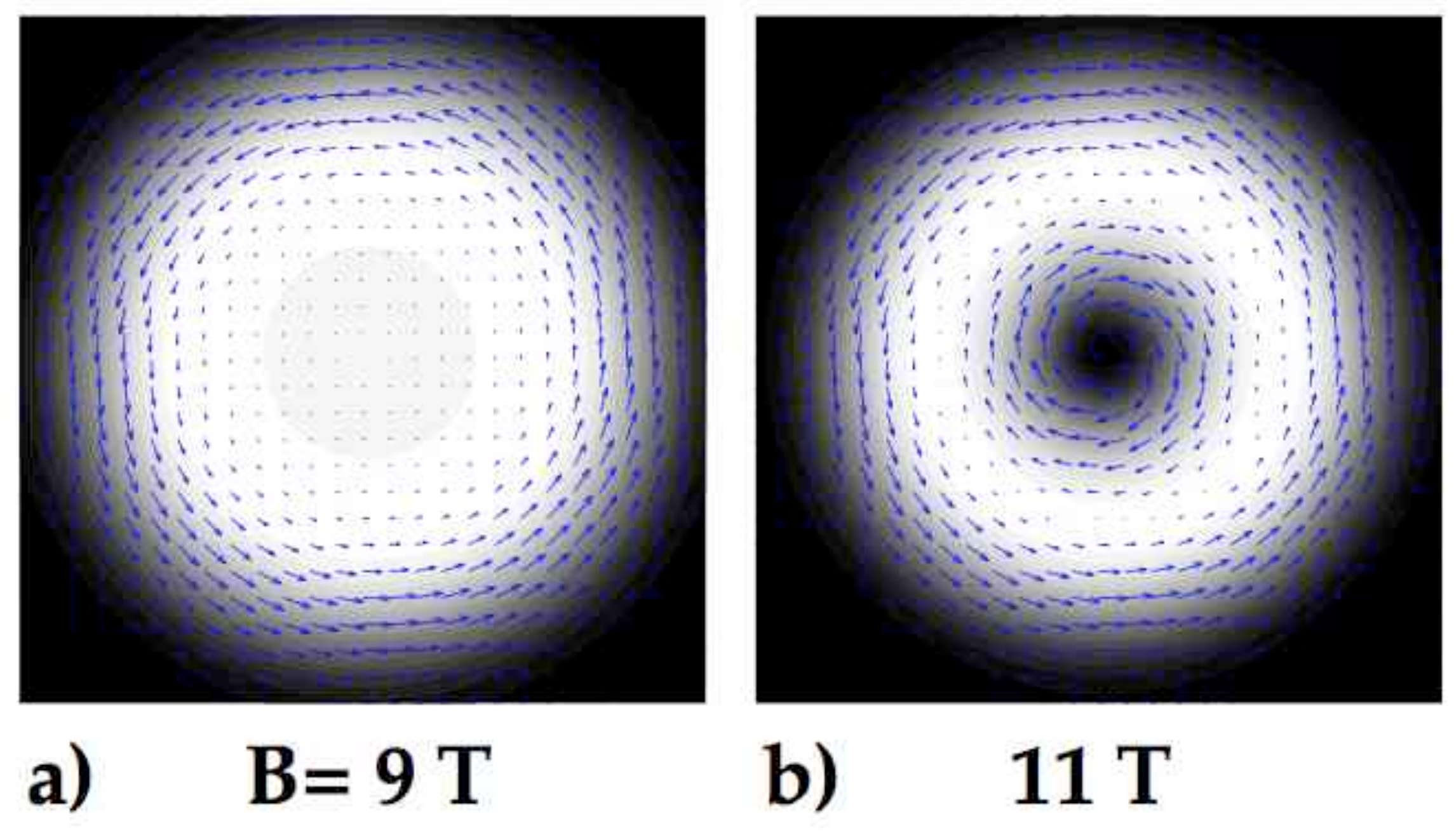} 
\caption{ 
a) Charge density (gray scale) and current density (arrows) 
in the maximum density droplet state of a 6-electron droplet 
at magnetic field $B=9\;{\rm T}$
calculated with the density-functional method.
The angular momentum is $L=15$, and   
the density inside the droplet is uniform. 
The solution shows also an edge current reminiscent of those 
in quantum Hall states. 
b) The single-vortex state in the same droplet at slightly increased 
magnetic field of $B=11~T$ with $L=21$.
It shows a pronounced vortex hole in the middle with 
a rotating current around it. Adapted from the results 
of~\textcite{saarikoski2004}. 
} 
\label{fig:mdd-single} 
\end{figure}

Holes in the charge density were identified as vortex cores 
in the density-functional studies of quantum dots~\cite{saarikoski2004} 
(see Fig. \ref{fig:mdd-single}). 
This work also directly showed with the configuration interaction method 
that for the $N=6$ case the nodal structure of the many-body wave 
function revealed an isolated vortex at the center of 
the dot.
%%%%
\begin{figure} 
\includegraphics[angle=90,width=.45\textwidth]{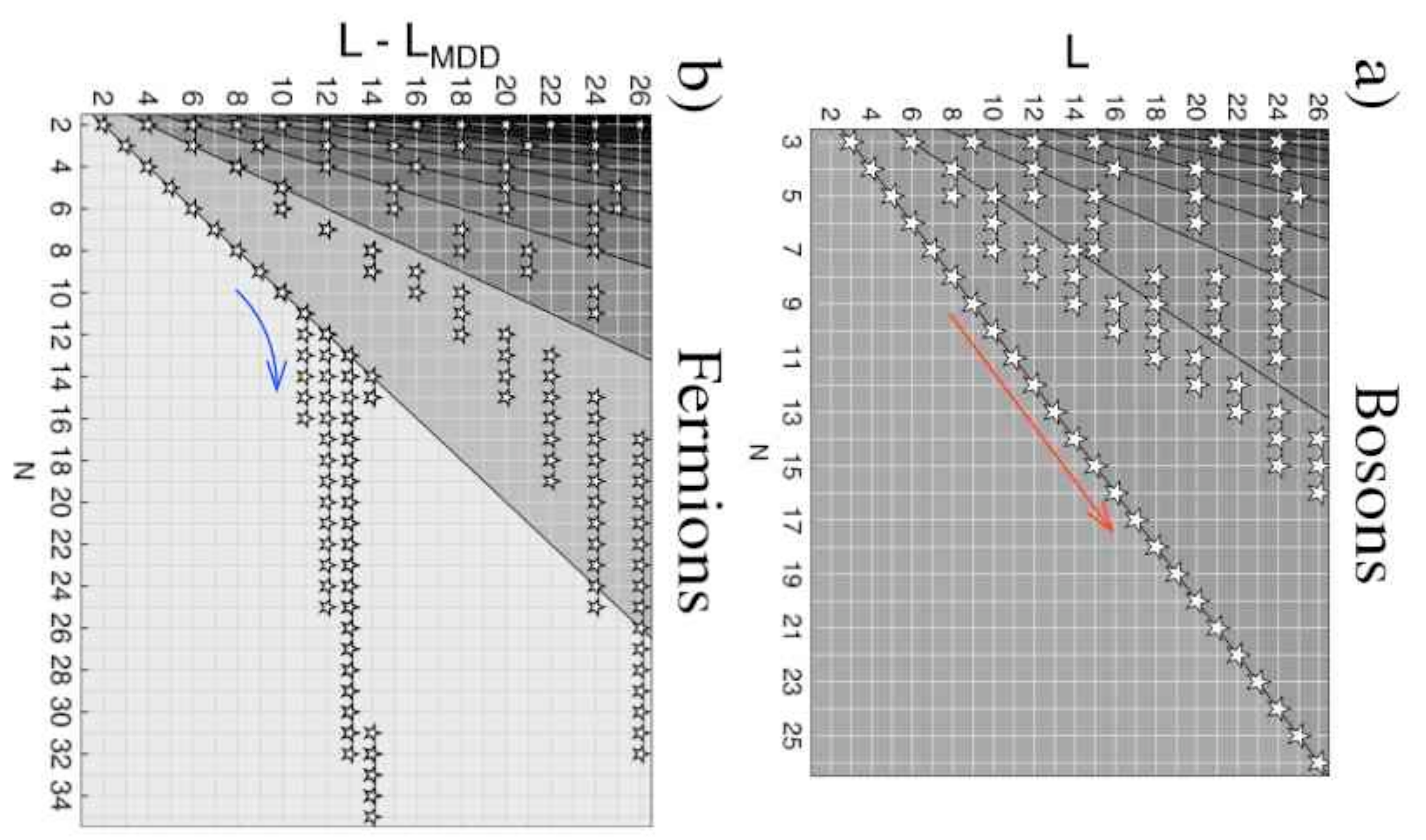} 
\caption{Systematics of boson and fermion ground states. 
When the external rotation $\Omega$ is gradually increased from zero, 
a droplet of $N$ particles goes through a series of ground states with 
increasing angular momentum $L$. 
Stars mark these $L$ values as a function of $N$ 
for a) boson droplets and b) fermion droplets. 
Calculations are done with the exact diagonalization method 
in the lowest Landau level approximation, and a harmonic confining 
potential Eq.~(\ref{eq:harmonic}). 
In the fermion results in b), the angular momentum of the maximum density 
droplet, $L_{\rm MDD}$, is substracted from $L$. 
The linear $N$ dependence of the first $(N,L)$-combination 
in bosonic systems (red arrow) indicates that the first $L$ above 
the non-rotating state has a central vortex. 
Fermionic systems with repulsive interactions 
show a similar behaviour only until $N=12$, where 
the breakdown mechanism of the MDD clearly 
changes (blue arrow), and a non-localized node 
emerges at a finite distance from the center. After~\textcite{harju2005} and \textcite{suorsa2006}. 
} 
\label{fig:angularmomentum} 
\end{figure} 
 
These results suggested that
the first magnetic flux quantum, which penetrates the electron droplet, 
is a free vortex and not bound to any particle as in the Laughlin 
wave function. Configuration interaction 
calculations for few-electron quantum dots provided further evidence for 
vortex formation in few-electron 
systems~\cite{toreblad2004,tavernier2004,manninen2005}. 
In the few-electron regime, the unit vortex can be 
localized at the center of the electron droplet, just like in the  
bosonic case discussed above.   
In this respect 
the vortex in few-electron droplets is a localized hole-like 
quasi-particle~\cite{saarikoski2004,manninen2005}. 
However, in the full quantum mechanical picture the vortex position 
in the electron droplet is always subject to fluctuations as shown 
by the above diagonalization studies. 
 
For bosonic systems, \textcite{bertsch1999}
suggested an ansatz (see Eq.~\ref{bertschpapenbrock}) 
to describe a single-quantized vortex at the center of the droplet at
$L/N=1$. 
Following~\textcite{manninen2001}  
a similar approximation for the corresponding single-vortex state 
in fermionic droplets can be defined with $L=L_{\rm MDD}+N$, 
\be 
\Psi_{1v}=\prod_{i=1}^N (z_i-z_c) \vert \mathrm{MDD}\rangle, 
\ee 
where $z_c$ is the center-of-mass coordinate, as defined above. 
When the number of electrons is large, the center-of-mass is, with a high 
accuracy, at the center of the trapping potential, and 
we can approximate $z_c=0$ and 
$\Psi_{1v}=\prod z_i \vert \mathrm{MDD}\rangle=\vert 0111\cdots
111000\cdots\rangle$ (for arbitrary $N$).  
For a single-vortex state where the hole is not located at the center, 
the wave function would be composed of single-determinants like 
$\vert 11011\cdots 11000\ldots\rangle$, where the position of 
the hole determines the average radius where the vortex 
is most likely to be found. 
The particle density has a minimum at the distance where the
amplitude of the empty single particle state has a maximum.
However, even in the LLL approximation  
the true many-body state is a mixture of all other determinants 
in the LLL subspace, and the exact  
vortex position is then subject to fluctuations. 
This effect can be captured by different trial wave functions. 
~\textcite{oaknin1995} constructed a nearly exact 
wave function for the single-vortex state.
\textcite{jeon2005} could describe the vortices in  
the composite fermion approach formulated for the hole states. 
This issue is discussed further in Sec.~\ref{sec:localization} 
which addresses vortex localization and fluctuations. 

\begin{figure} 
\includegraphics[width=.45\textwidth]{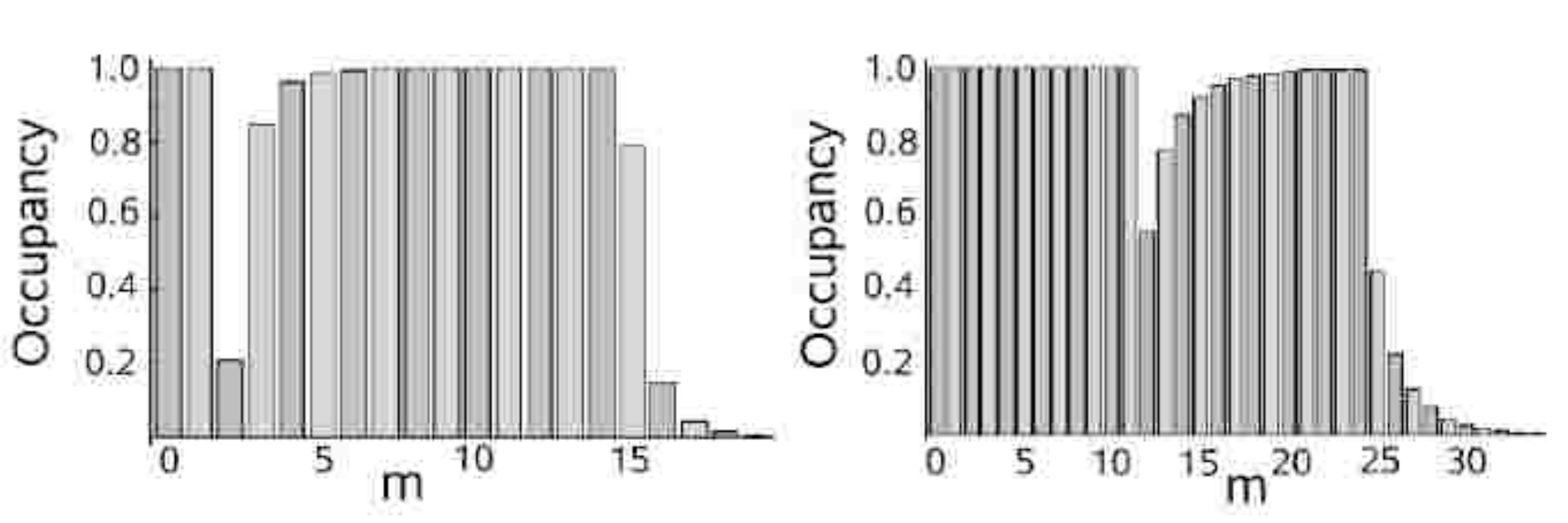} 
\caption{ 
Occupations of the single-particle (Fock-Darwin) eigenstates 
with angular momentum $m $ of fermions in a harmonic trap  
at ground states with $L=L_{\rm MDD}+13$ 
for $N=15$ (left) and $N=25$ (right). 
Since the mean particle distance from the center increases with $m $ 
the high-$N$ states resemble more edge excitations than central vortex 
states (cf. Fig. \ref{fig:angularmomentum}). 
After \textcite{suorsa2006}. 
} 
\label{fig:deltal13} 
\end{figure}

In a bosonic system, the yrast line has a pronounced cusp at angular momentum 
$L=N$ (see Fig.~\ref{fig:bosonyrastci}), 
corresponding to a state with a single-quantized vortex 
at the center of the trap. In a fermion system, however, the first cusp 
of the yrast line is not necessarily a central vortex state. 
\textcite{yang2002} have 
shown that a (vortex) hole is created at the center 
of the dot for low electron numbers. When $N > 13$ the hole 
locates at a finite distance from the center.  
In circularly symmetric systems, such a delocalized node would not be 
associated with the usual rotating charge current around a localized 
vortex core. A qualitatively 
similar regime of $N < 13$ for the central vortex 
was obtained within a spin-density-functional 
analysis~\cite{saarikoski2005c}. 
Calculations using the ``rotating electron molecule''-model 
reported a lower limit, $N<7$~\cite{li2006}. 
In the exact diagonalization studies in the LLL~\cite{harju2005} 
the ground-state angular momenta for 
the first cusp state beyond the MDD-state shows a marked change in 
the $N$-dependence above $N=12$ (Fig. \ref{fig:angularmomentum}b). 
For $N<12$ the node of the first cusp state is at the center of the 
electron droplet as indicated by its angular momentum $L=L_{\rm MDD}+N$.
These solutions can be readily identified as vortex states. 
However, for $N \ge 12$ the angular momentum increase 
is almost independent of $N$, which is an indication that the  
node can not reach the center but stays delocalized
close to the edge, as illustrated 
in Fig.~\ref{fig:deltal13}. This solution can also be interpreted as 
an edge excitation which helps to understand 
why different models and methods 
yield seemingly contradictory results for the MDD reconstruction, 
as discussed above. 
 
The intermediate angular momentum states between 
the MDD and the $\Delta L=N$ central vortex states 
show a node in the wave function at a finite distance from the 
center~\cite{oaknin1995,saarikoski2005b} that 
can be interpreted as a delocalized vortex, 
{\it i.e.}, a vortex approaching the center from the droplet surface 
as in the case of Figs. \ref{fig:vortexentryci}  
and \ref{fig:buttssinglevortex} for bosons. 
Note that these delocalized vortex states can be interpreted as center of 
mass excitations, 
as explained in connection with Eq. (\ref{eqcomex}).

For larger electron numbers it is energetically more favorable 
to generate two (or even more) vortices already at $L/N=1$. 
In other words, the wave function shows then two or more delocalized 
nodes at a finite distance from the center at $L/N=1$.
This is contrary to Bose systems, where the central
vortex state is the lowest-energy state at $L/N=1$ for any
particle number (see Fig.~\ref{fig:angularmomentum}a and b). 
Apart from this fact,  
vortices in both fermionic and bosonic systems are manifest in the nodal  
structure of the wave function in a very similar  
manner~\cite{toreblad2004,borgh2008}. 
 
\subsection{Vortex clusters and lattices} 
 
\label{sec:lattices} 
When the angular momentum of the quantum droplet increases with rotation, 
additional vortices successively enter the cloud of particles. 
Normally, in a harmonic trap  
these vortices are all singly-quantized and arrange in  
simple geometries, as it was observed for a rotating Bose-Einstein condensate  
in the early experiment by~\textcite{madison2000}, see  
Fig.~\ref{fig:madisonvortices}. 
With increasing system size and rotation, the vortices order in  arrays that  
resemble a triangular Abrikosov lattice  
\cite{aboshaer2001,ho2001}.  

\subsubsection{Vortex lattices in bosonic condensates} 
 
\begin{figure}[ptb]  
\includegraphics[width=0.45\textwidth]{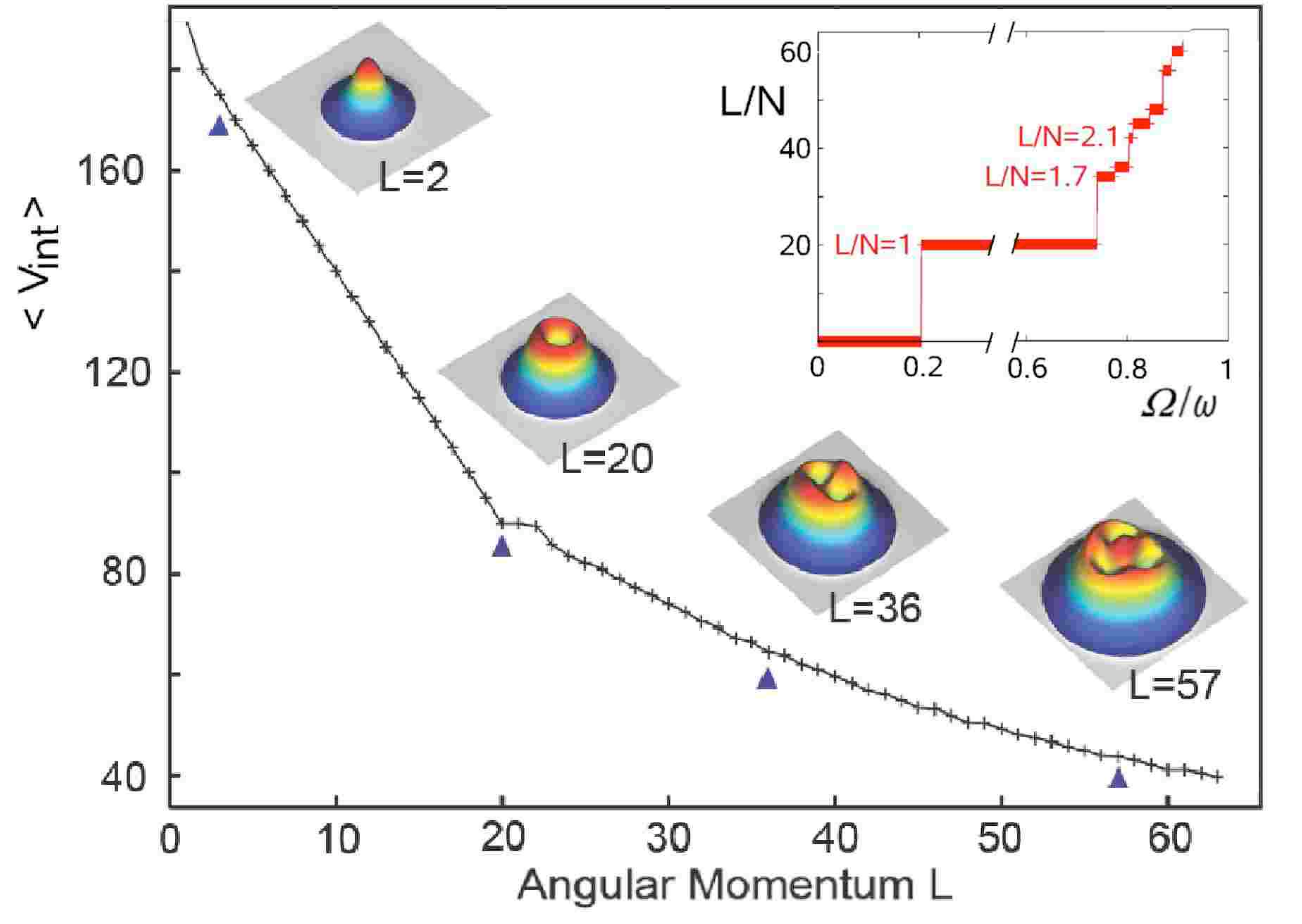} 
  \caption[]{({\it Color online}) Yrast line of $N=20$ {\it bosons}  
in a harmonic confinement, obtained by the CI method in the  
lowest Landau level and for 
contact interactions between the bosonic particles.  
The inset shows the  
total angular momentum of the ground state, plotted as a function  
of $\Omega /\omega$. 
Pair-correlated densities (renormalized in height) are shown  
for increasing angular momentum  
per particle, $l=L/N=0.1, 1.0, 1.8, $ and $2.85$ (as marked by the  
blue triangles). The reference point was chosen at high density  
for radii of order unity. After~\cite{christensson2008b}} 
\label{fig:bosonyrastci} 
\end{figure}

Let us begin by investigating the vortex structures along the yrast line, 
{\it i.e.}, let us 
study the states with highest angular momentum $L$ at a given 
energy. 
Figure~\ref{fig:bosonyrastci} shows the yrast line for $N=20$ bosons  
up to $L=3N$, calculated by exact diagonalization.
The vortex is located at the center when $L/N=1$. 
The inset at $L=20$ shows the pair-correlated density for that state, with a 
pronounced minimum at the origin. 
At angular momenta  $L>N$, the slope of the yrast line changes abruptly,  
and the spectrum is no longer linear beyond the first cusp at $L/N=1$. 
The inset in Fig.~\ref{fig:bosonyrastci} shows the angular momenta of the 
lowest-energy states  for a given rotational frequency $\Omega $ of the trap, 
that are obtained by minimizing the energy in the  
rotating frame, $E_{\rm rot} = E_{\rm lab} - \Omega L $.  
The pronounced plateaus correspond to stable states  
with vortices, that successively enter the bosonic cloud  with 
increasing trap rotation.  
Below a certain critical angular frequency, the cloud remains in the $L=0$ 
ground state. Beyond that frequency, the axially-symmetric single 
vortex  at the center
becomes the ground state, until more vortices penetrate  
the trap as the rotation increases.
In the exact results for small atom numbers, the vortices appear as  
clear minima in the pair-correlated densities, as here shown for the example  
of a two-vortex solution at $L/N=1.8$, and a three-vortex state, as here for   
$L=2.85$, see Fig.~\ref{fig:bosonyrastci}. 
Related results of vortex formation in small systems have
for example 
been studied by \textcite{barberan2006},~\textcite{dagnino2007} 
and~\textcite{romanovsky2008}. 

For weakly interacting bosons, many states between angular momenta 
$L=N$ and $L=N(N+1)$ can be described well with the composite particle 
picture~\cite{cooper1999,viefers2000}.
\textcite{cooper1999} have shown that for most states with a clear cusp in the
yrast line, the overlaps between the exact wave function and that of the 
Jain construction is in general very close to one for particle numbers 
$N\le 10$. \textcite{wilkin2000} furthermore showed that at some angular
momenta in this region, the so-called Pfaffian state is a good analytic 
approximation for the exact wave function. 

These findings are very similar to the results of the  
mean-field Gross-Pitaevskii method, where one finds successive transitions  
between vortex states of different symmetry. 
With increasing angular momentum, the arrays of singly-quantized vortices  are
characterized by a phase jump 
of the order parameter around the density minima at the vortex cores 
\cite{butts1999,kavoulakis2000}.   

\begin{figure}[ptb] 
\includegraphics[width=0.45\textwidth]{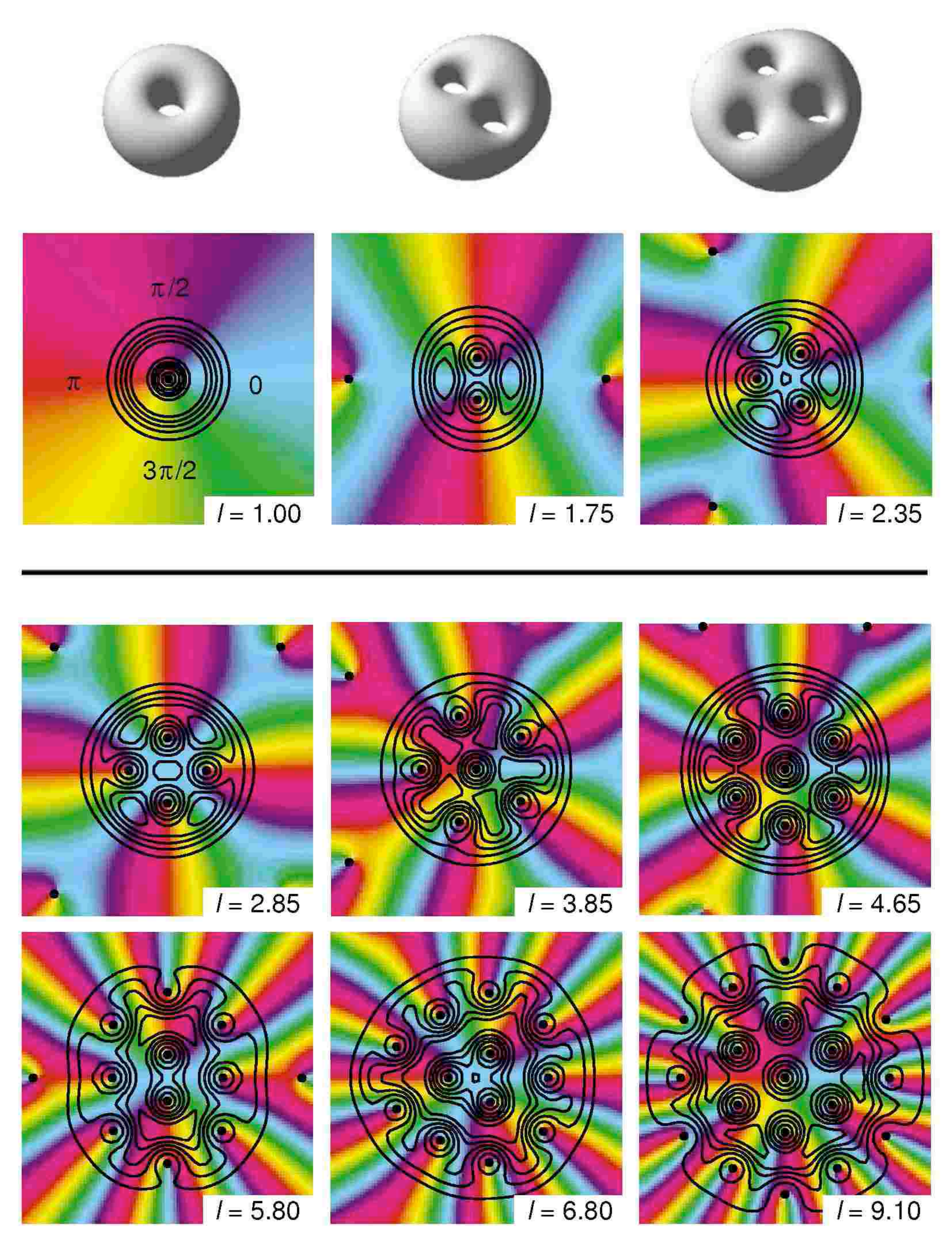} 
\caption{\baselineskip11pt  
({\it Color online}) Vortices in a rotating cloud of bosons.
Schematically shown are the vortex holes that  
penetrate the boson cloud with increasing  
angular momentum. 
The lower panel shows the phase of the order parameter, and its density  
contours. (The black dots indicate the vortex positions). After 
~\textcite{butts1999}. } 
\label{fig:vorticesgrosspit} 
\end{figure}

Figure~\ref{fig:vorticesgrosspit} shows a schematic picture of  
the equidensity surfaces for the unit vortex, a two-vortex and  
three-vortex state in the upmost panel, as well as  
the contours and the corresponding phase  
of the order parameter at higher ratios $l=L/N$, as demonstrated 
by~\textcite{butts1999}.  
At angular momenta beyond the unit vortex, the rotational symmetry of the  
mean-field solutions is broken.  At $L\ge 1.75 N$ the optimized  
Gross-Pitaevskii wave function shows a two-fold symmetry when the 
second vortex has entered the cloud, in much similarity to the aforementioned 
experimental results for  
$^{87}$Rb ~\cite{madison2000}, and in agreement with the  
pair-correlated densities in Fig.~\ref{fig:bosonyrastci} above.  
Higher rotational frequencies introduce new configurations of vortices. 
At $l\approx2.1$ there is a state with three vortices symmetrically 
arranged around the center of the trap. 
As $l=L/N$ increases, more and more vortices enter the 
cloud~\cite{butts1999,kavoulakis2000}, and eventually the 
vortices arrange in a pattern that 
resembles a triangular lattice~\cite{ho2001,baym2003,baym2005}. 
This is in agreement with the experiments which were able to reach and 
image the angular momentum regime 
where large vortex arrays emerge~\cite{aboshaer2001},  
reminiscent of the Abrikosov lattices in type-II superconductors. 
Stable {\it multiply-quantized} vortices with phase shifts larger  
than $2\pi $ were not obtained~\cite{madison2000} 
for a one-component Bose gas in the purely harmonic trap, 
in agreement with the theoretical results discussed above.  

As we discussed in detail in Sec.~\ref{sec:gp}, the effective mean-field 
potential in the Gross-Pitaevskii approach may break the rotational 
symmetry of the Hamiltonian to lower the energy. 
As a consequence, such a mean-field solution for the order 
parameter is not an eigenstate of the angular momentum operator 
and the solution may reflect the internal symmetry of the exact quantum 
state. Similar behavior has been observed also in density-functional 
studies of quantum dots~\cite{reimann2002}, 
and is further discussed also in the review by ~\textcite{cooper2008}. 
  
\begin{figure}[ptb]  
\includegraphics[width=0.5\textwidth]{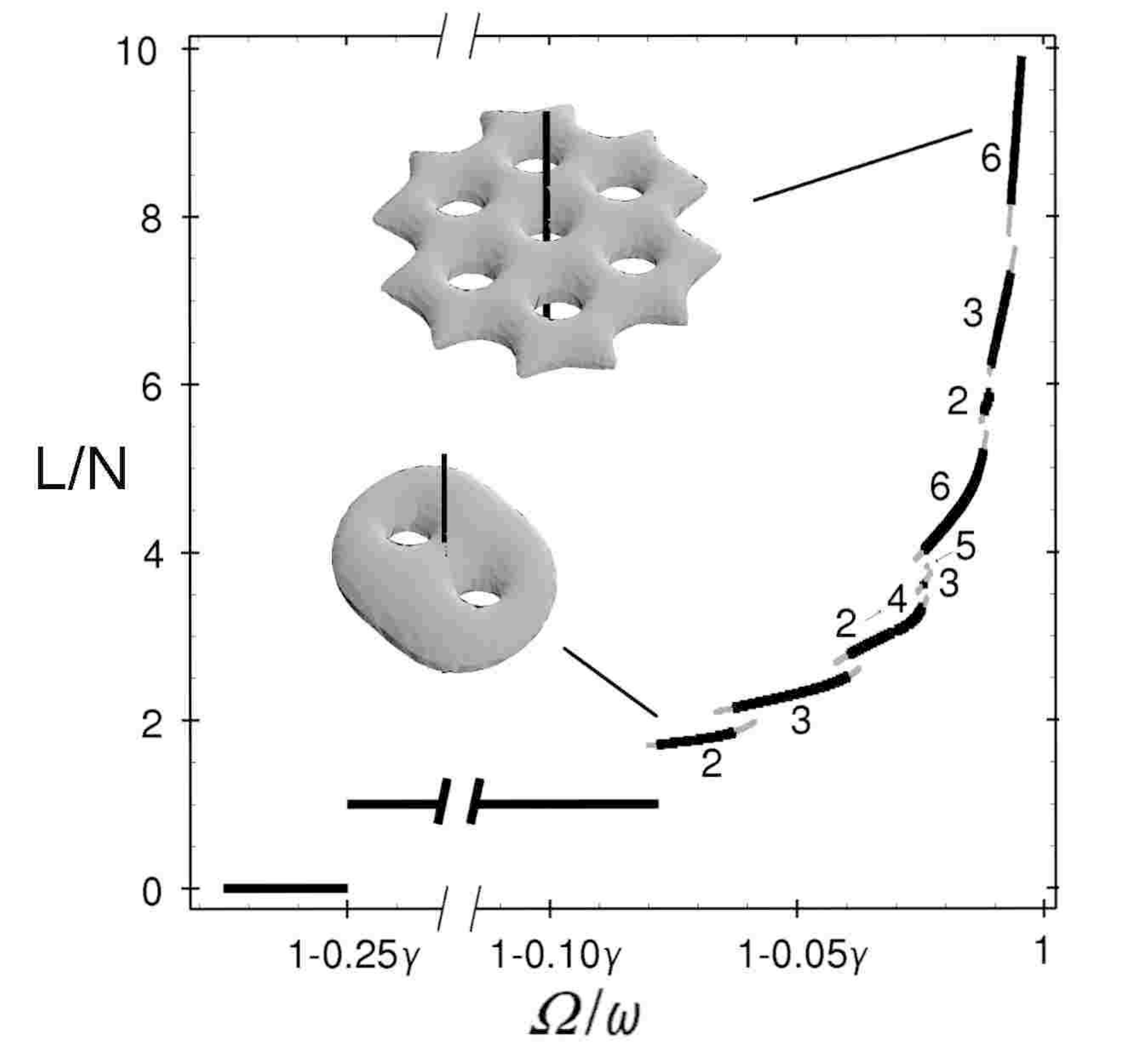} 
  \caption[]{Angular momentum per particle, $L/N$, 
as a function of the rotational frequency $\Omega /\omega $ of the trap.  
The discontinuities correspond to the transitions between different  
symmetries. The insets show the surfaces of constant density
in a spherical trap for states with two and six vortices. 
$\gamma=(2/\pi)^{1/2}aN/\sigma_z$, $a$ being the scattering length and 
$\sigma_z$ the axial width of the ground state of a
single particle in the trap. From ~\textcite{butts1999}. 
} 
\label{fig:buttsrokhsar} 
\end{figure}

Figure~\ref{fig:buttsrokhsar} shows the expectation value of the 
angular momentum of a bosonic cloud as a 
function of the angular velocity of the trap, as obtained 
from the Gross-Pitaevskii approach~\cite{butts1999}.  
The discontinuities in $l=L/N$ correspond to the 
topological transformations of the 
rotating cloud that are associated with the occurrence of additional 
vortices, as discussed above.   
 
In the purely harmonic trap,  
the oscillator  frequency $\omega $ limits the angular rotation  
frequency $\Omega $, see Eq.~(\ref{aspectratio}). 
When both quantities finally  
become equal, the condensate is no 
longer confined, and the atoms fly apart.

\subsubsection{Vortex molecules and lattices in quantum dots} 

\begin{figure} 
\includegraphics[width=.45\textwidth]{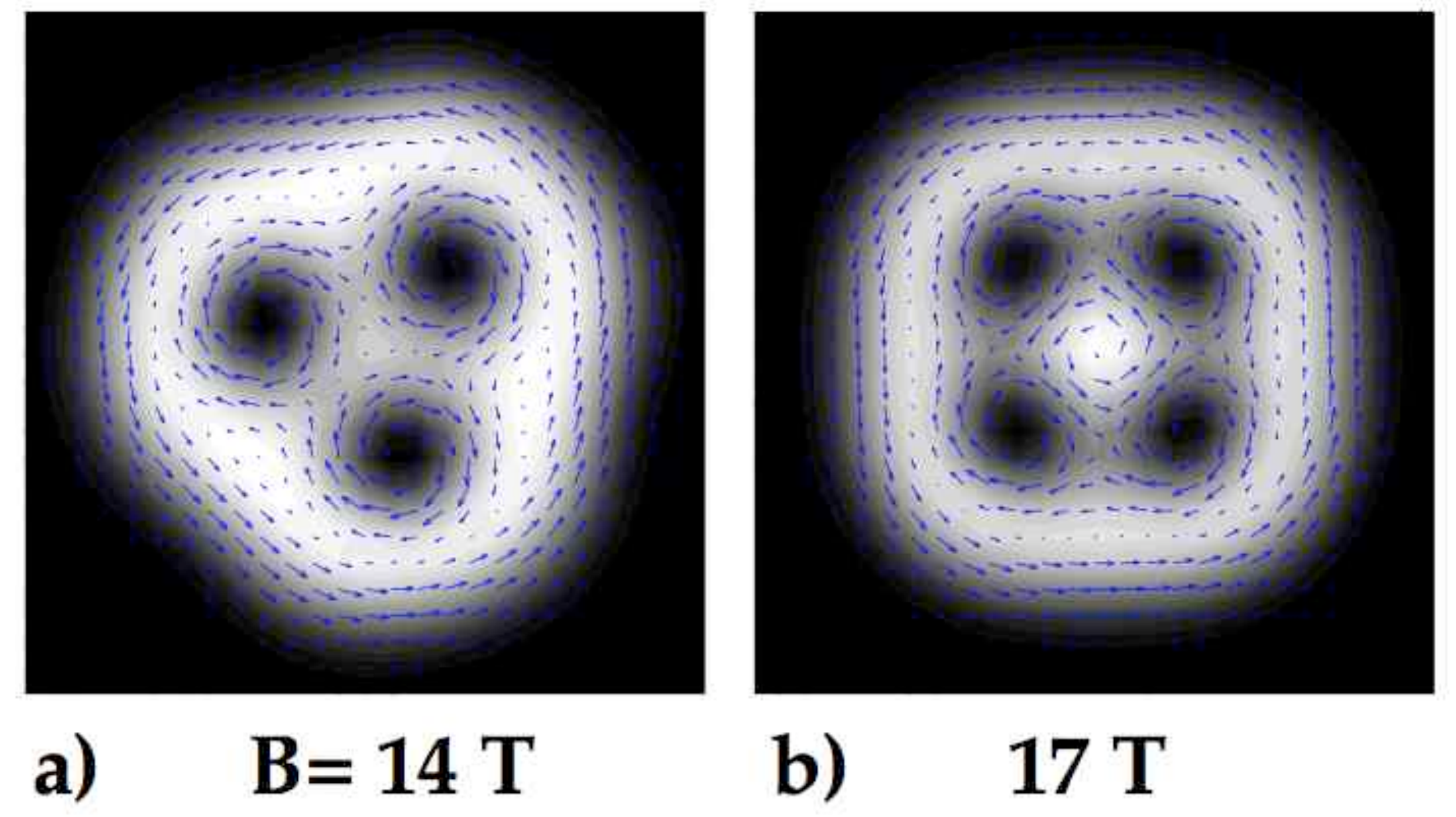} 
\caption{ ({\it Color online}) 
Vortex molecules in a 6-electron droplet. 
Charge density (gray scale) and current density (arrows) 
show rotating currents around a) three and b) four 
localized vortex cores in density-functional calculations. 
After~\textcite{saarikoski2004}. 
} 
\label{fig:few-vortex} 
\end{figure}

\begin{figure}[h!!] 
\includegraphics[width=.45\textwidth]{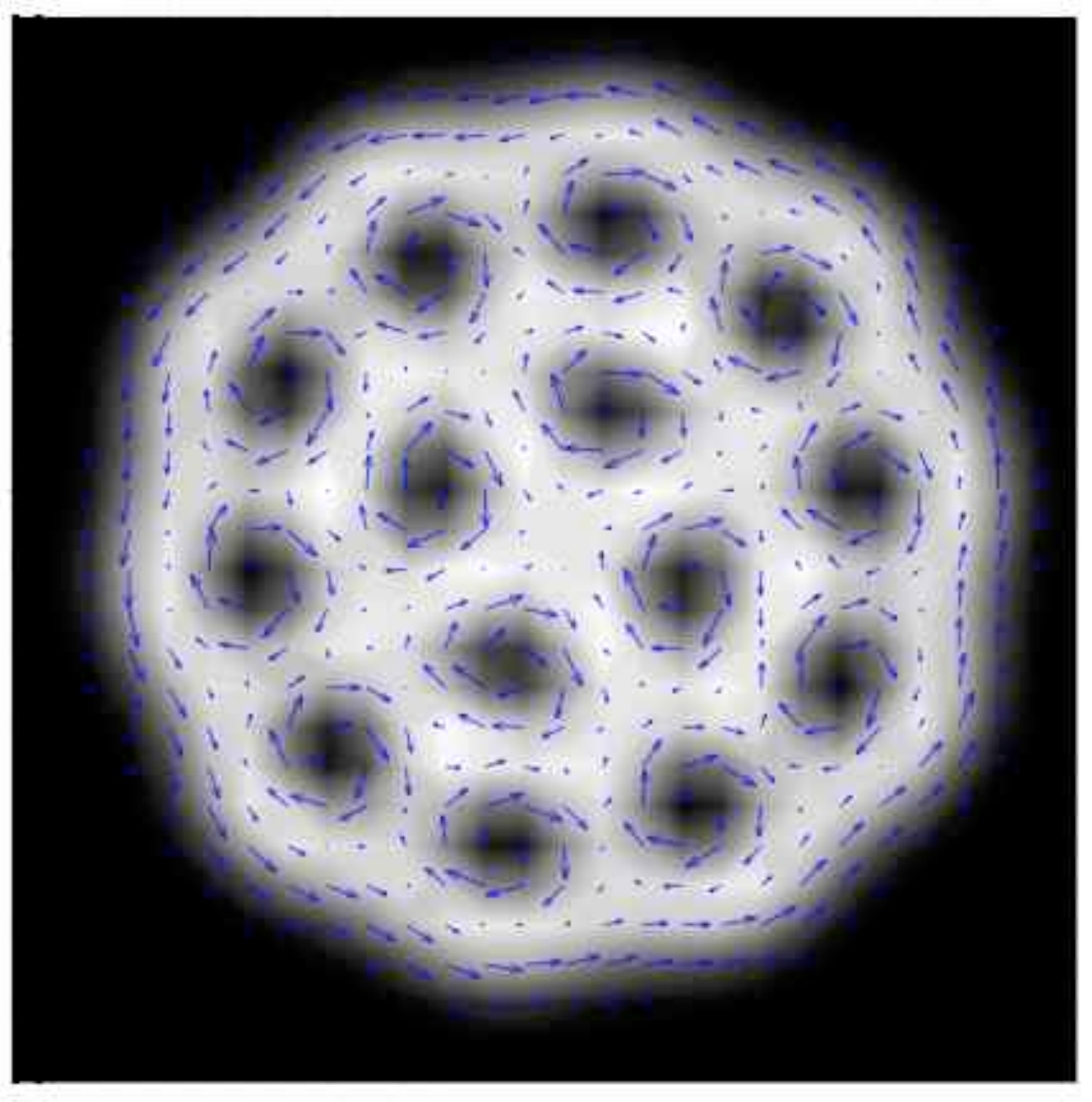} 
\caption{({\it Color online}) Electron density (gray scale) 
and current density (arrows) in a 24-electron quantum dot 
calculated with the density functional method. The solution shows a cluster of 
14 localized vortices arranged in two concentric rings.
After~\textcite{saarikoski2004}. 
} 
\label{fig:n24} 
\end{figure}

The close analogy between the bosonic ground state, $|N00000\cdots \rangle $ 
at $L=0$, and the fermionic maximum density droplet state, 
$|111\dots111000\dots\rangle $ at $L_{\rm MDD}=N(N-1)/2$, 
(see Sec.~\ref{sec:fermionboson}) suggests that vortex 
lattices may emerge also in fermionic systems to carry the angular momentum. 
Indeed, density-functional studies predicted the emergence of clusters 
or ``vortex-molecule''-like geometric arrangements of 
vortices inside small droplets of electrons in 
quantum dots~\cite{saarikoski2004} when the 
angular momentum increases beyond the MDD. This happens 
in a very similar way as in bosonic droplets at small rotation  
frequencies~\cite{toreblad2004}. An example of these vortex molecules 
in few-electron quantum dots is shown in Fig.~\ref{fig:few-vortex}.
Figure ~\ref{fig:n24} shows a cluster of 14 vortices 
in a 24-electron quantum dot calculated with the density-functional method 
in a local spin-density approximation (see Sec.~\ref{sec:dft} above). 
These vortices correspond to off-electron nodes. The filling factor 
of the state in Fig.~\ref{fig:n24} can be approximated 
as $\nu \approx 0.63$. As in the bosonic systems vortex clusters 
are composed of single-quantized vortices~\cite{saarikoski2005a}. 
Remarkably, the structure of the vortices that appear localized on two
concentric rings with four vortices on the inner, and ten vortices 
on the outer ``shell'', matches that of a classical 
Wigner molecule with 14 electrons at the verge of 
crystallization~\cite{bedanov1994}. 
This also holds for the three- and four-vortex solutions shown 
in Fig. \ref{fig:few-vortex}, where the triangle and square match the three- and 
four-particle classical Wigner-molecule configurations. 
 
The clustering of vortices has also been analyzed 
with the CI method using reduced wave functions 
(see Sec. \ref{sec:cond-wave}) in the case of few-electron 
circular~\cite{saarikoski2004,tavernier2004,tavernier2006,stopa2006} 
and elliptical~\cite{saarikoski2005a} quantum dots. 
In these studies the formation of few-vortex molecules 
has been found to follow a 
similar pattern in both the CI method and the density functional method. 
 
Using the idea of the Bertsch-Papenbrock ansatz~\cite{bertsch1999} 
and assuming $n$ fixed vortex sites, we can anticipate 
that the single determinant 
describing a vortex ring would be~\cite{toreblad2004} 
\be 
\Psi_{nv}=\prod_{j=1}^N \prod_{k=1}^n (z_j-ae^{i2\pi k/n}) \vert \mathrm{MDD}\rangle 
=\prod_{j=1}^N (z_j^n-a^n) \vert \mathrm{MDD}\rangle, 
\ee 
where $a$ is the radius of the ring of vortices. 
This wave function is not an eigenstate of the angular momentum, but 
it can be projected out by collecting the states with a given 
power of $a$ and symmetrizing the polynomial multiplying the  
$\vert \mathrm{MDD}\rangle$: 
\be 
\Psi_{nv}=a^{n(N-K)}{\cal S}\left(\prod_{j=1}^K  z_j^n\right)\vert\mathrm{MDD}\rangle, 
\label{vgener} 
\ee 
where ${\cal S}$ is the symmetry operator and $K$ determines the average 
radius of the vortex ring. For example, with $N=7$, $K=5$ and $n=3$, 
the most important configuration is  
$\vert 1100011111000\cdots \rangle$, in agreement with the   
CI calculations (in the LLL approximation) 
for vortex rings by \textcite{toreblad2004}.
% and describe 
%the state in a mean field model~\cite{manninen2006}. 
%Note that for   
%$n=1$, $K$ going from 1 to $N$ corresponds to the single vortex 
%moving from the surface of the cloud to the center, with increasing 
%angular momentum. 
We discuss localization and fluctuations of vortices further in 
Sec. \ref{sec:localization}. 

\begin{figure}
\includegraphics[width=0.8\columnwidth]{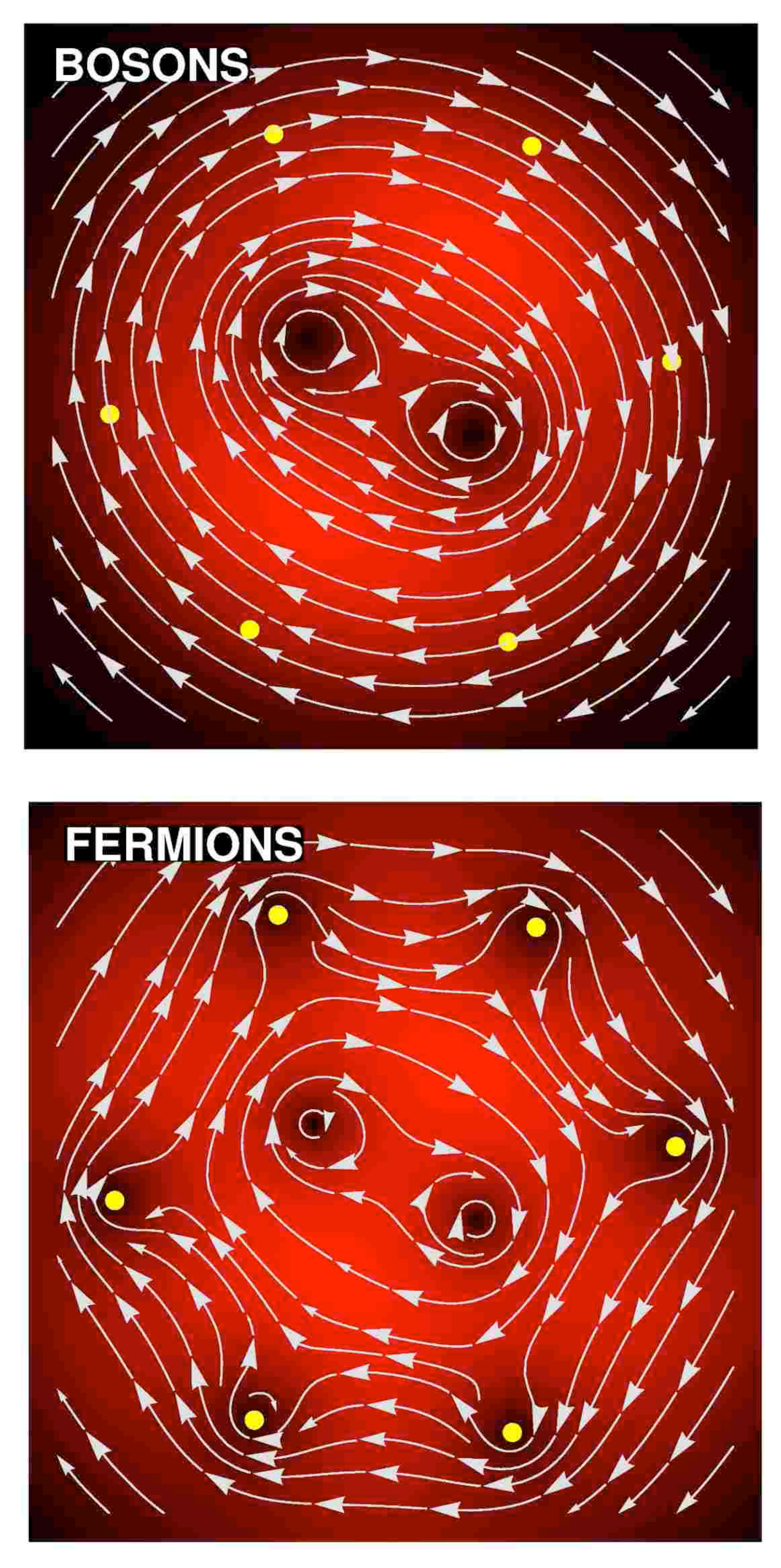}
\caption{({\it Color online}) Reduced wave function representation of
of the vortex structure of the model wave function Eq.~(\ref{vgener}) 
for bosons and fermions ($N=7$, $n=2$, $K=2$). 
The fixed particles are shown as light dots,
the current field with arrows (logarithmic scale) and the particle density as
shades of red (light color corresponding to high density).
}
\label{coverfig}
\end{figure}

Equation~(\ref{vgener}) also elucidates the origin of 
different vortex types and the similarity
of fermion and boson systems. The zeros of the symmetric polynomial
${\cal S}(\prod z_j^n)$ give the free vortices, while the zeros
of $\vert \hbox{MDD}\rangle$ give the Pauli vortices.
In a boson system, $\vert \hbox{MDD}\rangle$ is replaced with
the boson condensate $\vert 0\rangle$ which has no zeros, and only the
free vortices appear, as illustrated in Fig.~\ref{coverfig}.
 
Studies of electron-vortex correlations in 
quantum dots indicate that, at least in few-electron systems, 
the electron-vortex separation $d_{\rm e-v}$ can be approximated by a universal 
quadratic function of the filling factor,
$d_{{\rm e-v}} \sim d_{\rm e-e}\nu^2$, where $d_{\rm e-e}$ is the 
average electron-electron separation~\cite{anisimovas2008}. 
This shows that  
in the limit of high angular momentum (low $\nu$) electrons tend to 
attract vortices closer to electron positions, 
which eventually leads to the formation of 
electron-vortex composites and the emergence of finite-size 
counterparts of the quantum Hall states. 

The ``rotating electron molecule'' approach 
\cite{yannouleas2002,yannouleas2003} has also been 
used to analyze correlations between particles and vortices in electron 
droplets. However, this approach has been found to underestimate electron-vortex 
correlations~\cite{anisimovas2008} and vortex attachment 
to particles in the limit of high angular momentum~\cite{tavernier2004}. 
 
The vortex-molecule-like characteristics of the states are expected 
to vanish gradually with increasing vorticity.
However, exact diagonalization studies of few-electron systems with 
Coulomb interactions have suggested 
that the above-described vortex ordering into Wigner-molecule-like shapes
continues down to a filling factor $\nu={1\over 2}$, 
where the electron number equals the (off-electron) 
vortex number~\cite{emperador2006}. 
At $\nu={1\over 2}$ the structure of the state is complex~\cite{emperador2005} 
and possible electron pairing in this regime has been 
studied~\cite{harju2006,saarikoski2008}. 
This filling factor marks also the beginning of a regime $\nu<{1\over 2}$ 
where the vortex attachment to particles becomes 
pronounced~\cite{emperador2006}. 
We further discuss the breakdown of vortex molecules and the  
emergence of fractional quantum-Hall-liquid-like states in Sections 
\ref{sec:localization} and \ref{sec:rapid}, respectively.)

\subsubsection{Signatures of vortices in electron transport} 
 
\label{sec:transport} 
For quantum dots in the fractional quantum Hall regime, 
where vortices have been predicted to form, electron transport measurements 
have revealed a rich variety of transitions associated with charge 
redistribution within the electron droplet~\cite{ashoori1996,oosterkamp1999}. 
 
\begin{figure} 
\includegraphics[width=.49\textwidth]{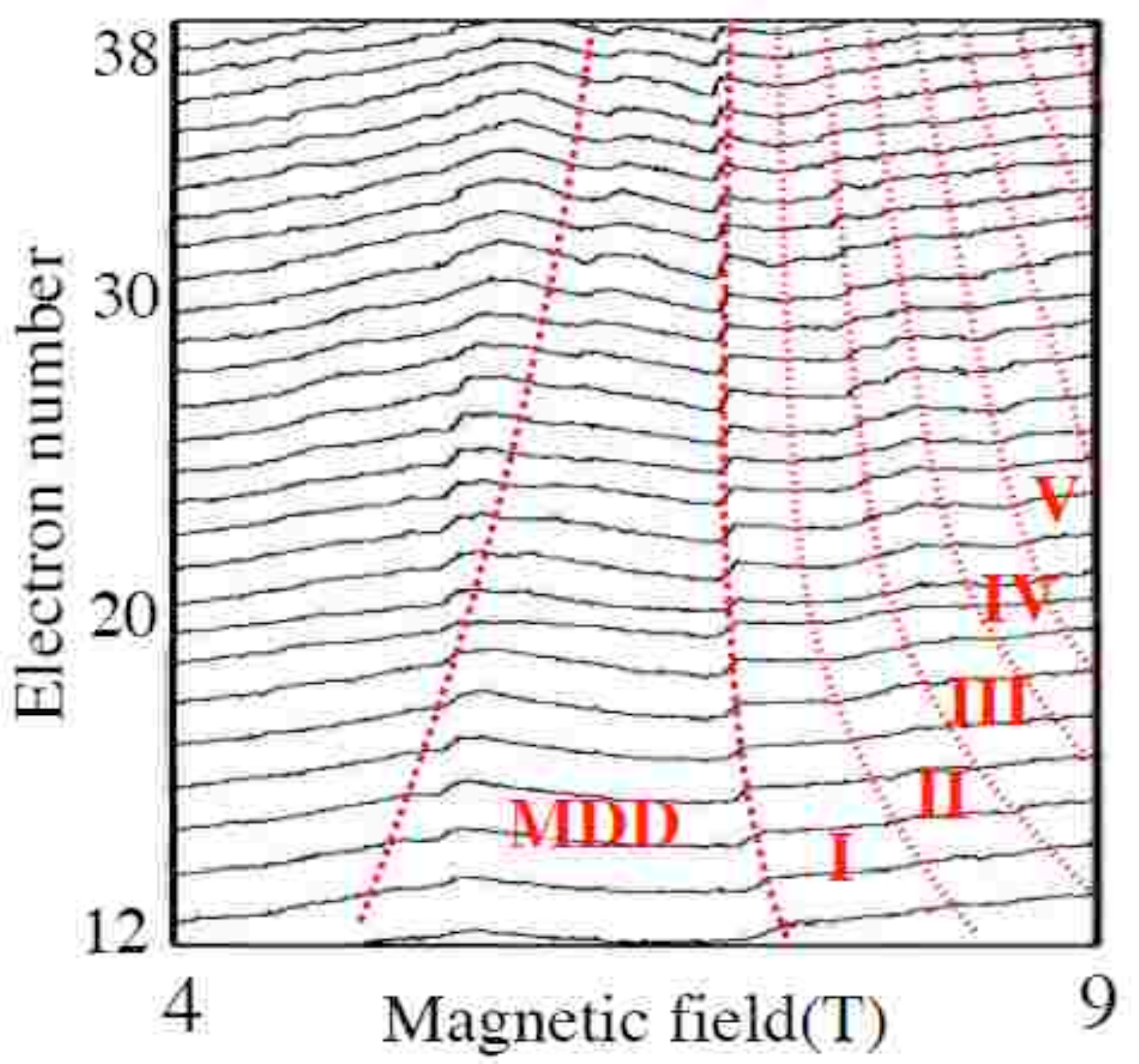} 
\caption{ 
Current peaks in the electron transport experiments and transitions 
in the spin-density-functional theory (red lines). 
The dashed lines denote the MDD boundaries and the roman numerals 
indicate number of vortices in the theory. 
After~\textcite{saarikoski2005c};  
the experimental data are from Fig.~2b in Ref.~\cite{oosterkamp1999}. 
} 
\label{fig:comparison} 
\end{figure}

Quantum dots contain a tunable and well-defined number of electrons. 
The electron transport experiments in the Coulomb blockade regime 
at low temperatures (around 100 mK) measure the chemical potential  
\begin{equation} 
\mu(N)=E(N)-E(N-1), 
\label{eq:chempot} 
\end{equation} 
which gives the minimum energy needed to add one more electron 
to the electron droplet. 
Transitions in the electron transport data can be seen as cusps or 
jumps in the chemical potential. Different quantum Hall regimes can be 
identified from these characteristic features
of the chemical potential as a function 
of both the electron number and the magnetic field, see 
Fig.~\ref{fig:comparison}.  

In experimental studies of vertical quantum dots, a harmonic external 
potential has been found to give a good approximation 
of the confining potential~\cite{matagne2002}. 
The harmonic confinement strength $\hbar 
\omega_0$ is determined by the size of the quantum dot device,  and 
usually depends on the number of electrons $N$ inside the quantum dot.
The area of the electron droplet has been found to increase with the 
gate voltage suggesting that the electron density in the 
droplet remains constant \cite{austing1999}. 
Confining potentials scaling as $\hbar \omega_0\sim N^{-1/4}$ 
in~\cite{koskinen1997} 
or  $\hbar \omega_0\sim N^{-1/7}$ in~\cite{saarikoski2005c} 
have been used in order to compare with experimental data. 

The MDD state in quantum dots is the finite-size counterpart of the 
$\nu=1$ quantum Hall state. Its existence has been firmly established 
in experiments since it gives rise to a characteristic shape in the chemical 
potential at $\nu=1$~\cite{oosterkamp1999}. 
The MDD state assigns one Pauli vortex at each electron position 
giving a total magnetic flux of $N\Phi_0$. 
As the rotation is further increased, the MDD 
reconstructs~\cite{chamon1994,goldmann1999,reimann1999,macdonald1993,toreblad2006}, 
and a vortex enters the electron droplet. 
This transition occurs approximately when the magnetic flux $\Phi=BA$ 
through the MDD of area $A$ exceeds $(N+1)\Phi_0$. 
Subsequent transitions involve an increasing number of such 
off-electron vortices~\cite{saarikoski2004,toreblad2004}.
Assuming a constant electron density in the droplet, 
the change in $B$ required for the 
addition of subsequent off-electron vortices 
in the droplet is approximately $\Delta B=\Phi_0n/N$. 
This result can be compared to density-functional calculations, 
which indicates a $1/N$-dependence of the 
spacing between the first major transitions after the MDD state. 
However, the limited accuracy of the available electron transport data 
at present does not allow to draw any more firm conclucions. 

The different ground states obtained within density-functional theory 
are compared to electron transport 
data in Fig. \ref{fig:comparison}. The transition patterns in theory 
and experiment show a narrowing of the stability domain of the MDD. 

\begin{figure} 
\includegraphics[width=.45\textwidth]{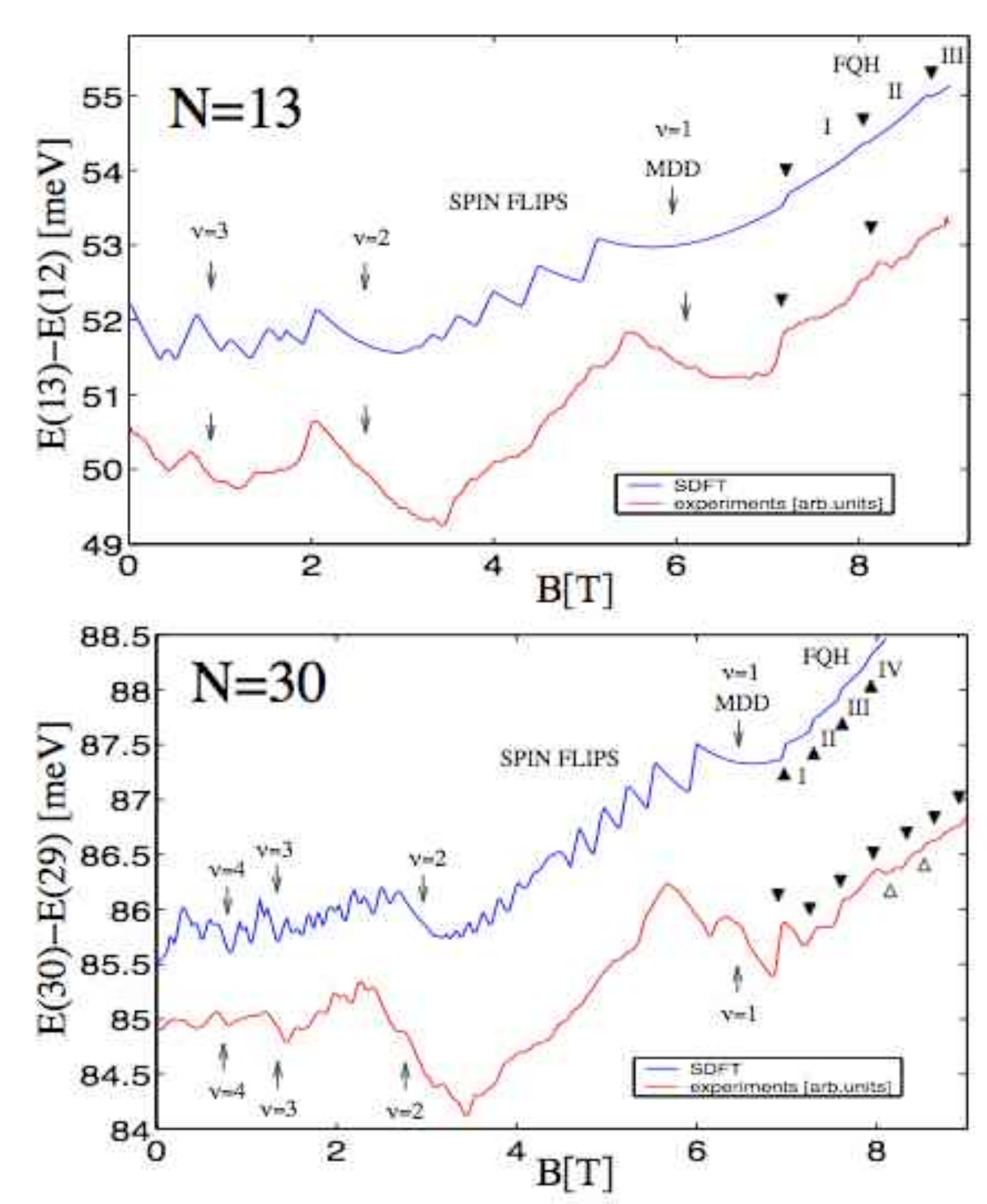} 
\caption{Chemical potential of a quantum dot 
device with $N=30$ ({\it upper panel}) 
and $N=13$ ({\it lower panel}) compared to the results from the
spin-density-functional theory. 
The experimental data are from~\textcite{oosterkamp1999}. 
Noise in the experimental data has been reduced by using a 
Gaussian filter. 
In the calculations the confining potential is assumed to be parabolic 
with $\hbar\omega_0$ being 4.00 meV for $N=13$ and 3.51 meV for $N=30$. 
Finite-size precursors of different quantum Hall states are identified. 
The roman numbers between the filled triangles indicate 
the number of vortices inside the electron droplet predicted by 
the density-functional calculations. 
The open triangles mark other possible transitions which 
are beyond the reach of density-functional theory. 
} 
\label{fig:chempiccy} 
\end{figure}

Closer examination of the chemical potential for different 
$N$ values and comparison with the mean-field results reveal different 
quantum Hall regimes as the magnetic field is increased. 
Fig. \ref{fig:chempiccy} shows the chemical potential for $N=13$ and $N=30$. 

The agreement with the electron transport data is best in 
the vicinity of the MDD domain. Experimental 
data show additional features not accounted for 
by the density-functional theory (open triangles in Fig. \ref{fig:chempiccy}), 
which could be attributed to correlation effects, especially a 
transition to partially polarized states~\cite{oaknin1996,siljamaki2002}. 
In a Quantum Monte Carlo study by \textcite{guclu2005a} 
the frequency of transitions per unit of magnetic field was calculated 
in the $\nu<1$ regime and it was found to roughly correspond to the 
frequency in experiments. 
However, many of the calculated transitions give 
rise to small changes in angular momentum and energy. A direct comparison with 
experiments is therefore difficult due to noise in experimental setups 
and inevitable imperfections in the samples. 
Nishi and coworkers have done experimental measurements 
and detailed modeling for few-electron quantum dots~\cite{nishi2006}. 
High-accuracy electron transport data that would go deep into the 
fractional quantum Hall regime, are still lacking for higher electron numbers.

Magnetization measurements of quantum dots could provide another 
way to probe for transitions caused by vortex formation inside 
electron droplets. Observed oscillations in the 
magnetic susceptibility $\chi = \partial M / \partial H$  
have been analyzed, showing the de Haas--van Alphen effect  
in large arrays of quantum dots~\cite{schwarz2002}. 
However, to resolve transitions in individual states in the regime of 
high angular momentum is challenging, because the shapes of 
the quantum dots in the ensemble must be sufficiently uniform,  
and the number of electrons in the samples has to be small. 
 
\subsection{Localization of particles and vortices} 
 
\label{sec:localization} 

We have seen above that localized vortices and vortex molecules 
have been observed in rotating bosonic systems,  
and very similar structures were predicted to occur in 
rotating fermion droplets. 
Vortex localization can be seen as analogous 
to particle localization within the framework of the particle-hole duality 
picture, discussed in Sec. \ref{sec:duality}. 
We start this section by a brief discussion of {\it particle}
localization in 2D systems. Insight and concepts derived from these 
studies are necessary as we proceed to discuss 
the analogy between particle and vortex localization.
 
\subsubsection{Particle localization and Wigner molecules} 
 
\label{sec:wigner} 
Wigner crystallization~\cite{wigner1934}
has been observed for electrons trapped 
at the surface of superfluid liquid helium~\cite{andrei1991} 
or in a two-dimensional electron gas in a semiconductor 
heterostructure~\cite{pudalov1993}. 
Recent addition energy measurements of islands of trapped 
electrons floating on a superfluid helium film have 
revealed signatures of a Wigner-crystalline state~\cite{rousseau2009}. 
In the low-density limit, the kinetic energy of the 2D electron gas 
becomes very small and the interparticle interactions dominate. 
The crystalline phase is 
expected to emerge at the density parameter 
$r_s\approx 37 a_B^*$, where $a_B^*$ is the 
effective Bohr radius~\cite{tanatar1989} ($r_s$ is a radius of a 
circle containing on average one electron). This estimate is in agreement
with more recent computations by \textcite{attaccalite2002,attaccalite2003}.
 
A finite system of  a few (nearly) 
localized electrons is commonly referred to as a ``Wigner molecule''. 
In small quantum dots, these Wigner molecules take the shapes of simple
polygons,  depending on the number of electrons that can be resolved 
by classical electrostatics~\cite{bolton1993,bedanov1994}.  
In the non-rotating case, the onset of electron localization occurs already at 
relatively high densities 
$r_s\approx 4 a_B^*$~\cite{jauregui1993,egger1999,reimann2000,yannouleas2007}. 
In this context it is interesting to note that in small systems most of the 
particles localize at the perimeter of the dot. For seven electrons, for
example, six particles localize at the vertices of a hexagon, with the seventh
particle at the dot center~\cite{bolton1993}:
the electrons along the perimeter 
essentially form a 1D system where the localization 
is even easier than in 2D~\cite{kolomeisky1996,viefers2004}. 
Localization in the radial direction takes place first 
followed by localization in the angular
direction~\cite{filinov2001,ghosal2006}.
\begin{figure}[h!!] 
\includegraphics[width=.45\textwidth]{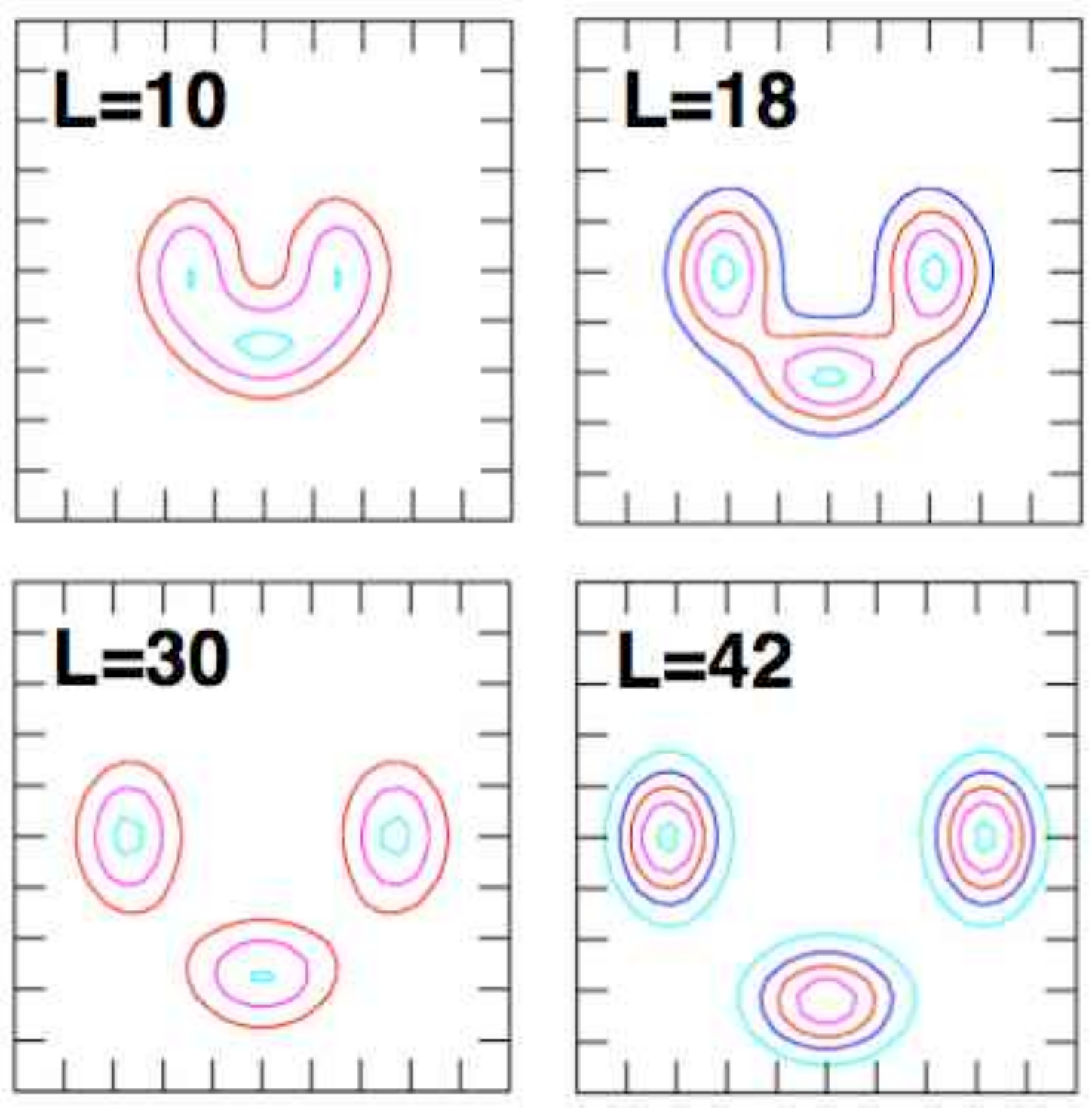} 
\caption{({\it Color online}) Pair-correlation functions of four fermions 
with repulsive Gaussian interactions 
at four different angular momenta 
$L=10$, $L=18$, $L=30$, and $L=42$, respectively, showing 
that localization increases with rotation. 
The contour plots are in the same scale to demonstrate  
the expansion due to the rotation. From~\textcite{nikkarila2008}. 
} 
\label{mattifloc1} 
\end{figure} 
In small electron systems there are no true 
phase transitions and the localization of electrons increases gradually  
with decreasing electron density~\cite{reimann2000}. 
Inelastic light scattering experiments 
have only been used to probe excitations of molecule-like states 
in few-electron quantum dots in the high-density regime 
where, however, localization has not 
yet occured~\cite{kalliakos2008}. 
Addition-energy spectra obtained from Coulomb 
blockade experiments~\cite{tarucha1996} 
have been proposed as a direct probe for signatures of  
localization~\cite{guclu2008}.
In large quantum dots, the crystallization occurs in ring-like patterns,
like the shells of an onion~\cite{filinov2001,ghosal2006}. 
A gradual rearrangement of addition energy spectra, which 
indicates a change in shell fillings, is then predicted to occur as 
the shell sizes of Wigner molecules differ from those of 
non-localized electrons. However, no experimental 
data yet exist in this regime. 
 
Quantum dots are often modeled as circularly symmetric and the  
associated quantum states and ground-state electron densities 
therefore also have the same symmetry. 
The localization of particles takes place in the internal  
frame of reference. In the laboratory frame the localization is seen
in the total density distribution only when using approximate  
many-particle methods which allow symmetry breaking, such as for example 
the unrestricted Hartree-Fock approach~\cite{yannouleas1999}.
Other possibilities are to break the symmetry of the confining potential,
as for example by an ellipsoidal 
deformation~\cite{manninen2001b,saarikoski2005a,dagnino2007,dagnino2009a,dagnino2009b}, 
or to analyze localization of the probing particle in the 
reduced wave function~\cite{harju2002b,saarikoski2004}. 
However, there are other 
straightforward methods to see the localization in  
exact calculation for circular confinement: 
Figure \ref{mattifloc1} shows the pair-correlation function 
(conditional probability) for four 
particles at different values of the total angular momentum. 
Clearly, when the angular momentum increases, the particles are 
further apart and the localization becomes more pronounced. 
Another possibility is to study the rotational 
many-particle energy spectrum, which is more intricate, but also more 
revealing. 

\subsubsection{Rotational spectrum of localized particles} 
 
\label{sec:spectrum} 
When the particles are localized, we may consider the system 
as a rotating ``molecule'' with a given point group symmetry. 
In the case of two identical atoms in a molecule 
the rotational spectrum shows a two-fold periodicity in the angular momentum,
which may be odd or even depending upon whether the atoms are bosons or  
fermions~\cite{tinkham1964}. Similarly, for $N$ identical particles 
forming a ring, 
only every $N$th angular momentum is allowed~\cite{koskinen2001} 
in a rigid rotation 
around the symmetry axis. For other angular momenta, the 
rotational state should be accompanied by an internal excitation. 
In the case of particles having no internal degrees of freedom (no spin), 
the only such excitations are vibrational modes of the molecule. 
Group theory can then be used to resolve the vibrational modes which are  
allowed to accompany a certain angular momentum 
eigenvalue~\cite{maksym1996,koskinen2001,viefers2004}. 
 
Plotting the energies of the many-body system as a function of the  
angular momentum, the lowest energy (yrast line) has 
oscillations with a period of the symmetry group.
The minima correspond to pure rotational states. 
Between the minima the states have vibrational excitations which increase 
the energy. Maksym showed that the energy 
spectrum of few electrons at high angular momenta 
can be quantitatively explained by a rotating and vibrating Wigner  
molecule~\cite{maksym1996} 
which is the basis for the molecular approaches 
to correlations in quantum dots~\cite{maksym2000} 
and quantum rings~\cite{koskinen2001}. 
Several other studies have later confirmed this  
observation, for a review see~\textcite{viefers2004}. 
This molecular approach for rotating particles has also been 
used by \textcite{yannouleas2002,yannouleas2003},
who introduced ``rotating electron molecule'' wave 
functions to describe rotating molecular states at high angular momenta.
These wave functions are available 
in analytic form, with their internal structure constructed 
by placing Gaussian functions at classical positions of electrons 
in high magnetic fields. 
 
Formulating a molecular model of a rotating system, 
we may approximate the many-particle spectrum (at zero magnetic field) by
\be 
E=\frac{L^2}{2I_L}+\sum_\nu \hbar\omega_{L\nu}\left (n_\nu+\frac{1}{2} \right), 
\label{mattile1} 
\ee 
where $I_L$ is the moment of inertia of the Wigner molecule and  
$\omega_{L\nu}$ the vibrational frequencies.
$I_L$ and $\omega_{L\nu}$ can be determined 
using classical mechanics in the rotating frame, and thus depend on
the angular momentum as indicated with the subscript $L$. 
The eigenenergies Eq.~(\ref{mattile1}) can be compared to those 
calculated from the exact diagonalization method. 
 
\begin{figure}[h!!] 
\includegraphics[width=.4\textwidth]{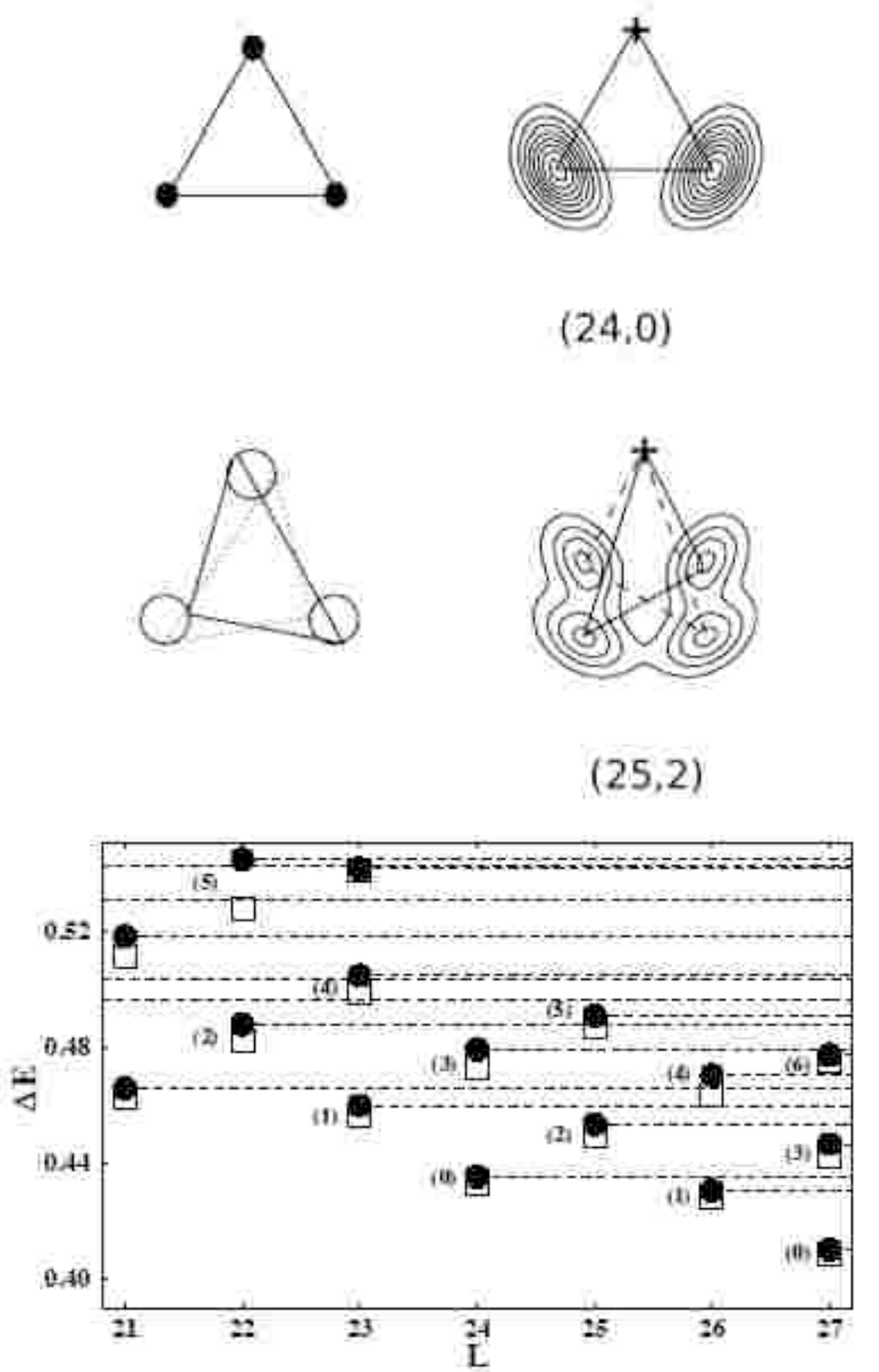} 
\caption{Classical orbits and pair-correlation functions of  
localized electrons in rotating frame {\it (upper panel)}. 
At angular momentum  
$L=24$ the lowest energy state is purely rotational while at 
$L=25$ a doubly excited rotational state is shown. In the  
rotating frame the classical motion shows a pseudo-rotation 
{\it (middle left)} while the pair-correlation shows maxima at the classical
turning points ({\it middle right}).The spectrum {\it (lower panel)} 
compares the exact energies {\it (dots)} with those of the classical model
{\it (squares)}. It shows  a periodicity of $\Delta L=3$ in angular momentum, 
which agrees with the localization in a triangular geometry 
(see upper panel). The horizontal dashed lines indicate the 
center-of-mass excitations which occur at all angular momenta. 
From~\textcite{nikkarila2007}.
} 
\label{mattifloc2} 
\end{figure}

To give an example for the signatures of localization in the many-body 
energy spectra, 
Fig. \ref{mattifloc2} shows the rotational three-particle spectrum.    
A broad range of low-lying states may be described quantitatively 
with the rotation-vibration model of Eq. (\ref{mattile1}). 
Figure \ref{mattifloc2} also shows examples of the pair-correlation 
functions for a purely rotating state and for a state including 
vibrational modes. Similar observations have been reported 
for other vibrational modes and particle  
numbers~\cite{maksym2000,nikkarila2008}. A more detailed 
quantum-mechanical analysis of the molecular states has recently also 
been reported by \textcite{yannouleas2009}.
 
Finally, we should consider what happens to the rotational 
energy spectrum when the particles have internal degrees of freedom, 
say spin. In the classical limit, the internal degrees of freedom separate 
from the spatial excitations (vibrations), since the Hamiltonian
is spin-independent. 
The different spin-states of the system will eventually become degenerate. 
However, the existence of the different spin states will give  
more freedom to satisfy the required symmetry (bosonic or fermionic) 
of the total wave function. Again, group theory can be used to  
determine the spin states which are allowed for a given angular momentum 
and a given vibrational state~\cite{maksym1996}. The energies agree well with
the classical model of Eq.~(\ref{mattile1})~\cite{koskinen2007}. 

The localization of the particles may in fact also be incomplete.  
This is indicated by the non-vanishing particle density in between the 
classically localized geometries, as well as small deviations in the symmetry
of the Wigner crystal. 
Especially the excited quantum Hall states (edge states) may show 
such structures, as discussed already in connection with 
vortex formation, see Sec.~\ref{sec:vortexformation}. 
The lowest-lying excitations of a large electron droplet above 
the MDD state have been predicted to show particle localization 
into rings of electrons around a compact non-localized core of the 
MDD electrons ~\cite{macdonald1993,chamon1994}. 
The wave functions of these states contain a single node at a finite distance 
from the center ({\it i.e.}, a non-localized vortex, see Sec.~\ref{sec:singlevortex}) 
which leads to a separated ring of localized electrons at the 
edge, often  referred to as the Chamon-Wen edge, that has been much 
discussed in the 
literature~\cite{goldmann1999,reimann1999,manninen2001b,reimann2002,toreblad2006}. 
The localized edge state appears when the MDD begins to 
break up with the entrance of the first vortex, 
but before further vortex holes penetrate the cloud 
(see Sec.~\ref{sec:vortexlocalization}). 
It should be noted that the current-spin-density functional 
theory~\cite{vignale1987,vignale1988} with the local density 
approximation~\cite{reimann1999} largely over-emphasizes the localization of 
electrons in the Chamon-Wen edge~\cite{toreblad2006}. 
 
\subsubsection{Localization of bosons} 
 
In a non-rotating condensate, all bosons may occupy the same quantum 
state. In the regime of high angular momenta, however, rotation 
may induce 
localization in bosonic systems in the same way as in fermionic systems. 
In both cases, the rotation pushes the particles
further apart, and the classical picture of a rotating and 
vibrating Wigner molecule~\cite{maksym1996} 
sets in. The similarity of bosons 
and fermions in reaching the classical limit was suggested 
by~\textcite{manninen2001}
on the basis of Laughlin's theory~\cite{laughlin1983} of the 
fractional quantum Hall  
effect, and has been subsequently 
studied more quantitatively:
a detailed comparison of few bosonic and fermionic particles in a harmonic 
trap~\cite{reimann2006b} 
indicated similar localization effects in both systems. 
Note that for small particle numbers in the LLL, the mapping  
between boson and fermion states, discussed in Section~\ref{sec:duality}, 
becomes increasingly accurate when the angular momentum 
increases~\cite{borgh2008}, in accordance with the classical 
interpretation of the spectrum. 
 
\subsubsection{Vortex localization in fermion droplets} 
\label{sec:vortexlocalization} 

There is an apparent analogy between vortex localization and 
particle localization: we have seen above that localized vortices 
cause minima in the electron density, with rotational currents 
around their cores. These ``holes'' arrange in vortex molecules, 
with shapes that indeed resemble those of Wigner molecules 
in the case of particle localization, discussed in Sec.~\ref{sec:wigner}. 
(Note that the Pauli vortices do not give rise to vortex structures in 
the electron density, since each electron carries one such vortex).  
 
\begin{figure}[h!!] 
\includegraphics[width=.45\textwidth]{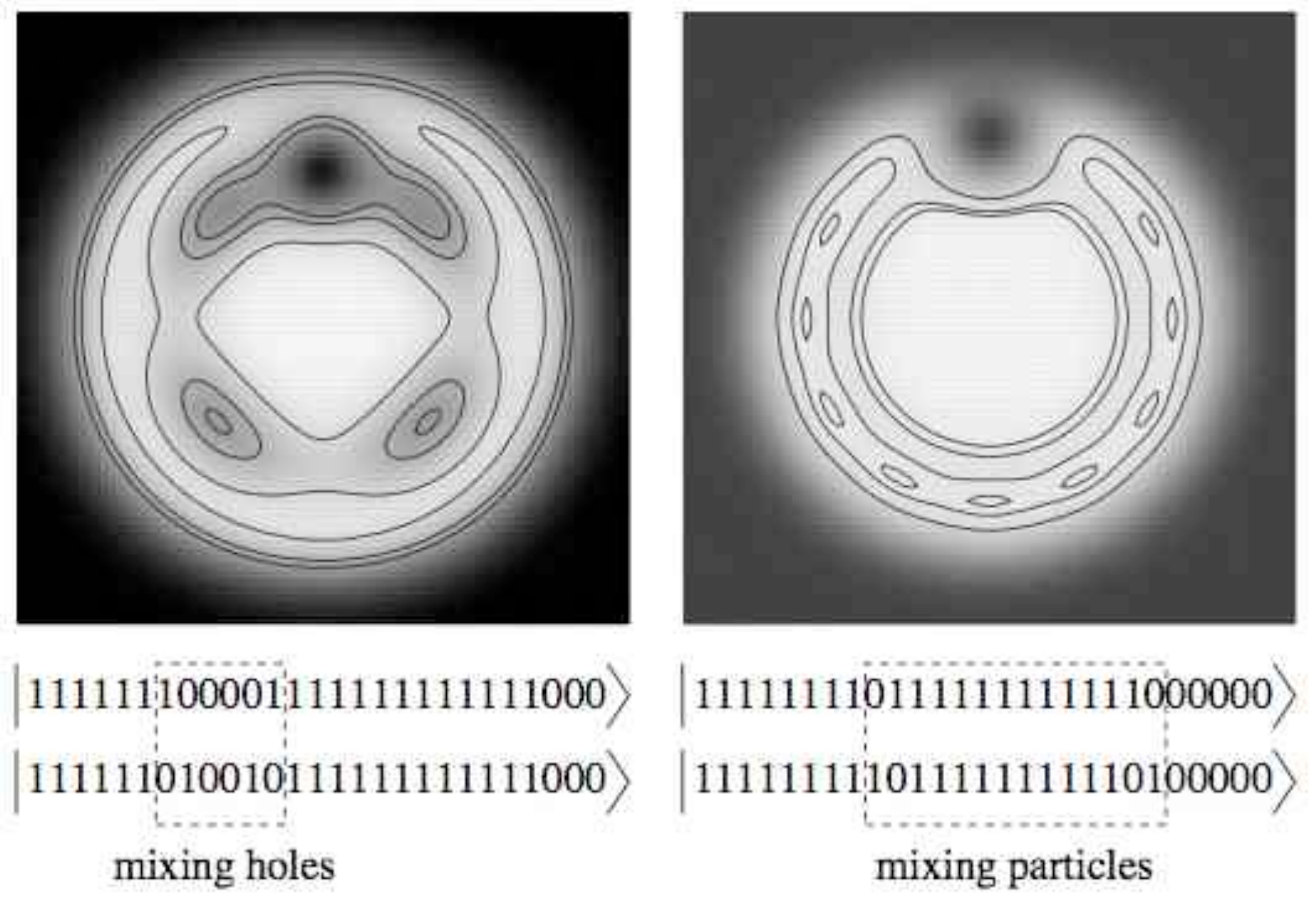} 
\caption{Electron-electron pair-correlation functions showing
localization of four vortices ({\it left}) and localization of 
electrons at the edge of the cloud in the case of one vortex ({\it right}).
White color means high density, and some constant-density contours are shown.  
The two most important configurations are given in each case, demonstrating 
that mixing of single-particle states close to holes leads to hole 
localization, while correspondingly the mixing of particles  
localizes particles. The results are calculated with the CI method for 
20 particles with angular momenta  
$L=202$ {\it (left)} and $L=242$ {\it (right)}. 
} 
\label{mattif2} 
\end{figure}

The vortex localization
can be illustrated by the configuration mixing of the exact quantum states. 
If the configuration has, say, 
four vortices and $\vert 11110000111111\cdots\rangle$ 
has the largest weight, other configurations with the same angular momentum, 
like $\vert 11101001011111\cdots\rangle$, have a finite weight. 
The CI method shows that the mixing of these states 
happens mostly around the holes in the filled Fermi sea, as 
indicated in Fig.~\ref{mattif2}.  
This means that the holes are strongly correlated and may localize. 
This can be directly compared to the localization of particles. 
As discussed further in Sec.~\ref{sec:wigner}, 
the lowest-lying excitations of a large electron droplet above 
the MDD state were predicted to show particle localization 
into ring-like geometries, with a single vortex hole at a finite distance 
from the center (see Sec.~\ref{sec:singlevortex}). 
In this case, the configuration mixing is shifted to 
the outer edge of the droplet where it leads to a ring of strongly 
correlated particles, as for example seen in Fig.~\ref{mattif2}. 
 
The localization of particles and vortices in a circular system 
breaks the internal symmetry (unless a single 
vortex is localized at the center). 
The density functional 
method, using a local approximation for the exchange-correlation 
effects, may show  the localization of particles and 
vortices directly in the particle and current densities 
(see Fig. \ref{fig:n24} and discussion of symmetry breaking 
in Sec.~\ref{sec:symmetry}), as discussed above. 
However, the true many-body wave function of the system must have the  
symmetry of the Hamiltonian.
Figure \ref{mattif2} already demonstrated that the   
localization of vortices can be seen in the pair-correlation 
functions by taking the reference point to be at the same 
radius as the vortices. 
Moreover, in a one-component fermion system, particle-hole duality 
(see Sec. \ref{sec:duality}) can be used to 
gain insight into correlations between vortices. 
Transformation of a bosonic wave function to a fermionic one can be used to 
illustrate the vortex localization. Any fermion state can be written 
as a determinant of the MDD times a symmetric polynomial, 
where vortex structures are included in the latter~\cite{manninen2005}. 
On the other hand, this polynomial 
is a good approximation to the exact boson wave function, as
discussed in Sec. \ref{sec:fermionboson}.
 
\begin{figure}[h] 
\includegraphics[width=.45\textwidth]{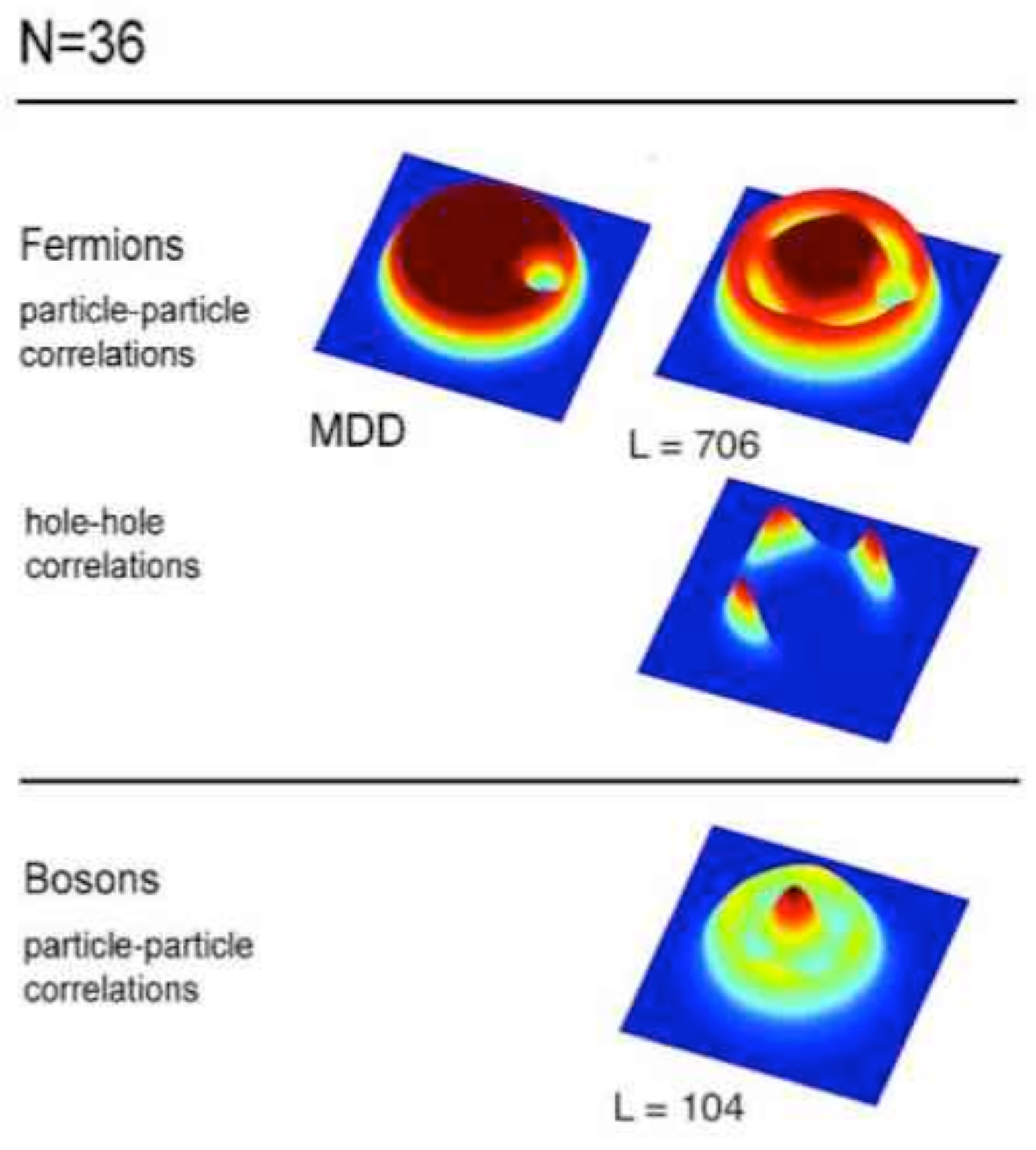} 
\caption{({\it Color online}): Pair-correlation functions for large fermion and boson  
systems with four vortices. The pair-correlation function of the MDD 
is displayed for comparison; it only shows the exchange-correlation hole 
at the reference point.} 
\label{mattif3} 
\end{figure}

Figure \ref{mattif3} shows examples of the particle-particle and hole-hole correlation  
functions which indeed reveal that vortices in both boson and 
fermion systems are well localized.
This can be understood by  
considering the angular momentum of the system of holes, and 
the corresponding filling factor of the LLL. For example, in the case 
of four vortices, the hole filling factor is as low as 
about 1/9, which corresponds to the value where the particles form  
a Wigner solid in an infinite system. 
In other words, when the electron filling factor approaches unity 
(from below), the hole filling factor approaches zero, for\-cing the holes
to be localized.
\begin{figure}[h!] 
\includegraphics[width=.4\textwidth]{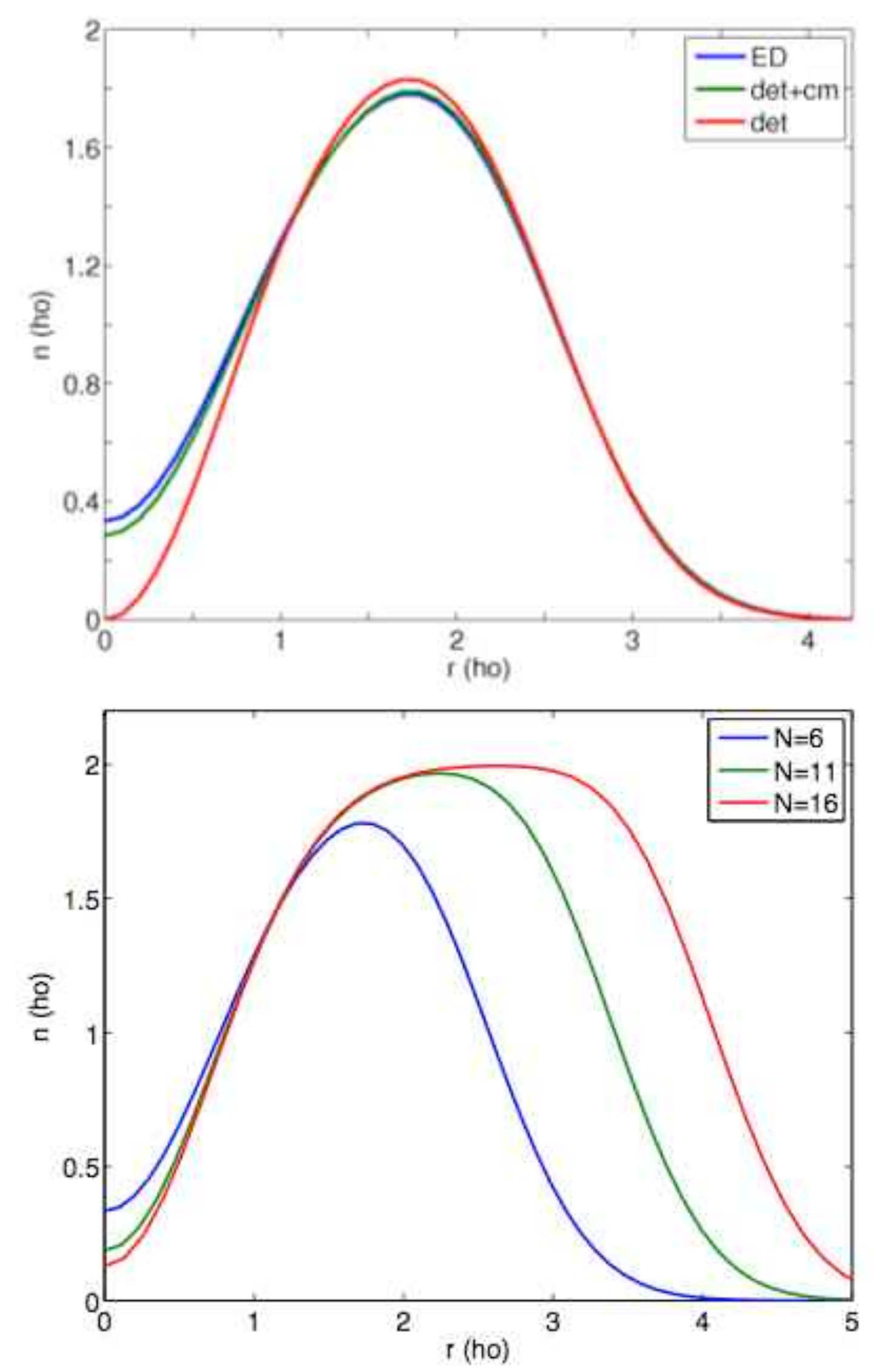} 
\caption{({\it Color online}) {\it Upper panel:} 
Radial electron densities 
in a harmonic trap ($\omega =1$) in a six-electron 
droplet with a central vortex at $L=21$ (in harmonic oscillator units).
The exact solution in the LLL 
is shown by the blue line, a single-determinant wave 
function which describes a central vortex is shown by the red 
line, and a single-determinant wave function in the center-of-mass (CM) 
transformed coordinates $z_i\to z_i-z_{\rm CM}$ is shown 
by the green line.
{\it Lower panel}: Radial electron densities for central vortex states 
$L=L_{\rm MDD}+N$, showing that vortex localization increases with 
electron number $N$ due to decrease in the center-of-mass motion. 
} 
\label{fig:cm-motion} 
\end{figure} 
 
\begin{figure}[h!!] 
\includegraphics[width=.45\textwidth]{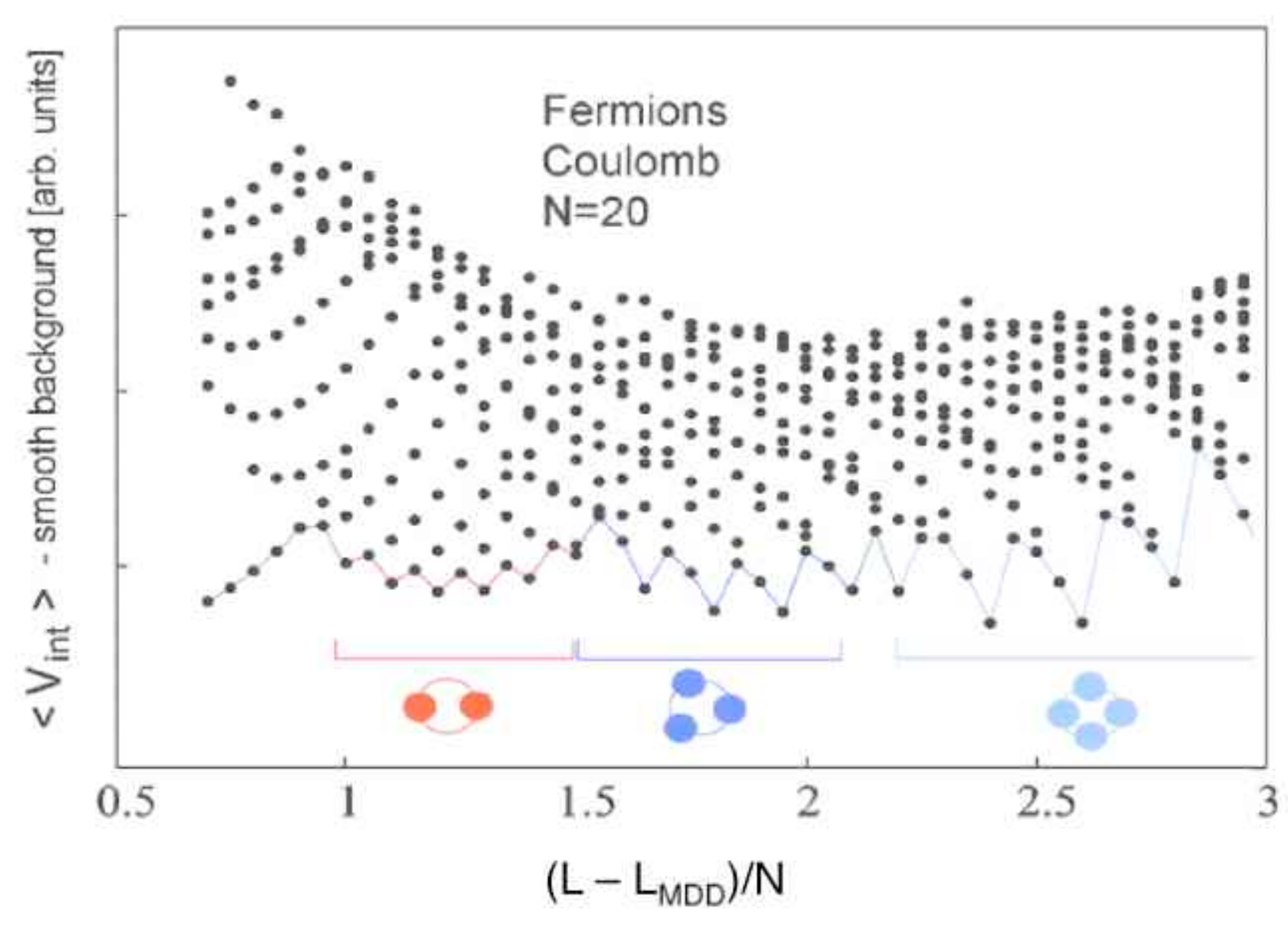} 
\caption{({\it Color online}) Fermion low-energy spectrum for 20 particles. 
The lowest energy many-particle states 
as a function of the total angular momentum (yrast states) are connected 
with lines to guide the eye. 
A smooth function of angular momentum was substracted 
from the energies to emphasize the oscillatory behavior of the yrast line. 
The periodicity of the oscillation reveals the number of localized vortices 
as schematically illustrated. From~\textcite{reimann2006a}.
} 
\label{mattif4} 
\end{figure} 
 
\begin{figure}[h] 
\includegraphics[width=.45\textwidth]{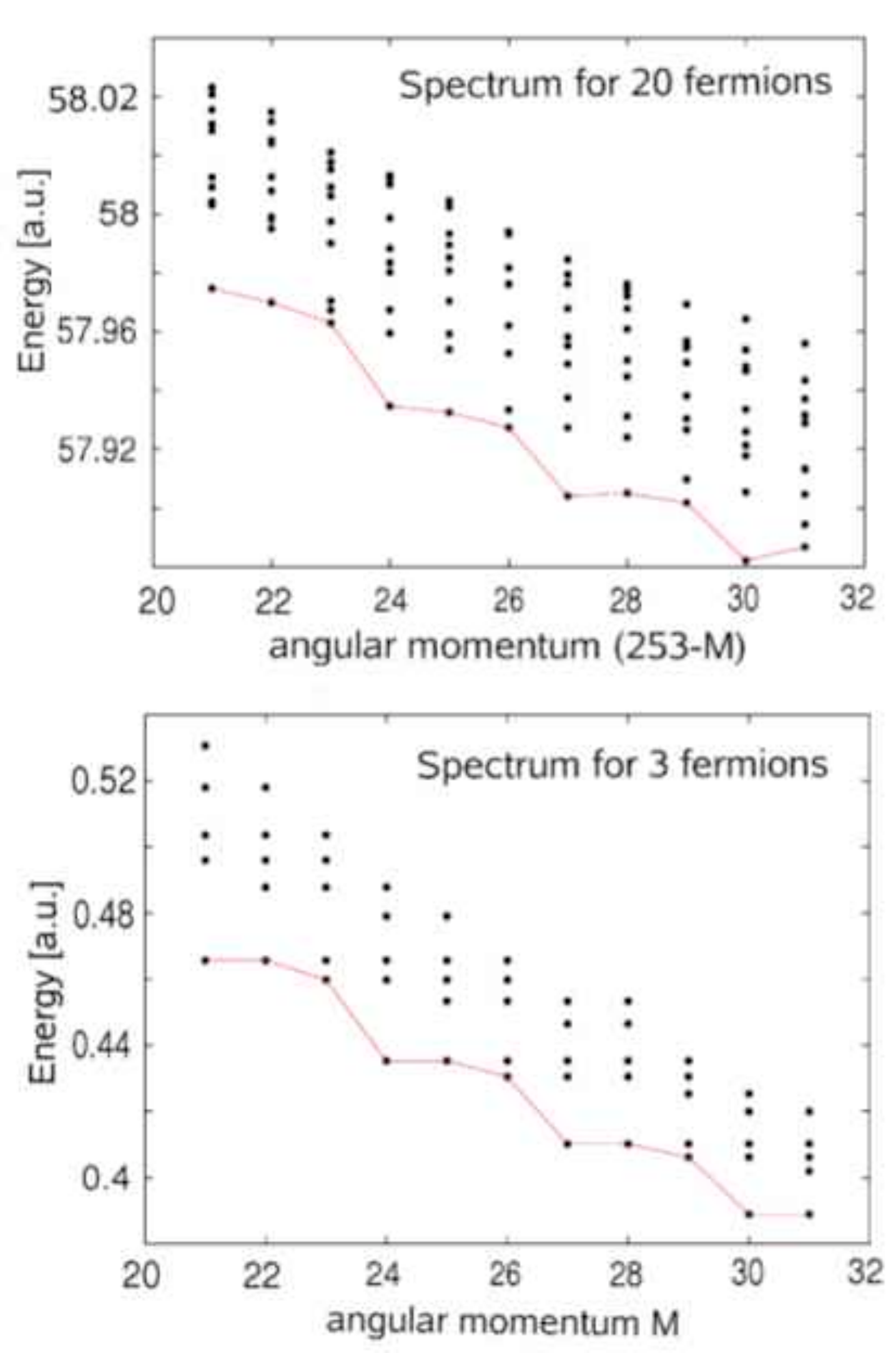} 
\caption{Fermion yrast spectrum for 
20 particles and three vortices (upper panel) 
and 3 particles (lower panel) show both 
the periodicity of $\Delta L=3$  
associated with the three-fold rotational symmetry of 
the vortex molecule in the former case and electron molecule 
in the latter case.
From~\textcite{manninen2006}. 
} 
\label{mattif5} 
\end{figure} 
 
Hole-hole correlations in Fig.~\ref{mattif3} 
show clearly the effect of the zero-point fluctuation in the vortex position. 
To examine this further in the case of fermions, 
let us as an example investigate the 
singly-quantized vortex for six electrons in a harmonic confinement.
As discussed earlier in Sec. \ref{sec:singlevortex} 
the MDD state in this case, with angular momentum $L=15$, 
is characterized by a relatively flat electron density.  
The electrons occupy the six lowest levels of angular momentum in 
the lowest Landau level with occupancies $|11111100\ldots\rangle$. 
When the angular momentum increases, at stronger magnetic fields 
the MDD state is  reconstructed and a vortex hole is created in the center. 
This state has angular momentum $L=21$. The single-particle determinant 
$|011111100\ldots\rangle$ with a weight 0.91 yields the 
largest contribution to the wave function in the lowest Landau level.
Due to fluctuations, the exact many-body wave function 
includes other single-particle determinants corresponding to $L=21$, 
such as $|1011110100\ldots\rangle$ and $|11011100100\ldots\rangle$. 
However, since their weights are relatively small, 
the state can be characterized by a rather flat maximum density droplet 
configuration with a vortex hole in the center. 
The electron density of this state, indeed, 
shows a deep hole in the center and a rotating current around it 
(upper panel of Fig. \ref{fig:cm-motion}). 
Fluctuations in the vortex position cause the particle 
density to remain finite in the center of the confining potential. 
A single-determinant wave function $|01111100\ldots\rangle$ 
transformed into the center-of-mass coordinates 
$z_i\to z_i-z_{\rm CM}$ shows a density profile 
that is very close to the exact results (upper panel of  
Fig. \ref{fig:cm-motion}). 
The quantum mechanical zero-point motion of 
the vortex hole leads to a finite density at the vortex core. 
The center-of-mass fluctuations decrease with electron number, which 
is reflected by localization increasing with particle number 
(lower panel of Fig.  \ref{fig:cm-motion}). 
 
\subsubsection{Vortex molecules} 
 
A section of the many-particle energy 
spectrum for $N=20$ 
electrons for different angular momenta $L$ is shown in Fig.~\ref{mattif4}
\cite{reimann2006a}.
The yrast line shows periodic oscillations, with the oscillation 
length (in units of $L$) equal to number of localized vortices in the system.
The reason behind these
periodic oscillations in the energy spectrum is deeply connected
with the above-mentioned particle-hole duality and
vortex localization: they are signatures of two, three, 
and four vortices, respectively, 
being localized at the vertices of simple polygons 
with $C_{2v}$ symmetry. 
For polarized fermions as in Fig.~\ref{mattif4}, the rigid rotation of the 
vortex ``molecule'' with $n$-fold symmetry is allowed only 
at every $n^{th}$ angular momentum, corresponding to a minimum (cusp) in the 
yrast line. At intermediate angular momenta, the rigid rotation is accompanied
by other excitations, such as vibrational modes, that result in higher
energies~\cite{nikkarila2008}. 
Figure~\ref{mattif5} compares a small part of the spectrum to that  
for three electrons. The marked similarity of these spectra demonstrates  
not only that the vortices are localized in a triangle, 
like the three electrons, 
but also that elementary excitations of the many-particle energy spectrum are 
vibrational modes of the vortex-molecule. 
 
\begin{figure} 
\includegraphics[width=.29\textwidth]{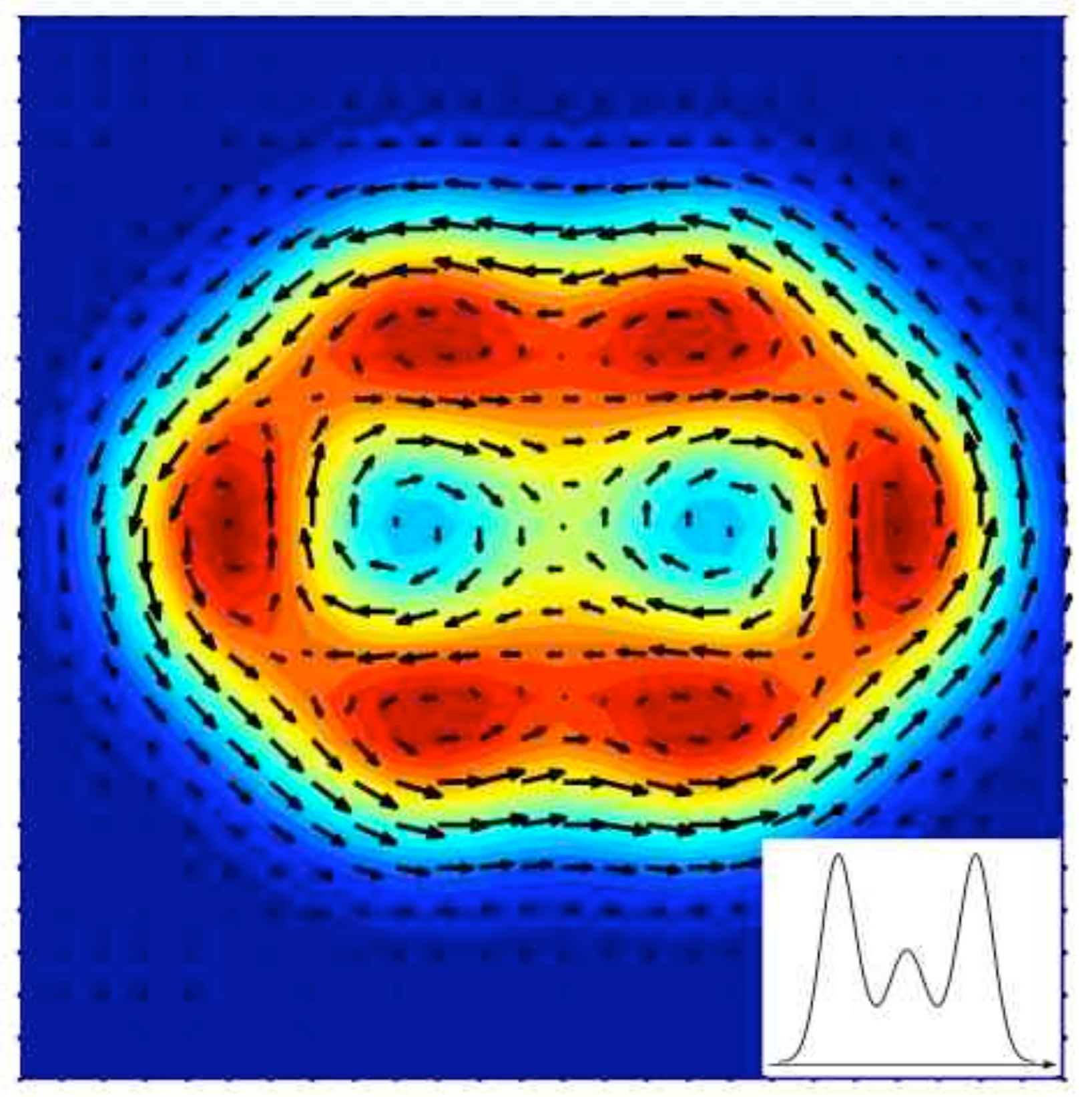} 
\caption{({\it Color online})  
Electron density (color, with red for maximum density) and 
current density (arrows) of an elliptically confined 
6-electron droplet with two localized vortices, 
calculated by the exact diagonalization method.  
The confinement strength is $\hbar\omega_0= 5.93\; {\rm meV}$, 
the eccentricity of the elliptic confining potential $\delta=1.2$ 
and the magnetic field is $B=17$~T. 
{\it Inset:} profile of the electron density at the longest major axis shows 
fluctuations in the vortex positions, which causes electron density to 
remain finite at the density minima. 
Adapted from Fig. 6 in \cite{saarikoski2005a}.} 
\label{fig:edellipse} 
\end{figure}

Under certain circumstances the particle and current densities 
of the (exact) many-body state may show directly the formation 
of vortex molecules. 
This may for example be the case for a broken 
rotational symmetry of the system, as for example
predicted for elliptically confined 
quantum dots \cite{manninen2001b,saarikoski2005a}. 
Fig. \ref{fig:edellipse} shows the electron density of an 
elliptical 6-electron 
quantum dot calculated by exact diagonalization. 
Two localized vortices can be identified as minima in the charge 
density, around which the current shows the typical loop structure. 
In highly excentric confining potentials, vortex 
structures containing three and more localized vortices were also 
predicted to form~\cite{saarikoski2005a}. 
The effect of fluctuations in the vortex positions is clearly seen 
also in this case. 
To some extent, electron localization is observed as well. 
In this case, the wave function can be characterized by two hole-like 
quasi-particles at the center of a ring of six electrons. 
It should be noted that Fig.~\ref{fig:edellipse} 
shows the {\em exact particle density},
and not the mean-field particle density. 
Since elliptically deformed quantum dots have been realized 
experimentally~\cite{austing1999b} this may be the most direct 
way to image vortex structures in quantum dots. 
Localized vortex structures have been predicted to emerge also in 
other quantum dot geometries~\cite{saarikoski2005a,marlo2005}. 

\begin{figure}[ptb] 
\includegraphics[width=0.5\textwidth]{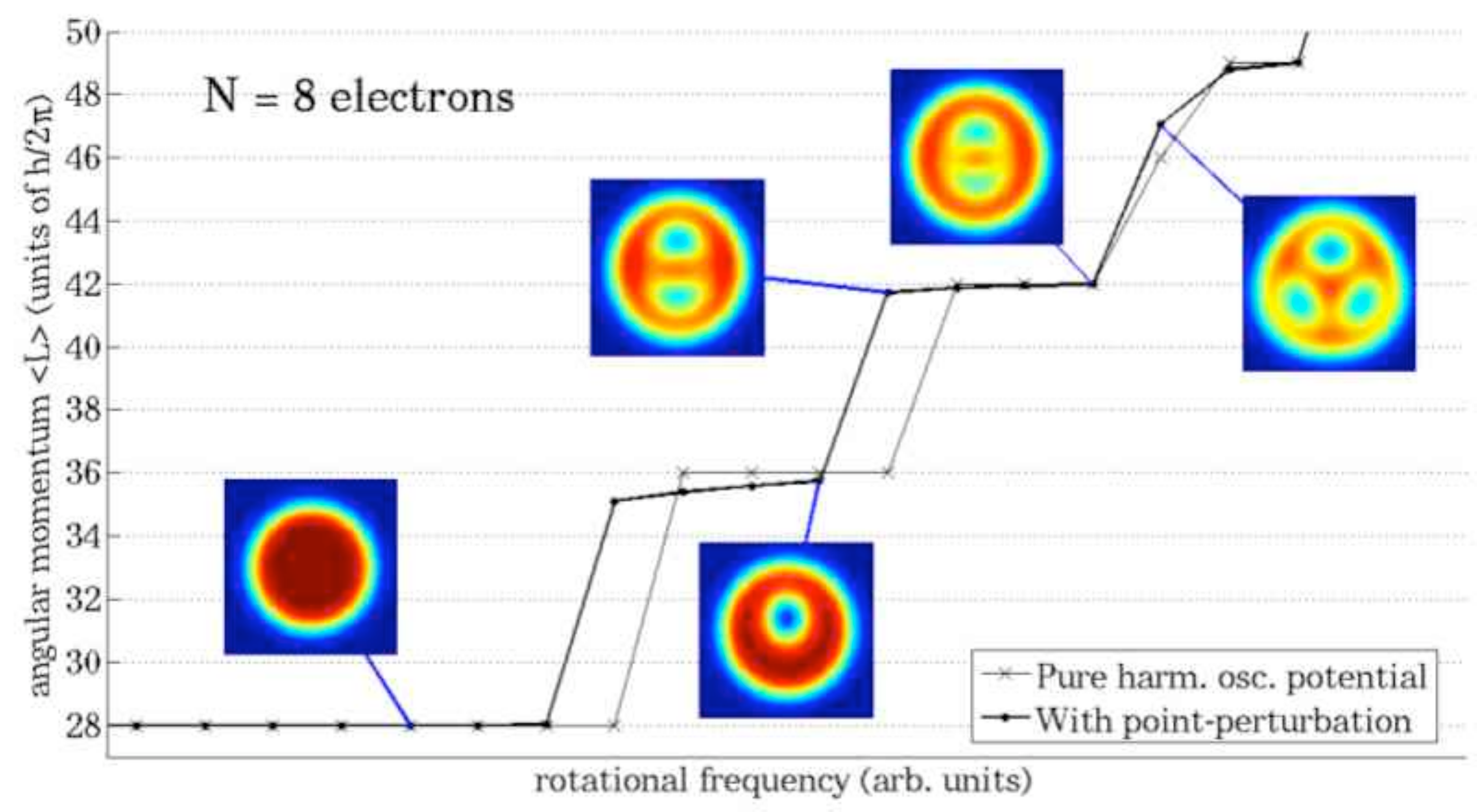} 
\caption{({\it Color online})  
Angular momentum as a function of the rotational frequency of the  
parabolic trap with $N=8$ electrons in the lowest Landau level.  
The unperturbed result ({\it thin line}) 
is comparable to the expectation value of angular momentum in the 
presence of an added point perturbation which breaks the rotational symmetry
({\it thick solid line}). 
The insets show the densities in the perturbed system.  
The vortices appear as pronounced minima in the density distribution,  
their number increasing with the trap rotation. 
Results are calculated with the exact 
diagonalization method~\cite{christensson2008b}.}  
\label{fermivortices} 
\end{figure}

A perturbative approach to visualize  vortices
in the particle density is to include a point-perturbation in 
the external potential~\cite{christensson2008b}, which can 
pin the vortices.  The resulting particle 
density clearly shows the vortex localization.
An example is shown in Fig.~\ref{fermivortices} for a system 
of 8 electrons. 
With this small perturbation, the expectation value of the  
angular momentum still has a  
nearly similar dependence on the rotational frequency than the
unperturbed system. It is thus expected that each angular momentum
jump in the nonperturbed system corresponds to addition of
one vortex as seen in the perturbed system.
 
\subsection{Melting of the vortex lattice} 
\label{Sec:Melting} 

After single vortex lines in rotating condensates were experimentally 
realized by phase imprinting techniques~\cite{matthews1999},  
many experimental studies concerned the formation of lattices of 
vortices in bosonic cold-atom gases 
in the regime of high particle-to-vortex ratio (filling factor) 
$\nu_{pv}=N/N_v$ ~\cite{chevy2000,madison2000,madison2001}. 
The modes of the vortex lattice~\cite{baym2003,baym2004} 
as well as the structure of the vortex cores were 
analyzed~\cite{coddington2003,schweikhard2004,coddington2004}.
When the vortex density increases with the angular momentum, 
it is expected that for rapid rotation, the vortex density may 
finally become comparable to the particle 
density~\cite{ho2001,fetter2001,cooper2001}. 
An interesting issue is then how the system changes with the increasing 
particle-to-vortex ratio~\cite{baym2005}. 
At rapid rotation, 
strongly correlated states analogous to fractional quantum Hall 
states may emerge~\cite{wilkin1998,viefers2008,cooper2008}. 
These states are quantum liquid-like states of particles and vortices
where correlations may give rise to the formation of particle-vortex 
composites. 
It is believed that a phase transition occurs 
with a vortex density somewhere between the rigid vortex lattice and
the quantum liquid of vortices. 
This transition is often refered to as ``melting''. However, the process is 
not fully understood and calculations yield different estimates for 
the critical vortex density. 
Moreover, in present day experiments 
the particle-to-vortex density is usually very high,  
$\nu_{pv}\gtrsim 500$ \cite{schweikhard2004}. 
 
\subsubsection{Lindemann melting criterion} 
 
The vortex density at the transition from localized vortex lattice 
states to liquid-like states can be approximated by assuming 
that the melting process is analogous to the melting of solids when atomic 
vibrations increase above a threshold amplitude. 
In the Lindemann model the melting 
point of solids is determined from the condition that when thermal vibrations 
reach a critical amplitude, melting of the material occurs~\cite{lindemann1910}. 
This amplitude in solids is often approximated to be around 10\% to 20\% of 
the lattice spacing. 
Using an analogous idea, the melting point of the vortex lattice 
can be approximated from the condition that thermal and 
quantum zero-point vibrations reach a critical 
threshold amplitude~\cite{blatter1993}. 
 
~\textcite{rozhkov1996} studied the vortex lattice melting 
in superconductors at zero temperature 
to obtain an estimate for the vortex density 
where zero-point fluctuations become large enough 
to melt the vortex lattice.
Their study was motivated by the presence of large quantum fluctuations in 
high-$T_c$ materials but their results give also an estimate of 
the vortex lattice melting in ultra-cold rotating Bose-Einstein 
condensates. Using the Lindemann criterion they approximated that 
melting takes place at particle-to-vortex filling factor $\nu_{pv}\sim 14$ 
at a presumed threshold zero-point vibration amplitude of 
14\% of the nearest-neighbour inter-vortex distance. 
 
Other calculations using the Lindemann criterion 
have given comparable estimates of the filling factor at 
the vortex lattice melting (see also the discussion in the reviews by 
\textcite{cooper2008} and \textcite{fetter2008}). 
\textcite{sinova2002} reported that  
the critical density in their model system of rapidly rotating bosons 
corresponds to $\nu_{pv} \sim 8$.
\textcite{baym2003,baym2004,baym2005} analyzed normal 
modes of vortex lattice vibrations in the mean-field 
limit and found that the vortex lattice melts 
at $\nu_{pv} \sim 10$.

\subsubsection{Transition to vortex liquid state} 
 
The predictive power of the Lindemann model is 
poor because melting in solids is known to be a co-operative phenomenon, 
and the process therefore cannot be accurately described 
in terms of the mean vibration amplitude of a single particle. 
However,~\textcite{rozhkov1996} obtained another estimate $\nu_{pv}\sim 11$ 
for the melting point of the vortex lattice by comparing the energy of 
a Wigner crystal model wave function to the energy of a 
Laughlin-type  wave function.
These wave functions were assumed to correspond to the ordered 
vortex-lattice state and the vortex-liquid state, respectively. 
Exact diagonalization calculations 
with contact interactions in a periodic toroidal geometry 
showed that the excitation gap collapsed at $\nu_{pv}\sim 6$, 
which was interpreted as a lower bound for a vortex lattice 
melting~\cite{cooper2001}. The associated vortex-liquid states 
at integer and half-integer $\nu_{pv}$ were in this work shown
to be, in general, well described with so-called parafermion states
studied by \textcite{read1999}.
 
In contrast to bosonic systems, 
the vortex lattice melting has not been studied 
theoretically in fermion systems. However, we can obtain an estimate for 
a corresponding transition using the particle-hole duality 
(Sec.~\ref{sec:duality}). There is a transition 
from the fractional quantum Hall liquid to  localized electrons 
({\it i.e.} the formation of a Wigner crystal) 
when the filling fraction of the LLL decreases below 
$\nu\approx 1/7$~\cite{lam1983,pan2002}. There are about 
6 to 8 vortices per particle, not counting the Pauli vortices, 
at the transition point.
Using the particle-hole duality we can now reverse 
the role of particles and vortices. In the dual 
picture a lattice of localized vortices then 
melts to a quantum Hall liquid when 
the particle-to-vortex ratio decreases to a value between 6 and (about) 8.   
This corresponds to a filling factor between $\nu \approx 0.8$ and $ 0.9$, 
where a vortex lattice is expected to melt in a 2DEG. 
The close relation between boson and fermion states in the LLL, 
Eq. (\ref{mapping}), would suggest that also in 
boson systems the vortex lattice should melt when the particle-to-vortex 
ratio decreases to about $8$, which is not too different from 
the values mentioned above. 
 
In conclusion, even though the results of different calculations 
show a considerable variation for the melting point, they  
all  indicate vortex lattice melting 
well before the number of vortices in the system becomes comparable 
to the particle number. However, much of the details are not understood,  
and experiments do not yet reach the transition regime. 
The transition may happen gradually and go through several intermediate 
states with increasing vortex delocalization, or, 
as the name explicitly suggests, it may occur through 
an abrupt loss of vortex ordering. 
 
\subsubsection{Breakdown of small vortex molecules} 
 
As discussed earlier, rotation in the intermediate angular momentum regime 
in small quantum droplets may give rise to formation of vortex molecules 
which are analogues of vortex lattice states of infinite systems. 
However, in finite-size systems, edge effects may play an important role.
This was noted also in the context of Wigner crystallization 
in quantum dots, where the  
onset of localization occurs at electron densities which are much 
higher than the corresponding values for the infinite 2D electron gas. 
The importance of edge effects has been pointed out also for
bosonic systems~\cite{cazalilla2005}.

Partly, localization effects account for 
the fact that in small systems also the $\nu=1/3$ state appears 
localized, as for example visible in the 
pair-correlation functions. The same applies to 
 vortices, and in very small systems it is difficult to make a difference 
between a vortex molecule and a vortex liquid, since both show similar 
short-distance correlations. 
 
\begin{figure} 
\includegraphics[width=.5\textwidth]{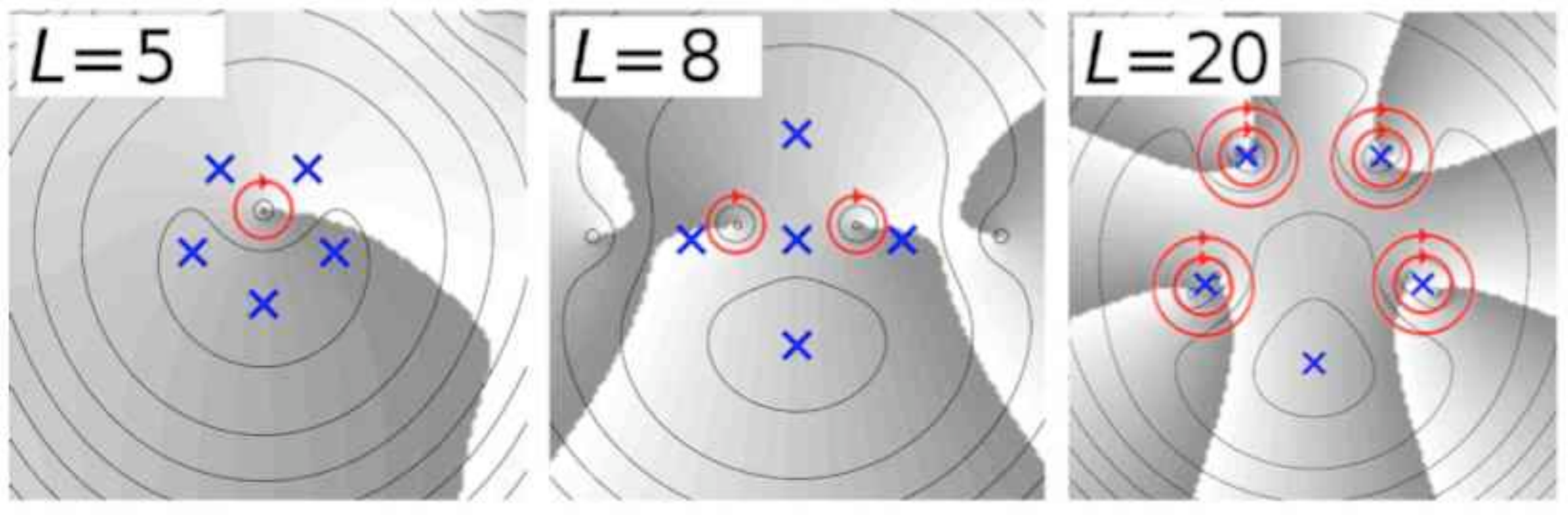} 
\caption{({\it Color online})  Reduced wave functions of 
a bosonic 5-particle system in a harmonic trap, showing the formation of 
one and two free vortices in the region of high particle 
density (marked as circles) at low angular momenta $L=5$ and $L=8$, 
respectively {(\it left and middle}). When the angular momentum increases, 
two vortices are finally captured by each particle to 
form a state which is approximated by the bosonic Laughlin 
state $m=2$ (two concentric circles) at $L=20$ ({\it right}). 
Particle interactions are Coulombic here and the probe particle 
is at the bottom. After~\textcite{suorsa2006}.} 
\label{fig:bosoniclaughlin} 
\end{figure}

The analysis of few-electron quantum dots using the exact diagonalization 
method has shown 
that the final break-up of vortex molecules and the transition into the 
fractional quantum Hall regime of electrons is associated with 
the formation of composites of particles and vortices~\cite{saarikoski2004}. 
Electrons ``capture'' free vortices, breaking up the vortex molecules.  
Similar processes have been reported also for 
bosons in the LLL by analysing the vortex attachment with reduced wave 
functions (see Fig.~\ref{fig:bosoniclaughlin}). 
These calculations suggest, however, that vortices 
continue to show ordering at surprisingly low 
particle-to-vortex filling factors, well below the 
obtained stability limits of vortex lattices in bosonic condensates. 
This is also evident for fermions, as shown in Fig.~\ref{mattif2},  
where hole correlations show vortex molecules at very high angular momentum 
and large zero-point fluctuations. In the case of fermions, 
vortex localization may continue to filling factors down to $\nu={1 \over 2}$ 
where a transition from prominent vortex localization into 
particle localization occurs~\cite{emperador2006}. 
These calculations showed signs of vortex-hole bunching and 
the formation of concentric rings of localized vortices, 
until the number of (free) vortices was equal to the number of particles.
Below $\nu={1\over 2}$, no such signatures 
are seen. Instead, this regime is characterized by particle 
localization. The conditional probability densities begin to  
show prominent localized structures~\cite{koskinen2001,yannouleas2007}. 
The corresponding bosonic case has not been studied, but 
due to close analogies of bosonic and fermionic states, 
similar results are expected to hold also for small bosonic droplets where 
vortex localization should disappear at $\nu_{pv}=1$. 
 
These results suggest that signatures of vortex localization 
in small systems disappear at a particle-to-vortex ratio which is 
an order of magnitude lower than the value where vortex lattice melting 
occurs in large bosonic condensates. However, as mentioned before,  
in small systems the separation of liquid and solid is difficult,  
and the observed transition is also related to the formation 
of composite particles (see Sec.~\ref{sec:rapid}).  
 
\subsection{Giant vortices} 
 
\label{sec:giant} 

In multiply-quantized vortices, the phase changes several integer 
multiples of $2\pi $ when encircling the singularity. 
However, they are not stable in a purely harmonic confinement potential.
The existence of many singly-quantized vortices is energetically prefered,  
and the effective repulsive interaction between the vortex cores
leads to a lattice of singly-quantized 
vortices~\cite{butts1999,castin1999,lundh2002}.
The instability of multiply-quantized vortices in harmonic potentials, 
and the break-up into singly-quantized vortices was further
discussed by~\textcite{mottonen2003} and \textcite{pu1999}.
Disintegration of a multiple quantized vortex 
has also been observed experimentally~\cite{shin2004}. 

\begin{figure} 
\includegraphics[width=.5\textwidth]{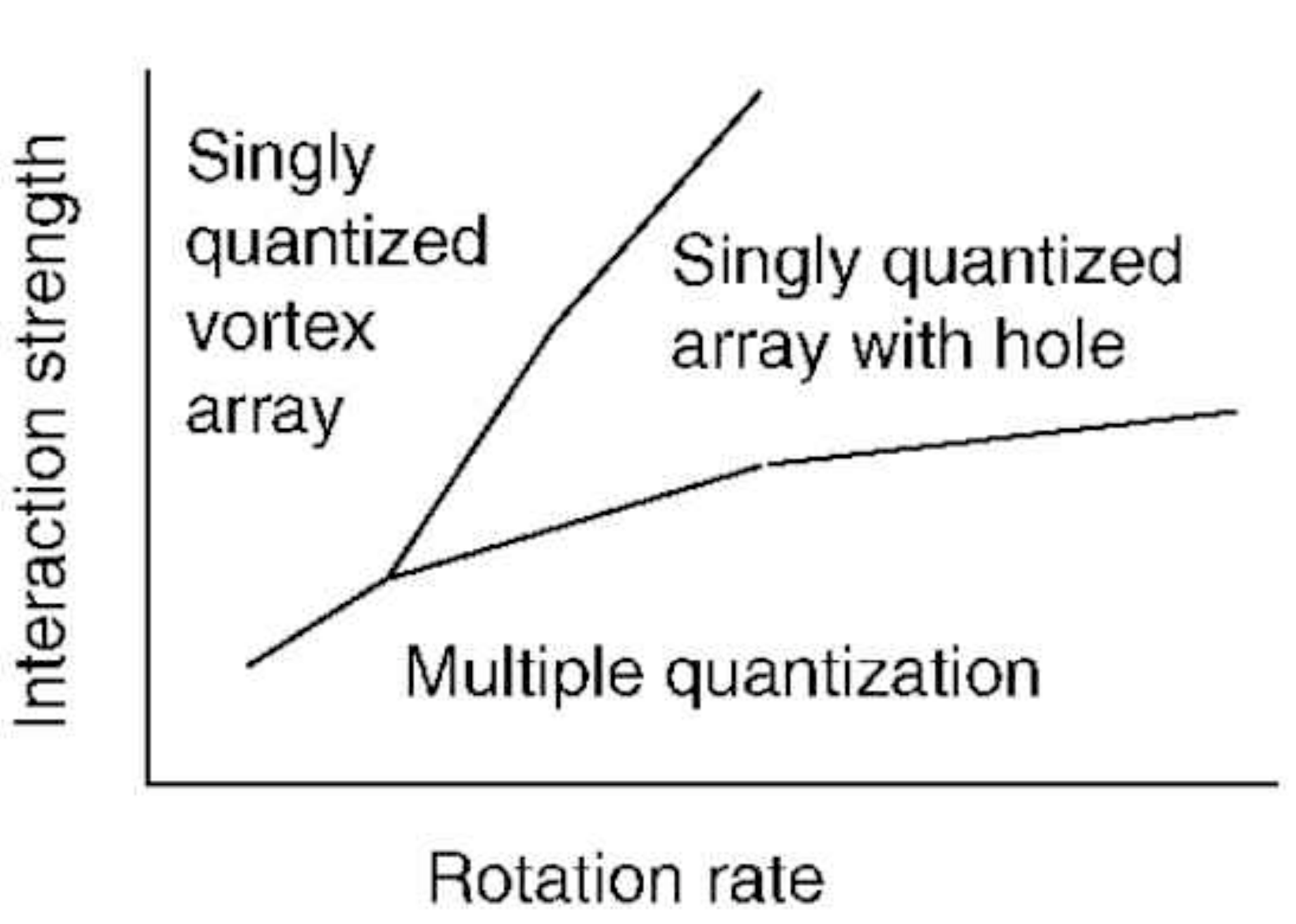} 
\caption{Schematic phase diagram of the ground states of a bosonic cloud in an
  anharmonic confinement. From~\textcite{kavoulakis2003}.} 
\label{fig:anharmphase} 
\end{figure}

Rotating condensates in {\it anharmonic} potentials
that rise more rapidly than $r^2$, however, show a behavior that is very
different from purely harmonic traps. Most commonly, a quartic perturbation
is added to the oscillator confinement\footnote{See {\it e.g.},  
~\cite{fetter2001,lundh2002,kasamatsu2002,kavoulakis2003,fischer2003,jackson2004a,jackson2004b,fetter2005,fu2006,blanc2008}}. 
Due to the anharmonicity it is possible to rotate the system 
sufficiently fast such that the centrifugal force may create a large density hole at the trap center.
So-called ``giant'' vortices with a large core at the center may exist
that originate from multiple quantization. 
Singly-quantized vortices may also form a close-packed ensemble 
inside a large density core. In addition, for certain parameter ranges, the
usually-quantized lattice exists. 
\textcite{kavoulakis2003} found a very rich phase diagram, for which 
a schematic picture is given in Fig.~\ref{fig:anharmphase},
showing the different possible phases as a function of the 
interaction strength and the trap rotation.  
In the following, we discuss the formation and structure of such ``giant'' vortex 
states in both bosonic as well as in fermionic quantum droplets.

\subsubsection{Bose-Einstein condensates in anharmonic potentials} 
 
\begin{figure} 
\includegraphics[width=.5\textwidth]{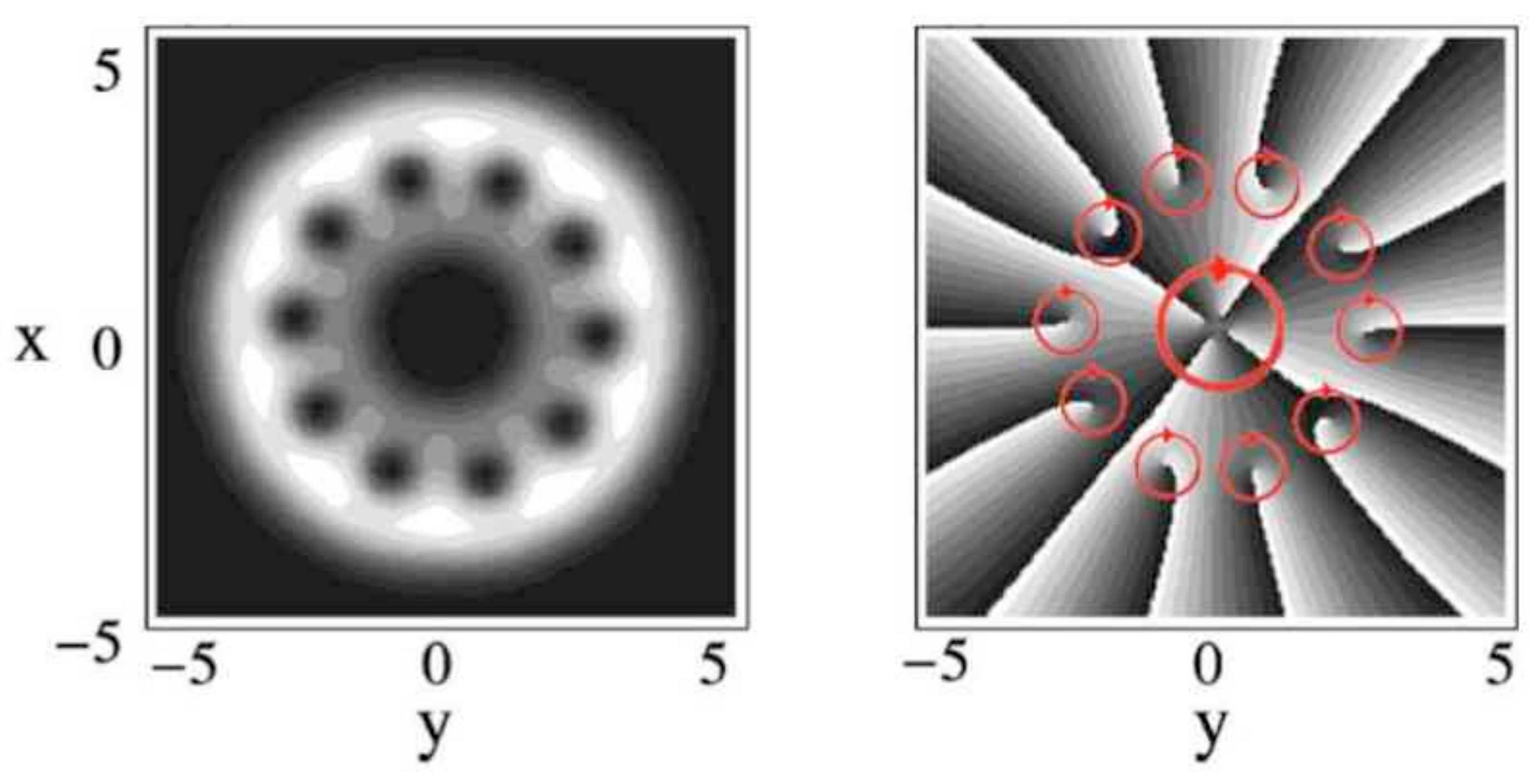} 
\caption{({\it Color online})  A rotating Bose-Einstein condensate in a mixed state with 
a giant vortex in the center, surrounded by ten single-quantized vortices. 
The giant vortex is composed of four phase singularities. 
The left panel shows the particle density (white is high density and 
black is zero density) 
and the right panel shows the phase profile. Locations of phase singularities 
are marked with red circles. The large circle marks the  
ensemble of four phase singularities in the core of the giant vortex. 
After~\textcite{kasamatsu2002}, who applied the
Gross-Pitaevskii method.} 
\label{fig:giant1} 
\end{figure}

\textcite{lundh2002} proposed that 
in the presence of anharmonicity of the confining trap potential, 
multiply-quantized vortices with a giant vortex core could 
exist in a rotating condensate, and calculated the ground-state vortex 
structures within the Gross-Pitaevskii formalism.
In fact, vortices in these states are not truly multiple-quantized 
vortices but rather dense-packed ensembles 
of single-quantized vortices~\cite{fischer2003,kasamatsu2002}. 
Phase singularities do not completely 
merge into the same point 
because the residual interaction between phase singularities 
is logarithmic as a function of intervortex separation in the 
region of low particle density surrounding 
the cores. 
Despite this fact, the composite core has a large and uniform spatial 
extent. Therefore, the name ``giant vortex'' was coined for 
these structures. 
Depending on the strength of the anharmonicity, the condensate can exist 
in a phase where only single-quantized vortices occur, in a state 
where all vortices form a giant vortex, and in a mixed phase where 
both giant vortices and single-quantized vortices 
exist~\cite{kasamatsu2002,kavoulakis2003,jackson2004a,jackson2004b}. 
An example 
of the latter is shown in~Fig. \ref{fig:giant1}. 
 
We further note that anharmonicity, which is required for giant 
vortex formation, may be induced also via the presence of another,  
distinguishable particle component.
The interaction between the particles would then 
create an {\em effectively} anharmonic potential for the particle components 
which may induce giant vortex 
formation~\cite{bargi2007,christensson2008a,yang2008}. This is discussed in 
Sec.~\ref{sec:multi-component} in the context of multi-component 
quantum droplets. 
 
\subsubsection{Giant vortices in quantum dots} 
 
\begin{figure} 
\includegraphics[width=.49\textwidth]{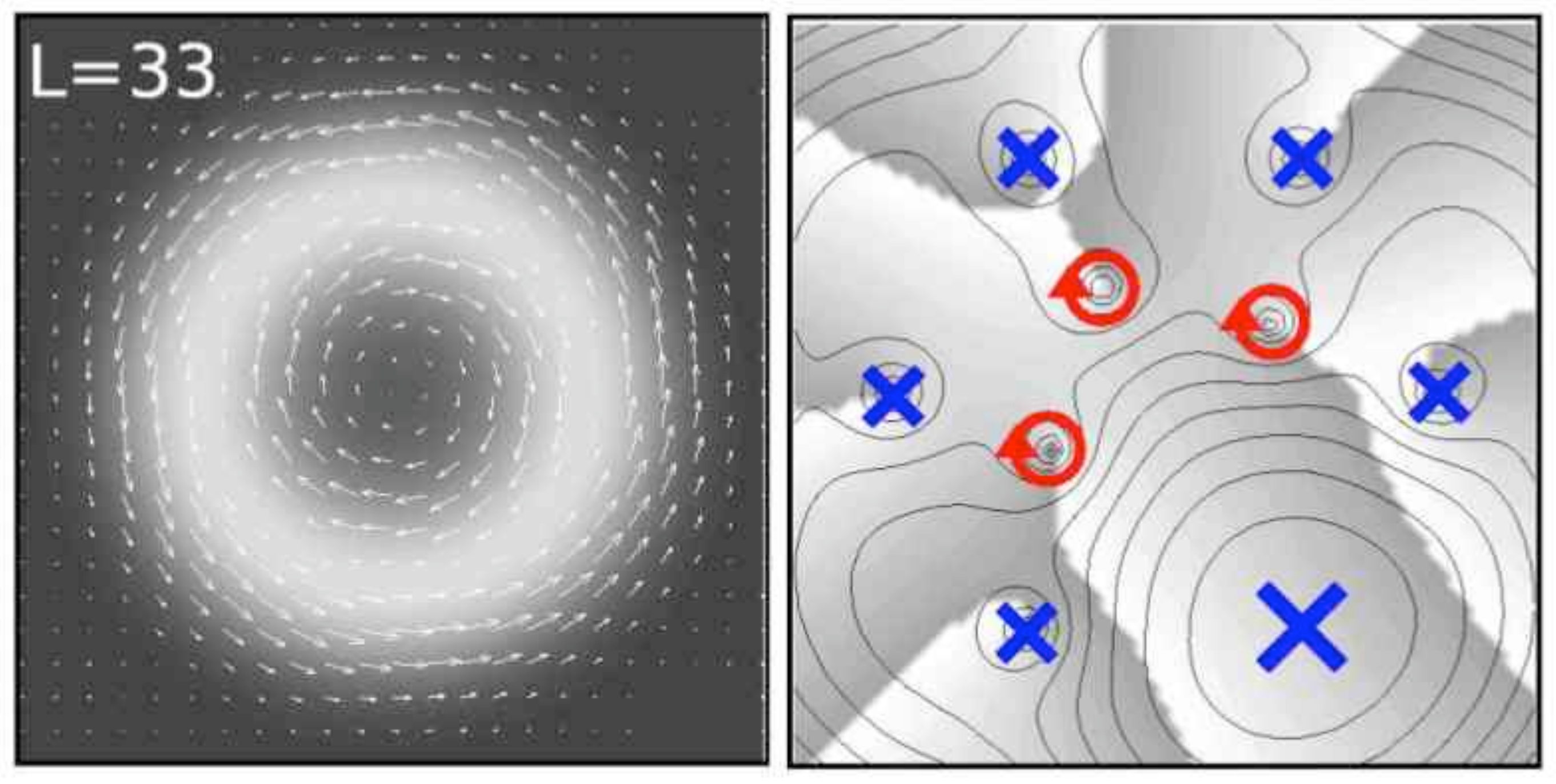} 
\caption{({\it Color online})  A giant vortex in a six-electron quantum dot 
calculated with the exact diagonalization method. 
The left panel shows the particle density (black is low density) 
and current density (arrows), and the right panel shows 
the reduced wave function, 
where phase singularities are marked with red circles 
and electron positions with crosses. 
The giant core in this case comprises three phase singularities. 
Interactions and fluctuations keep the phase singularities separated. 
The probe electron is on the bottom-right. 
From~\textcite{rasanen2006}.} 
\label{fig:giant2} 
\end{figure}

Giant vortex structures are predicted to form also in fermionic droplets with
repulsive interactions, as it was 
shown by exact diagonalization calculations for 
few-electron quantum dots~\cite{rasanen2006}. 
Similarly to the bosonic case, giant vortices emerge in anharmonic 
confining potentials 
and their structure shows a large core with multiple phase singularities. 
It was found that even a slight  
anharmonicity in the confining potential is sufficient for  
these giant vortex states to become energetically favorable. 
In addition to the particle interactions, 
fluctuations tend to keep phase singularities separated,  
broadening the charge deficiency in the core to a larger area 
(see Fig.~\ref{fig:giant2}). 
The electron density of a central giant-vortex state shows 
a ring-like distribution.

Unlike bosonic systems, giant vortices with repulsive fermions
were only found in the limit of small numbers of particles.
This could be seen as another manifestation of the 
tendency of vortices to drift 
towards the edge of the droplet in the limit of large 
particle numbers (see  Sec.~\ref{sec:singlevortex}), 
breaking apart the giant vortex pattern at the center. 
In electron droplets interacting via 
Coulomb forces, density-functional calculations predicted that 
giant vortex formation  is generally limited to systems with 
less than 20 fermions~\cite{rasanen2006}. 
 
\subsection{Formation of composite particles at rapid rotation} 
 
\label{sec:rapid} 
In the regime of high vorticity, electron-vortex correlations 
are particularly strong and cause vortices to be bound to electrons. 
This regime is ultimately linked with the fractional quantum Hall effect
in the 2D electron gas.
Actually, the early works aiming to explain this effect used a
disk geometry~\cite{laughlin1983,girvin1983} and are in fact  
more relevant for quantum dots than for the bulk
properties of quantum Hall systems.

\begin{figure} 
\includegraphics[width=.49\textwidth]{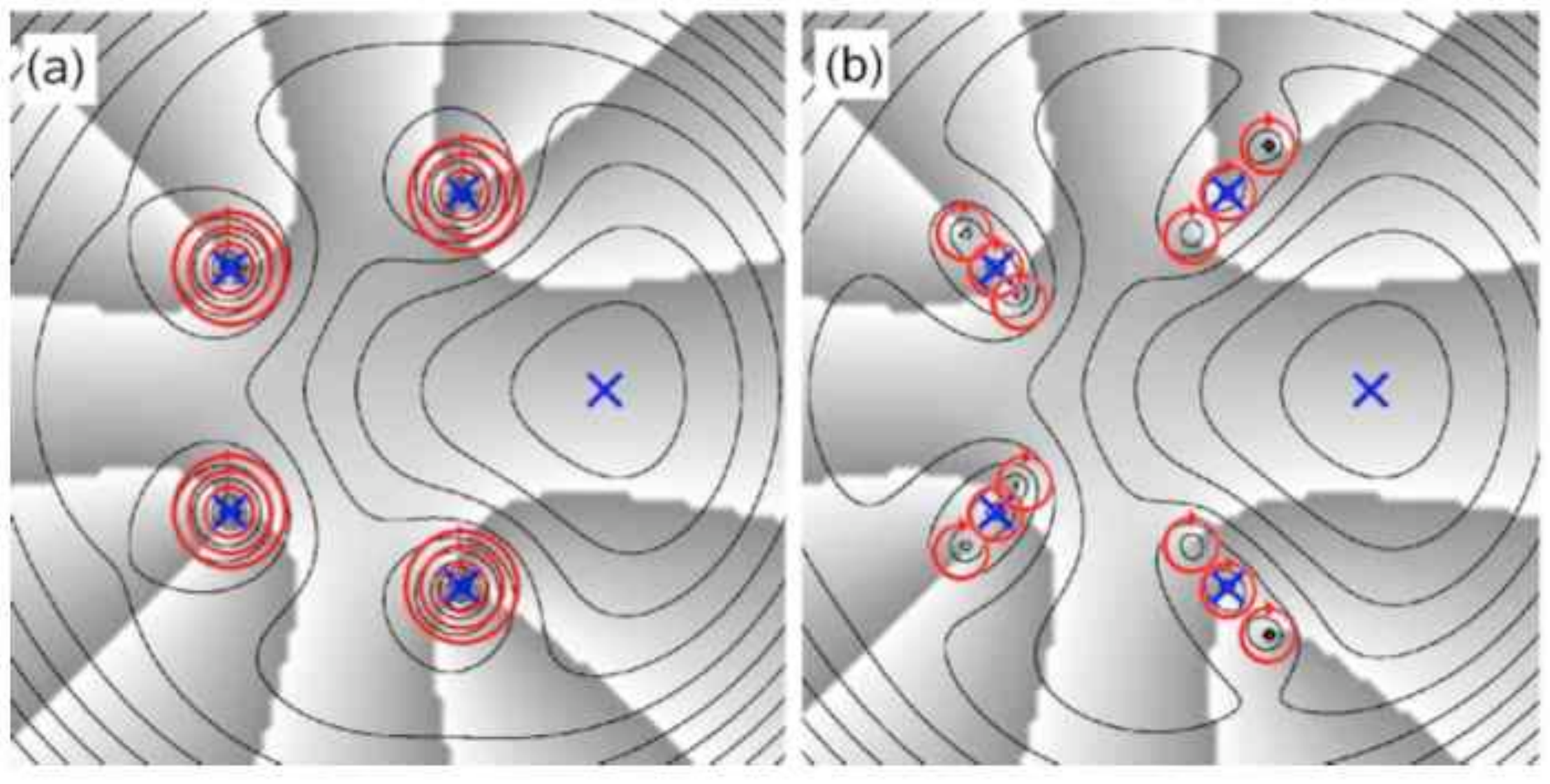} 
\caption{({\it Color online}) The reduced wave functions for 
(a) the approximate Laughlin state $\nu={1 \over 3}$ and 
(b) the exact $L=30$ ground state for five electrons in a parabolic 
external potential. The Laughlin state fixes a triple-vortex 
(concentric rings) on each electron position (crosses). 
In the exact solution there are clusters of three vortices 
near each electron.} 
\label{fig:laughlinexact} 
\end{figure}

Figure~\ref{fig:laughlinexact} 
shows the nodal structure of the reduced wave function for the 
Laughlin state state for $N=5$ electrons as well as the corresponding 
$L=30$ state obtained with the CI method.
In the Laughlin $\nu={1\over 3}$
state, there are three vortices on each electron position, one Pauli
vortex and two extra vortices, as shown in
Fig. \ref{fig:laughlinexact}(a).  In the exact wave function, 
there are clusters of three vortices near each electron 
(except near the probe electron). 
There is one vortex on top of each 
electron position, as required by the Pauli principle, but, 
in addition, there are two vortices very close-by,  
separated by their mutual repulsion to opposite sides. 
Calculations show that small changes in the position of one of 
the fixed electrons in the reduced wave function causes the vortex 
to be dragged along with the electron, which indicates 
vortex attachment to the electron.
The overlap between the exact state and the Laughlin approximation 
is 0.98. The state can be interpreted as a finite-size 
precursor of the $\nu={1\over 3}$ fractional quantum Hall state, 
for which the Laughlin wave function yields an accurate description.
However, in contrast to the Laughlin state, the attachment of nodes 
to particles in the exact wave function 
shows a small spatial separation.  

\begin{figure} 
\includegraphics[width=.24\textwidth]{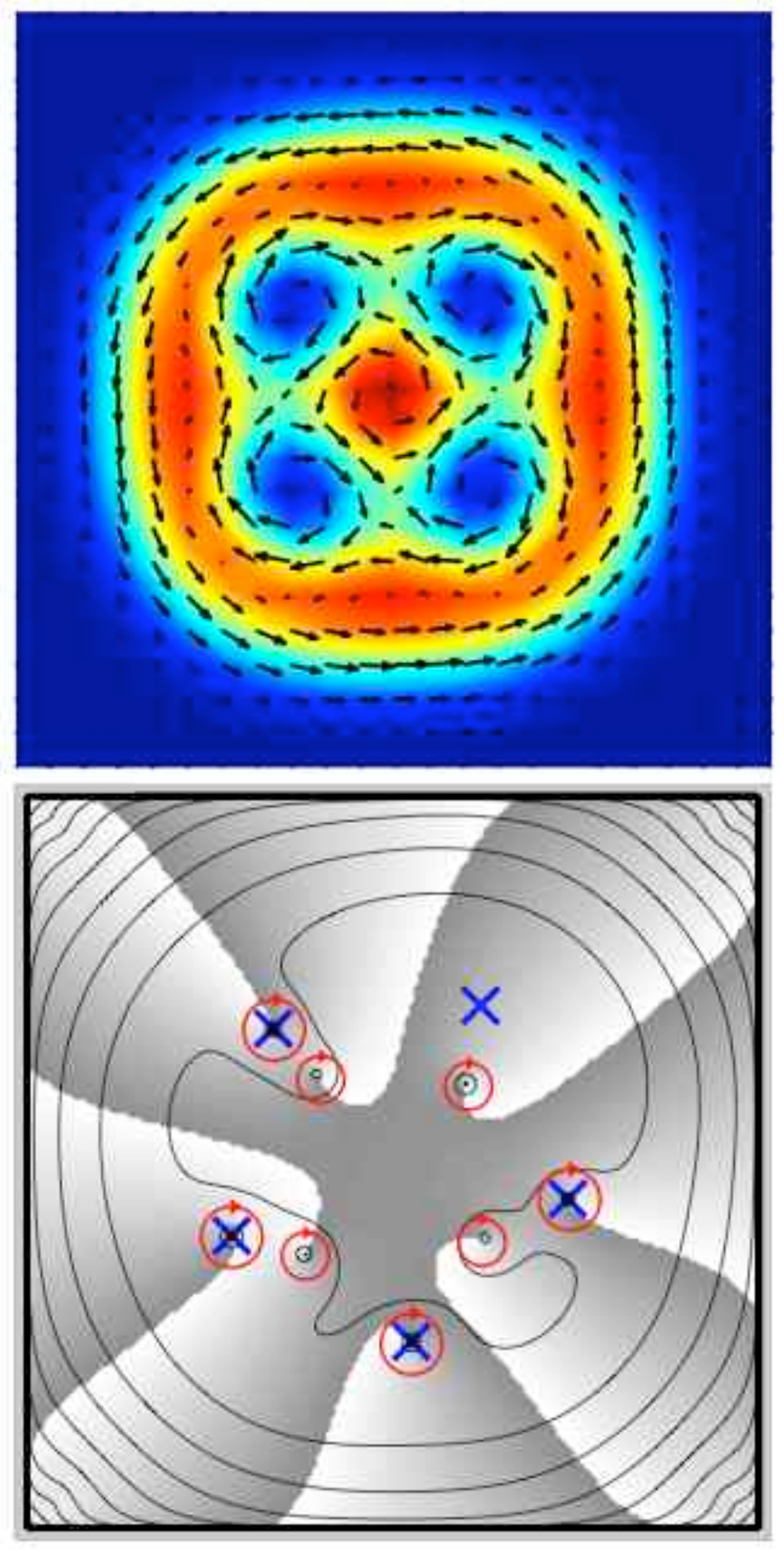} 
\caption{({\it Color online})  A $\nu={1\over 3}$ state of five electrons in a harmonic trap.
{\it Left:} electron density (color) from the density-functional 
method and current density 
(arrows). The confinement strength is $\hbar \omega_0=5 \; {\rm meV}$ and 
the magnetic field is $B=18 T$. {\it Right:} reduced wave function 
for the same state constructed from the Kohn-Sham single-particle states. 
The probe electron is at the top-right. 
} 
\label{fig:sdftnu13} 
\end{figure}

\begin{figure} 
\includegraphics[width=.45\textwidth]{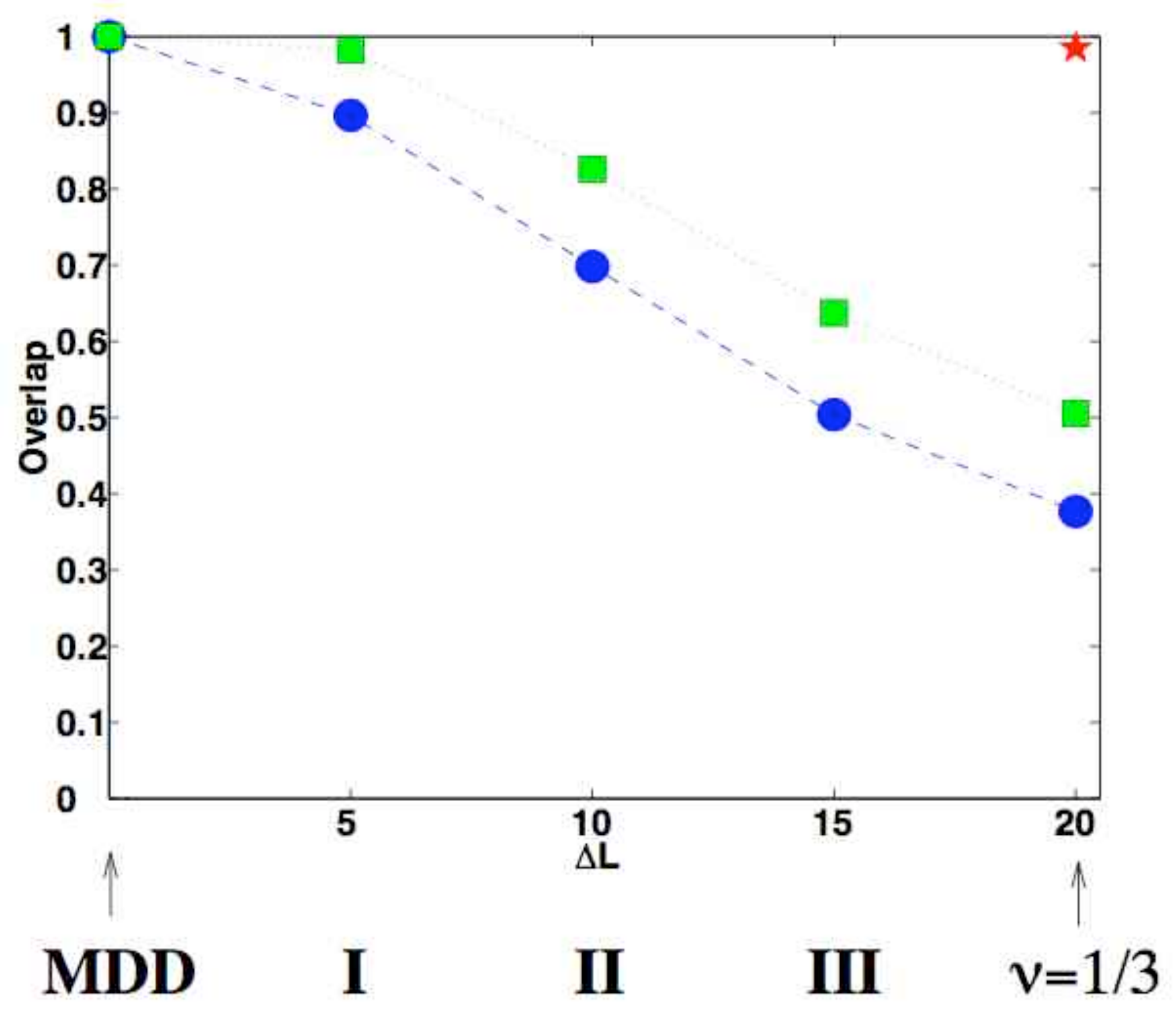} 
\caption{({\it Color online}) 
Overlap of a single-determinant wave function with the exact one 
as a function of the angular momentum increase with respect to the 
MDD state $\Delta L = L - L_{\rm MDD}$ for 5-electrons in parabolic 
confinement. The higher points {\it (squares)} are obtained with a coordinate 
transformation to the center-of-mass, 
$z_i \to z_i - z_{\rm CM}$ and the lower ones {\it (circles)} 
without it. The star at $\Delta L=20$ shows the overlap with 
the Laughlin $\nu=1/3$ state. 
The roman numbers count the vortices inside the electron 
droplet. From~\textcite{harju2005}. 
} 
\label{fig:overlaps} 
\end{figure}

The attachment of vortices to particles explains also the absence of vortices 
for the probe electron in the exact many-body state, see 
Fig.~\ref{fig:laughlinexact}. 
In the fractional quantum Hall regime, the density-functional  
method failed to reveal the correct 
nature of the ground state. The solutions of the spin- 
as well as current-spin-density-functional theory  
show only a cluster of vortices inside the electron 
droplet, but these methods are unable to associate two extra vortices 
to each electron~\cite{saarikoski2005b} (see Fig.~\ref{fig:sdftnu13}). 
The density-functional approach fails to properly include 
these correlations. A single-determinantal 
wave function constructed from the self-consistent Kohn-Sham orbitals
yields an approximate description for few-vortex 
states near $\nu=1$, but the overlaps with the exact wave functions 
diminish as the angular momentum of the system increases. 
Fig. \ref{fig:overlaps} shows that for a five-electron system at $\nu=1/3$ 
the overlap is only of the order of 0.5~. Compared to this, the overlap 
with the Laughlin $\nu=1/3$ wave function that amounts to 0.98 is very high. 
\begin{figure}
\includegraphics[width=.5\textwidth]{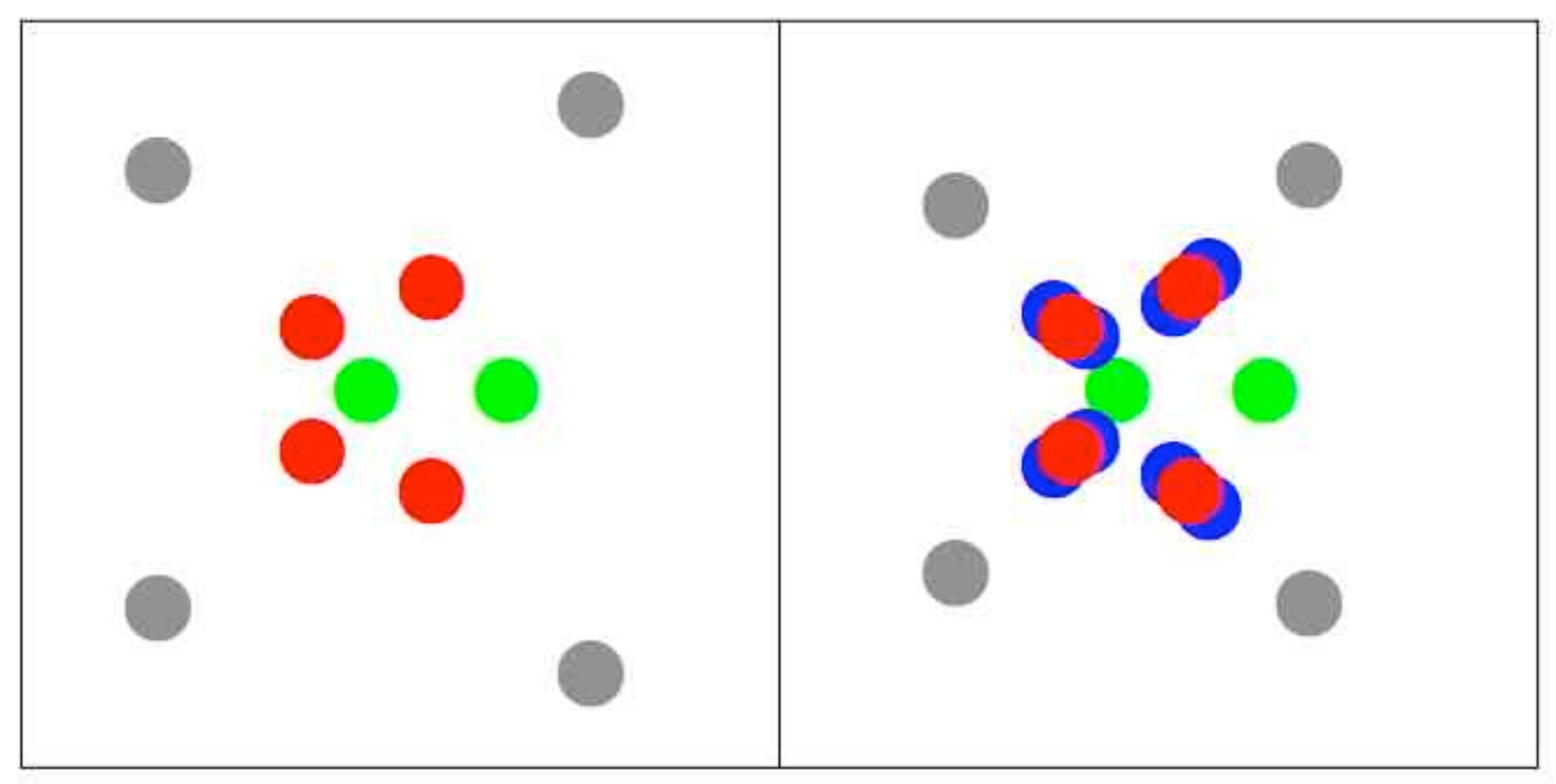}
\caption{({\it Color online})  Schematic view of the sites of vortices,  
determined from the reduced wave function of an exact diagonalization
for five electrons with angular momentum $L=16$ (6 above the MDD, left) and
$L=36$ (6 above the state with filling factor 1/3, right). Fixed electron
positions with Pauli vortices are
denoted by red bullets, 
vortices attached to electrons making composite particles 
by blue, and free vortices inside the electron droplet by green bullets. 
Free vortices outside the droplet are shown by gray bullets.
}
\label{matticomp}
\end{figure}
When the angular momentum of the droplet increases further,
additional vortices appear in the Laughlin-like state and 
the filling factor decreases below $\nu=1/3$. 
These vortices are not bound to composite particles,
rather they correspond to the Laughlin excitations with fractional
charge. The pattern of vortex formation is expected to be similar 
to that after the MDD: first a single vortex enters from the surface and moves
towards the center until it is energetically favorable 
to have two vortices, and so on. This is illustrated in Fig.~\ref{matticomp}, 
which shows the vortex sites for a five-electron system determined from the
reduced wave function. Again, a similar behavior is
expected in the case of bosonic particles. 
However, despite the recent progress in realizing BEC's at 
extreme rotation~\cite{lin2009}, an analysis of these states  
appears still to be beyond the current experimental capabilities.

There are two basic mechanisms to unbound the vortices from the 
particles, namely the softening of the interaction potential by 
{\it e.g.} the finite thickness of the system, and secondly by impurities. 
When the system has a finite thickness, the incompressible $\nu=1/3$ 
Laughlin state breaks down as the vortices are gradually less bound to 
the electron coordinates. This effect is in contrast to the 
screening of the Coulomb interaction energy
whereby, in the strong-screening limit, the zeros are 
exactly localized to the electron positions~\cite{tolo2009}.  
We should also mention that repulsive impurities attract vortices
at the impurity position~\cite{baardsen2009}. 

%%%%%%%%%%%%%%%%%%%
%+++++++++++++++++++++++++++++++++++++++++++++++++++++++++++++++++++++

\section{Multi-component quantum droplets}
\label{sec:multi-component}

Multi-component quantum droplets are composed of
different particle species, 
that may for example be different atoms, different
isotopes of the same atom, different spin states of an atom or electron, 
or even different hyperfine states of an atom.
In such systems inter-component 
interactions can modify the many-body wave function
significantly. 

The properties of multi-component BEC's
have been much discussed, both experimentally 
and theoretically, over the past few years. 
For recent reviews on multi-component BEC's, see~\textcite{kasamatsu2005b} 
and parts of the article by~\textcite{fetter2008}.
We do not attempt to cover the vast literature on binary or spinor BEC's,
but instead set our focus mainly on structural
properties and vorticity of few-particle droplets
and the analogies between bosonic and fermionic two-component systems.
Only a brief outlook on spinor condensates with more components
is given at the end of this chapter. 

Theoretical studies of multi-component quantum liquids 
were performed already in the 1950s for superfluid helium mixtures, 
see for example the early 
works by~\textcite{guttman1953}, \textcite{khalatnikov1957}, 
and~\textcite{leggett1975}. 
Examples for vortex patterns 
include the Mermin-Ho vortex~\cite{mermin1976} and the Anderson-Toulouse 
vortex~\cite{anderson1977}.
These vortices are non-singular and the order-parameter
is continuously rotated by superposing a texture on it (see below).
More recently, doubly-quantized vortices in the $A$-phase of $^3$He were 
found~by~\textcite{blaauwgeers2000}.
With ultra-cold atoms, condensate mixtures may be achieved 
by using different atomic species, such as $^{87}$Rb 
and $^{41}$K~\cite{modugno2002}, 
or for example the different isotopes of $^{87}$Rb~\cite{burke1998,bloch2001},
in the same trap. 

Another possibility to create multi-component condensates 
is given by the different hyperfine states of the same 
atom, as for example  $^{87}$Rb with the hyperfine states 
$\mid F=1, m_f = -1 \rangle $ and $\mid F=2, m_f = 1\rangle $
~\cite{myatt1997,matthews1998,matthews1999,hall1998a,hall1998b}.
The atoms in the two states have nearly equal 
inter- and intra-component scattering lengths, and  
the spin flip rate is very small due to weak hyperfine
coupling, which yields a stable two-component system with a long
lifetime~\cite{julienne1997,kasamatsu2005b}.
In fact, the first experiment by~\textcite{matthews1999} 
creating vortices in a BEC 
made use of these internal spin states, following a suggestion
by~\textcite{williams1999}: 
they proposed a phase-imprinting technique, where 
an external coupling field was used to control independently the two 
components of the quantum gas. 
In this way, angular momentum could be induced in one component, that 
formed a quantized vortex around the non-rotating core of 
the other component, when the coupling was turned off.
Since the magnetic moments for the $^{87}$Rb atoms in the two 
hyperfine states are nearly equal, they could be confined by the 
same magnetic trap. 

Optical traps have the advantage that one is not restricted 
by certain hyperfine spin states. 
Already in 1998, experimentalists at MIT could 
create a BEC of $^{23}$Na~\cite{stamperkurn1998,stenger1998}
where different ``spinor'' degrees
of freedom of the atomic quantum gas can be trapped simultaneously. 
Other examples are $^{39}$K, and $^{87}$Rb ~\cite{barrett2001}. 
In these alkali systems one can trap the three projections of the   
hyperfine multiplet with $F=1$, 
adding three (internal) degrees of freedom to the system. 
However, population exchange (without trap loss) 
among the hyperfine states may occur due to
spin relaxation collisions~\cite{stenger1998}. 
The dynamical loss of polarization of a BEC due to spin flips
was examined by~\textcite{law1998}. 
Larger atom spins can also be realized, as 
for example with $^{85}$Rb and $^{133}$Cs. 
Such condensates show a wealth of quantum phenomena 
that do not occur in simple scalar condensates~\cite{ho1998,ohmi1998}.
The interactions between the different components of the trapped 
cold-atom gas  may lead to topologically interesting, new quantum states. 

Rotating two-component fermion droplets may be realized with 
electrons in quasi-two-dimensional quantum dots~\cite{reimann2002} 
with a spin degree of freedom.
Usually, the magnetic field causes polarization of the droplet
due to the Zeeman coupling.
However, in 2D electron systems, the Zeeman splitting can be
tuned by applying external pressure~\cite{leadley1997} or by
changing, {\it e.g.}, the Al-content in a
${\rm GaAs}/{\rm Al}_x{\rm Ga}_{1-x}{\rm As}$-sample
~\cite{saliskato2001,weisbuch1977}.
In systems with low Zeeman coupling the regime of vortex formation
beyond the maximum density droplet is associated
with various spin polarization states~\cite{siljamaki2002}.
These states occur in much analogy to those in two-component bosonic
systems~\cite{saarikoski2009}.
In the regime of rapid rotation, some of the many-electron states
can also be identified as finite-size counterparts of non-polarized
quantum Hall states, such as the much studied
$\nu={2 \over 3}$ and  $\nu={2 \over 5}$ states~\cite{chakraborty1984,xie1989}.

\subsection{Pseudospin description of multi-component condensates}

For a bosonic condensate with $n$ components, the order parameter
$\Psi $ becomes  of vector type $(\psi _1, \psi _2, \ldots , \psi _n)$.
One may interpret this 
as a ``pseudospin'' degree of freedom~\cite{kasamatsu2005a,kasamatsu2005b}.
As an example, for $n=2$ distinguishable particles of kind $A$ or $B$
the order parameter is then a spinor-type function, 
$\psi =(\psi _A, \psi _B)$,  and  
the pseudospin ${\bf T}$ points ``up'' ($T=1/2$) or 
``down'' ($T=-1/2$) for either of the two components in the absence of the other.
This concept straightforwardly extends to higher half-integer, 
as well as integer pseudospins. 

When rotation is induced in the multi-component or ``spinor'' 
system, vortex formation becomes much more complex due to the increased 
freedom of the system to carry angular momentum. 
Spatial variations in the directions of
the atomic spins may lead to very different patterns, such as 
the aforementioned spin textures.
For atomic quantum gases, these structures were extensively investigated
theoretically\footnote{See {\it e.g.}~\cite{ho1998,ohmi1998,yip1999,chui2001,
stoof2001,alkhawaja2001a,alkhawaja2001b,alkhawaja2002,isoshima2001,
isoshima2002,mizushima2002,mizushima2002b,mizushima2002c,martikainen2002,
kita2002,reijnders2004,mueller2004a}.}.
Many theoretical studies applied the spin-dependent 
Gross-Pitaevskii formalism.
The Thomas-Fermi approach has been
used to determine the density profiles 
of ground state and vortex structures for two-component mixtures of 
bosonic condensates~\cite{ho1996}.
This approach was later simplified to describe segregation of components
in the presence of vorticity~\cite{jezek2001,jezek2005}.

In their most general form, the two-body interactions are
often parameterized by 
$V_{ij} = [c_0 + c_2 ({\bf T}_i \cdot {\bf T}_j)] \delta ({\bf r}_i-{\bf r}_j)$
with the usual contact interactions of strengths $c_0$. 
For $c_2>0$, {\it i.e.}, repulsive spin-dependent interactions, 
as for example for $^{23}$Na, the system  minimizes 
the total spin. Consequently, this parameter regime is 
called the ``antiferromagnetic'' one, while for $c_2 <0$, as for example for 
$^{87}$Rb, the spin-interactions are called ``ferromagnetic''
\cite{ho1998,stamperkurn1998,stenger1998,miesner1999}. 
Typically, the ratio of the spin-dependent and spin-independent 
parts of the contact interaction is of the order of a few percent.
In the following we set $c_2=0$
and restrict the discussion to the special case of SU(2) symmetry,
unless otherwise stated.

\subsection{Two-component bosonic condensates}
\label{sec:bosetwocompGP} 

Let us now consider a bosonic gas of atoms that is a mixture of two 
distinguishable species $A$ and $B$ with fixed numbers of atoms 
$N_A$ and $N_B$. The majority of experimentally studied two-component gases
has very similar interactions between the like and unlike species.
Similar $s$-wave scattering lengths yield a very small inelastic spin 
exchange rate \cite{julienne1997}, providing a stable two-component system
with a long lifetime~\cite{kasamatsu2005b}. 
Therefore, the case $g_{AA}\approx g_{BB}\approx g_{AB}$ 
(with interaction strengths as defined in Sec.~\ref{sec:gp} above)
appears as the most relevant one.  Thus, we first assume equal and 
(pseudo)spin-independent coupling
strengths $g$ between all particles, and also choose the harmonic 
trapping potentials for the two components to be identical.
As mentioned above, the two-component Bose gas is then described by 
a pseudospin $1/2$ and 
the order parameter is a vector, $(\psi _A, \psi _B)$.

We have seen in Section~\ref{sec:single-component} above that 
for repulsive interactions, a condensate with only one kind of atoms
that is brought to rotation, develops first a single vortex at the trap center 
at $L/N = 1$. With increasing angular momentum, the 
single-component, so-called 
``scalar'' condensate nucleates an increasing number of vortices 
inside the condensate, until the triangular Abrikosov vortex lattice is 
formed~\cite{madison2000,aboshaer2001}, in agreement with the results of 
Gross-Pitaevskii mean-field theory~\cite{butts1999,kavoulakis2000}.
The case of a two-component gas 
is more complex since the system may divide its
angular momentum between its components.
One possibility is that one component is at rest, while another carries 
all the angular momentum. The component at rest may then fill 
the core of the first vortex in the other component, creating a so-called 
{\it coreless} vortex state. 
When the rotation increases, the Abrikosov lattice of the scalar condensate
now may become a lattice of such coreless vortices.
The vortex lattice geometry depends crucially on the interactions between the
components, as well as the sizes and numbers of components. 

\subsubsection{Asymmetric component sizes}

\label{sec:asymmetry}
\begin{figure}
\includegraphics[width=.45\textwidth]{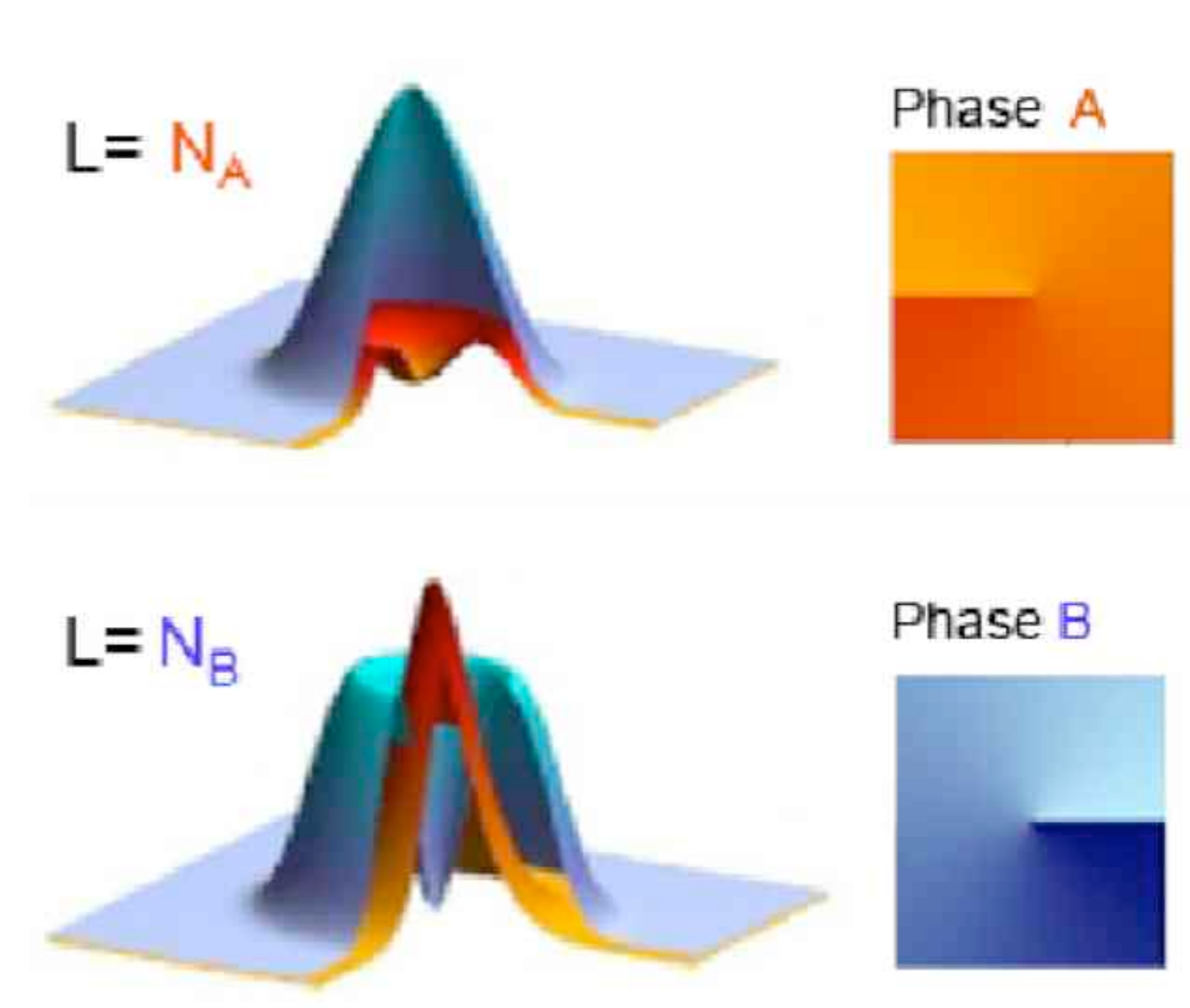}
\caption{({\it Color online})  Densities {\it (left)} and phases {\it (right)} in a two-component 
rotating Bose-Einstein condensate, as obtained from the Gross-Pitaevskii 
equations, for a ratio of atom numbers in the two components of  
$N_A/N_B=0.36$ and equal coupling strengths $g_{AA}=g_{AB}=g_{BB}=50$~a.u. 
(The densities are cut in one quadrant in order to visualize 
them for both components in one diagram). The rotational frequency
is $\Omega =0.45$.
The {\it upper panel} shows a coreless vortex at angular momentum  
$L=N_A$, where the smaller component $A$ shows a unit vortex 
at the center, as it is clearly seen from the phase of $\Psi _A$ plotted to the
right (from dark to light shading), 
changing by $2\pi$ when the center is encircled once.
The phase singularity is absent in $B$ component.
The {\it lower panel} shows the case $L=N_B$, where now the larger component
encircles the smaller one, filling the unit vortex at the center. 
The phase singularity consequently now occurs in the 
order parameter of  $B$ component, as shown to the right 
(from dark to light blue). After data from~\textcite{bargi2008}.
}
\label{twocomponentGP_1v}
\end{figure}

Figure~\ref{twocomponentGP_1v} shows the mean-field (Gross-Pitaevskii)
densities and phases of the order parameters $\psi _A$ and $\psi _B$ 
for a two-component condensate with unequal particle populations
$N_B>N_A$~\cite{bargi2007,bargi2008}.
At $L/N_A=1$, the system forms a single vortex in the smaller component $A$, 
which is clearly seen in the phase plot of the order parameter in
Fig.~\ref{twocomponentGP_1v}. 
The phase jump is $2\pi $ along any closed path encircling the origin.
The larger component rests at the origin $L_B=0$ with no vorticity 
(and, correspondingly, a  flat phase profile in the order parameter). 
When the angular momentum reaches $L=N_B$, a singly-quantized 
coreless vortex is formed in the larger 
component $B$, while the component $A$ now is 
stationary at the origin.

Referring back to the work of Skyrme
in the context of nuclear and high-energy
physics~\cite{skyrme1961,skyrme1962}
such coreless vortices were also called ``skyrmions'', 
see the review 
by~\textcite{kasamatsu2005b}.\footnote{This terminology has also been used for analogous textures in 
liquid $^3$He-A~\cite{mermin1976,anderson1977,salomaa1987}, 
and in quantum Hall states~\cite{lee1990,sondhi1993,barrett1995,aifer1996,oaknin1996}.}.
\begin{figure}
\includegraphics[width=.4\textwidth]{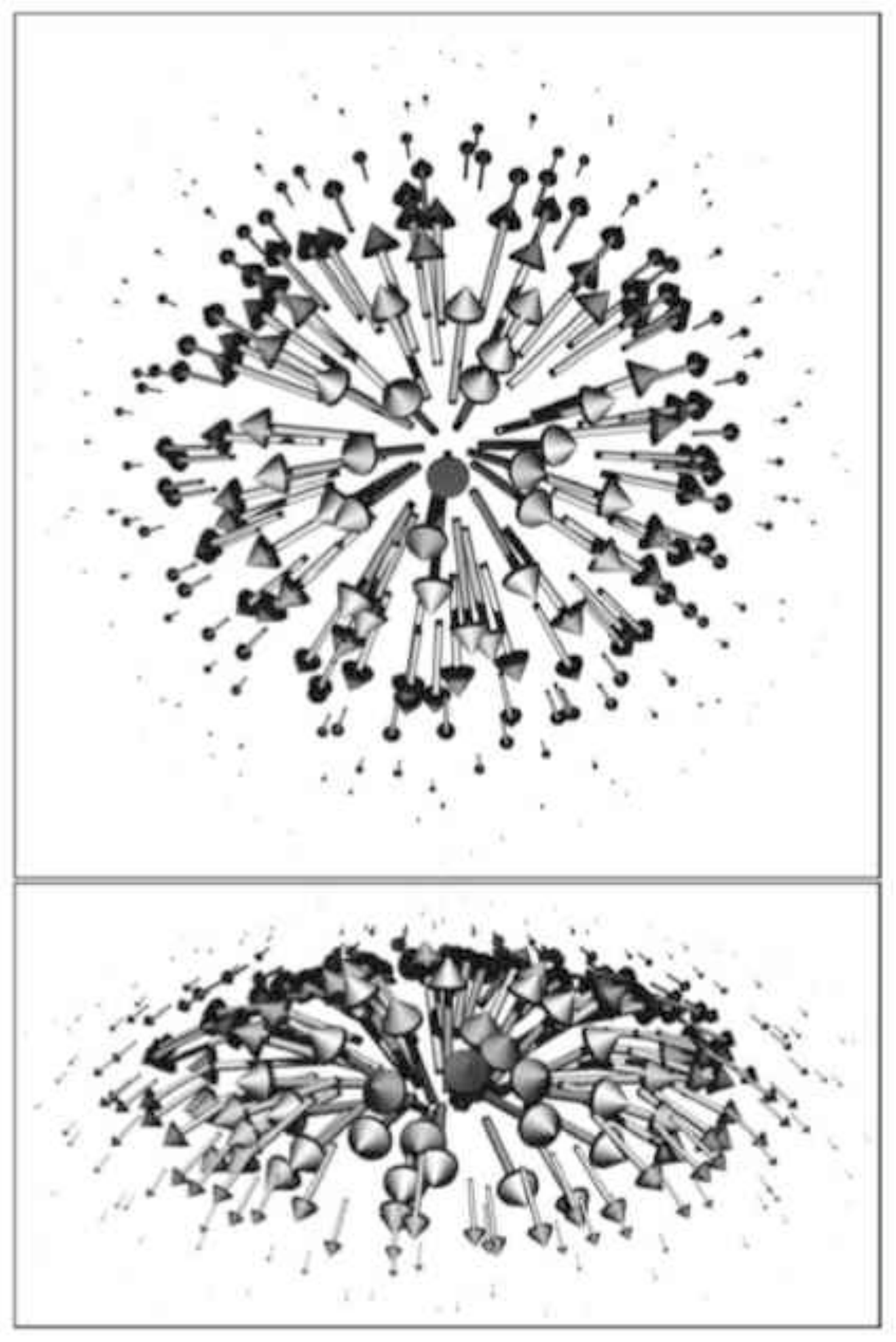}
\caption{Schematic view of a skyrmion (top and perspective).
The spin tilts from ``up'' for one component at the
center, where one component shows a maximum density filling the vortex in the
other component, to ``down'' towards the edge. From \textcite{mueller2004a}.} 
\label{mueller2004_fig1}
\end{figure}
A very graphic illustration of the 
pseudospin behavior in a single coreless vortex state 
is given in Fig.~\ref{mueller2004_fig1}~\cite{mueller2004a}, showing the 
top and perspective view of such a skyrmion in a two-component system. 

\begin{figure}
\includegraphics[width=.5\textwidth]{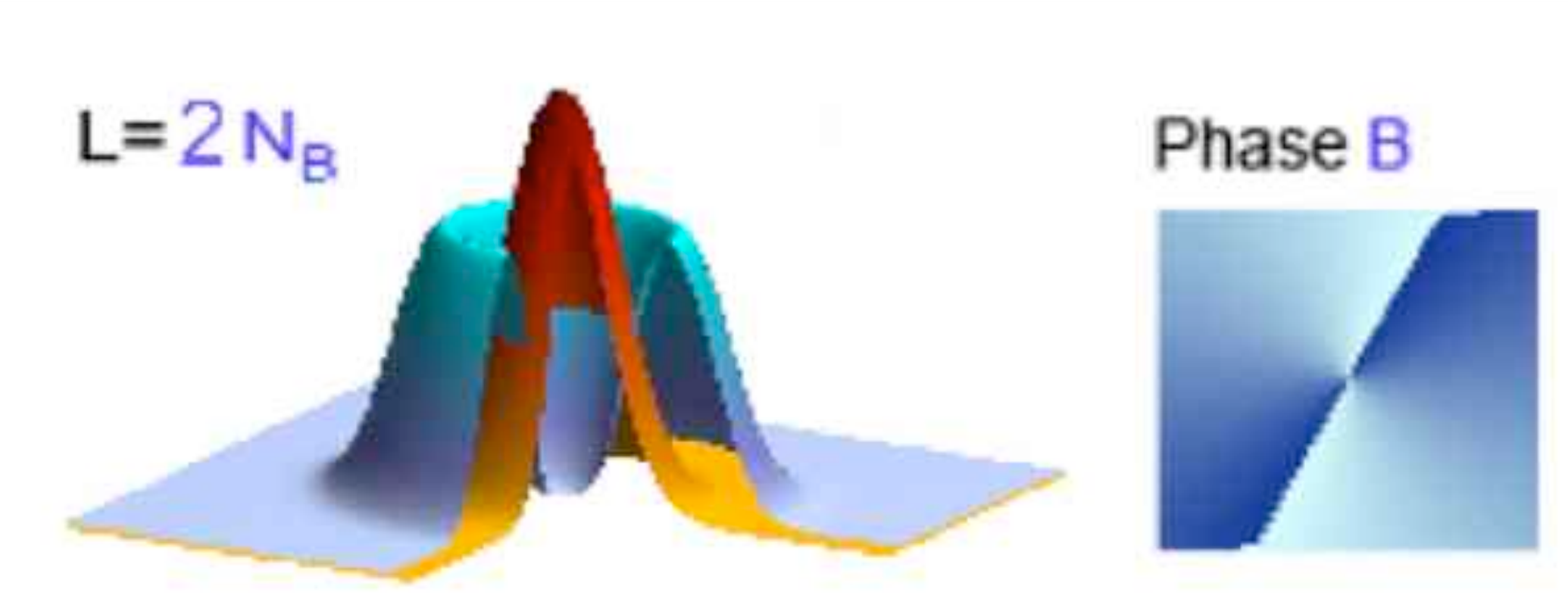}
\caption{({\it Color online})   
Densities {\it (left)} and phases {\it (right)} 
of the Gross-Pitaevskii order parameters 
in a two-component rotating Bose-Einstein condensate with a
coreless vortex with a double phase singularity. For notation see
Fig.~\ref{twocomponentGP_1v} above. After data from~\textcite{bargi2008}.
} 
\label{twocomponentGP_2v}
\end{figure}

As $L$ increases, beyond $L=N_B$, a second vortex enters the larger 
component $B$, merging with the other vortex at $L=2N_B$. 
The smaller component $A$ remains localized at the center, and the system as a
whole has a two-fold phase singularity at the center.
An example is shown in 
Fig.~\ref{twocomponentGP_2v}. The central minimum in the 
density of the larger component $B$ expands with increasing angular 
momentum. It encircles the smaller
one, that is non-rotating and localized at the trap center. 
A phase change of $4\pi$ in a closed path around the center
indicates a vortex that is two-fold quantized.
At $L=3N_B$ a triple phase singularity emerges at the center,
but eventually the scenario breaks down with increasing rotation
frequency.

In single-component quantum liquids, multiply-quantized 
vortices are not favored in parabolic potentials.
However, any external potential that grows more rapidly than quadratically
may give rise to these giant vortex 
structures~\cite{lundh2002,kavoulakis2003} discussed in Sec.~\ref{sec:giant}
before. 
In two-component systems, it was found that
the smaller, non-rotating component at the trap center may effectively act as
an additional potential to the (harmonic) trap confinement, 
rendering the potential effectively
anharmonic close to the trap center for the rotating 
component~\cite{bargi2007}.
With increasing rotation, both components carry a finite fraction of
the total angular momentum, and multiply quantized or ``giant'' vortex states
are no longer energetically favorable. 

\begin{figure}
\includegraphics[width=.5\textwidth]{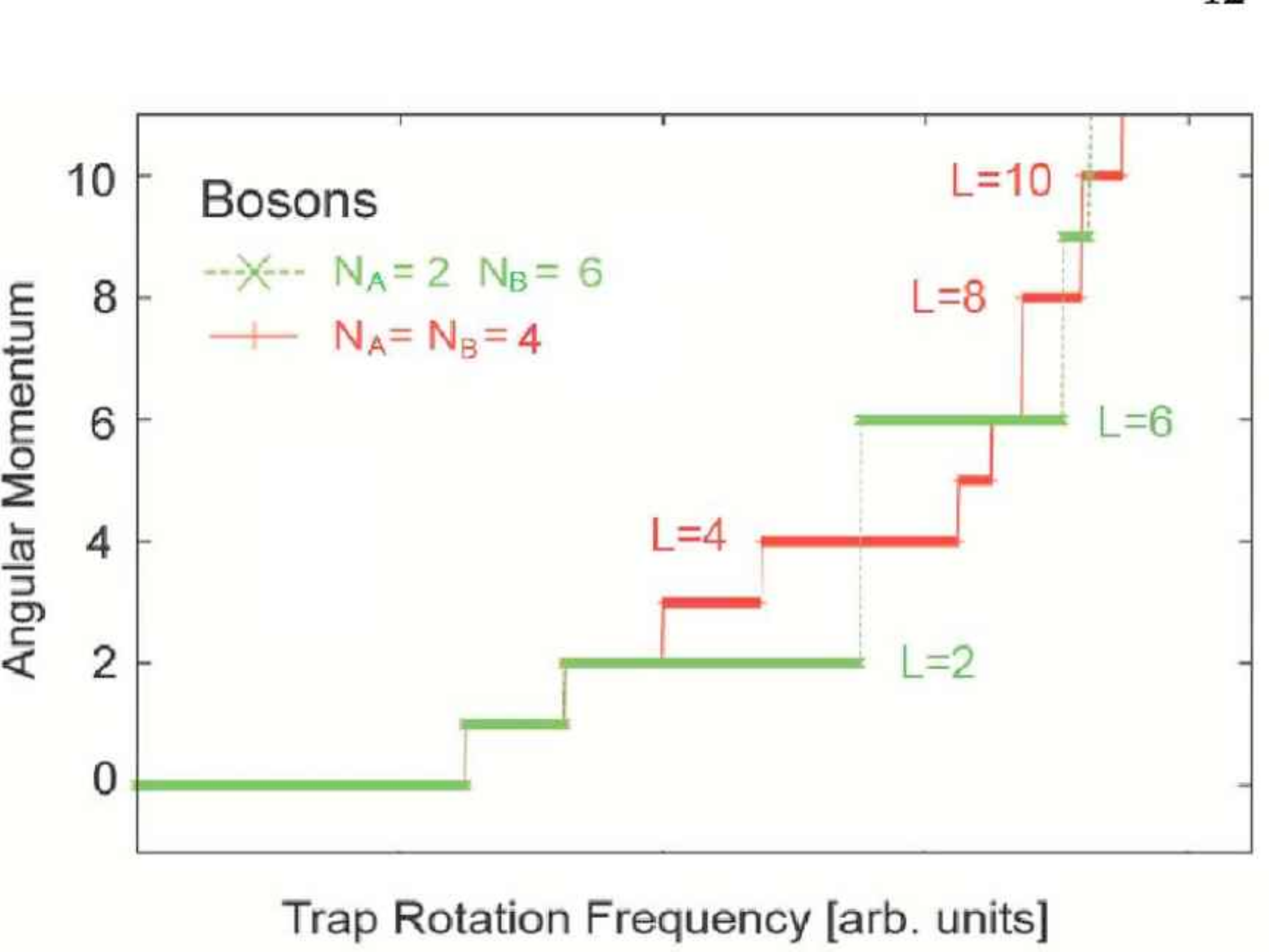}
\caption{({\it Color online}) Angular momentum as a function of the trap rotation frequency (in
arbitrary units) for $N=8$ bosons with equal masses and interactions, in a
harmonic trap, for equal population ($N_A=N_B=4$) (red line) and 
unequal population ($N_A=2$ and $N_B=6$) (green line).
From \textcite{bargi2009}.}
\label{lomega_bosons2comp}
\end{figure}

In exact diagonalization studies of multi-component systems, 
the additional degree of freedom through
the pseudospin increases the dimension of the Hamiltonian matrix 
significantly, which leads to severe restrictions in the particle numbers 
or angular momenta that can be studied. 
Nevertheless, the results obtained for few-particle systems confirm
the existence of Anderson-Toulouse and Mermin-Ho types of coreless 
vortices, as they were obtained within the Gross-Pitaevskii approach. 
For a two-component system with $N_A+N_B =8$ 
bosons with contact interactions in a harmonic trap,  
Fig.~\ref{lomega_bosons2comp} shows the 
total angular momentum $L$ as a function of the rotational frequency 
$\Omega /\omega $. As in the case of scalar Bose gases 
(see Fig.~\ref{fig:bosonyrastci} in Sect.~\ref{sec:lattices} 
above), 
plateaus with increasing $\Omega $ can be associated with 
vortices that successively
enter the bosonic cloud with increasing trap rotation
\cite{butts1999,kavoulakis2000}. 
These plateaus correspond to 
cusp states along the yrast line in the two-component 
system~\textcite{bargi2009}.

\begin{figure}
\includegraphics[width=.45\textwidth]{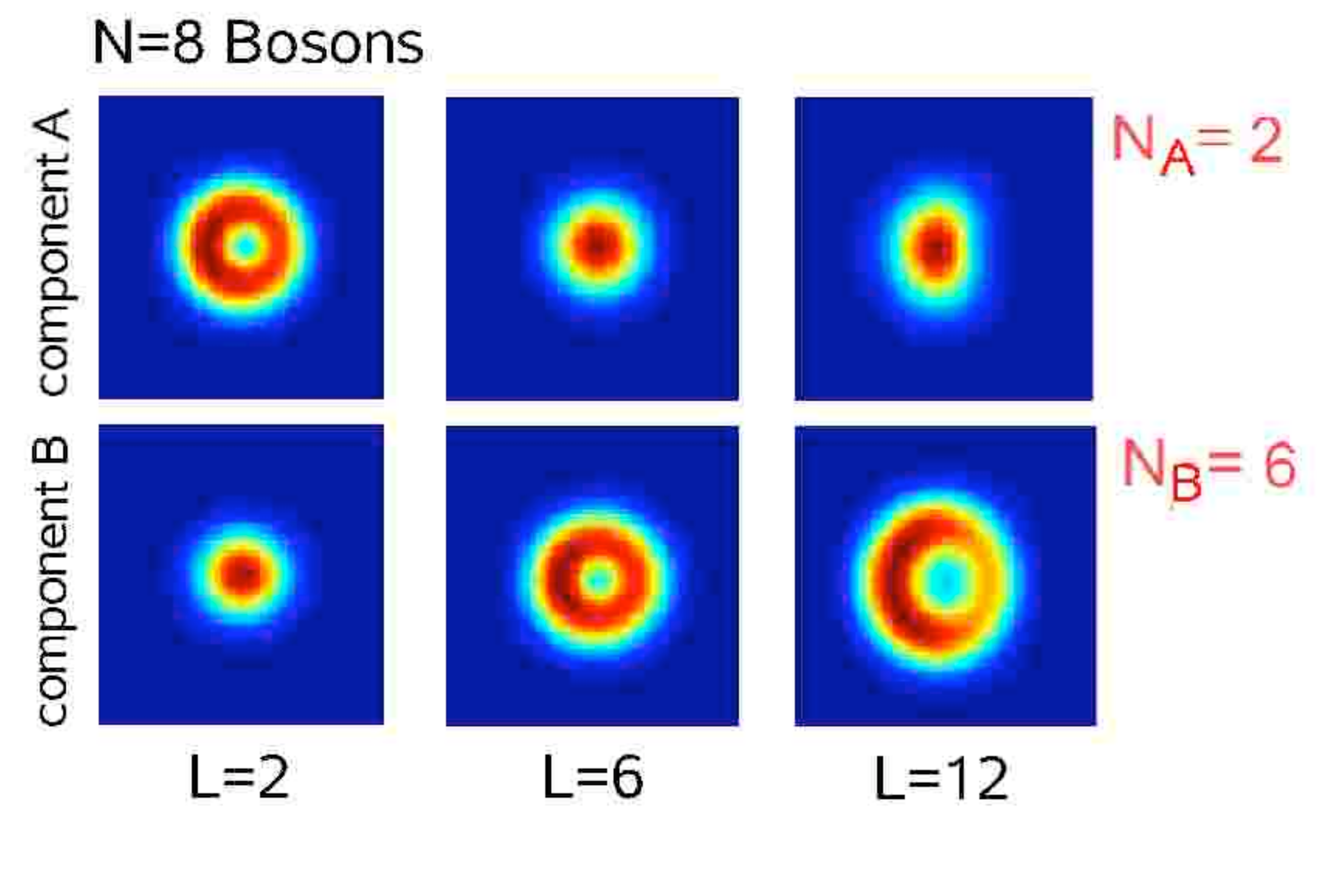}
\caption{({\it Color online}) Density plots of conditional probabilities for a two-component 
few-boson system in a harmonic trap, with two bosons in component $A$ and six
bosons in component $B$. The reference point was chosen in component $B$ 
at an off-center position close to the maximum of the probability density. 
Axes are from $(-4,4)$ in atomic units. The color scale in the density plots
is from blue (zero) to red (maximum). (To increase the visibility, the plot
range of the conditional probablities in the two components was re-scaled to
the same constant in all panels.) 
The charge deficiency of the vortex cores causes deep minima 
to appear in the pair-correlation functions.
After \textcite{bargi2009}.}
\label{bosonspairNA2NB6}
\end{figure}

The exact quantum states retain the symmetry of the 
Hamiltonian, and thus one must turn to conditional probability densities
(pair-correlation functions) and reduced wave functions 
to map out the internal structure of the wave
function, as discussed in Sec.~\ref{sec:wave-function}. 
For unequal populations of the two species, here 
$N_A=2$ and $N_B=6$, at those angular momenta where the pronounced 
plateaus occur in the $L$-versus-$\Omega $-plot in Fig.~\ref{lomega_bosons2comp}, 
the pair-correlations are shown in 
Fig.~\ref{bosonspairNA2NB6}.
At $L=2$ a vortex is seen as a hole at the center in the smaller
component, encircling the larger component that forms a Gaussian at the 
trap center. At angular momentum $L=6$ a single vortex is created in the
larger component, as seen in the middle panel. Twice this angular momentum
creates a two-fold quantized vortex structure
in the larger component.
The existence of coreless vortices, as predicted by
the Gross-Pitaevskii equation in the mean-field limit
(Sec.~\ref{sec:bosetwocompGP}), is accurately reproduced by
the exact solutions in the few-body regime. 

\subsubsection{Condensates with symmetric components}

\begin{figure}
\includegraphics[width=.4\textwidth]{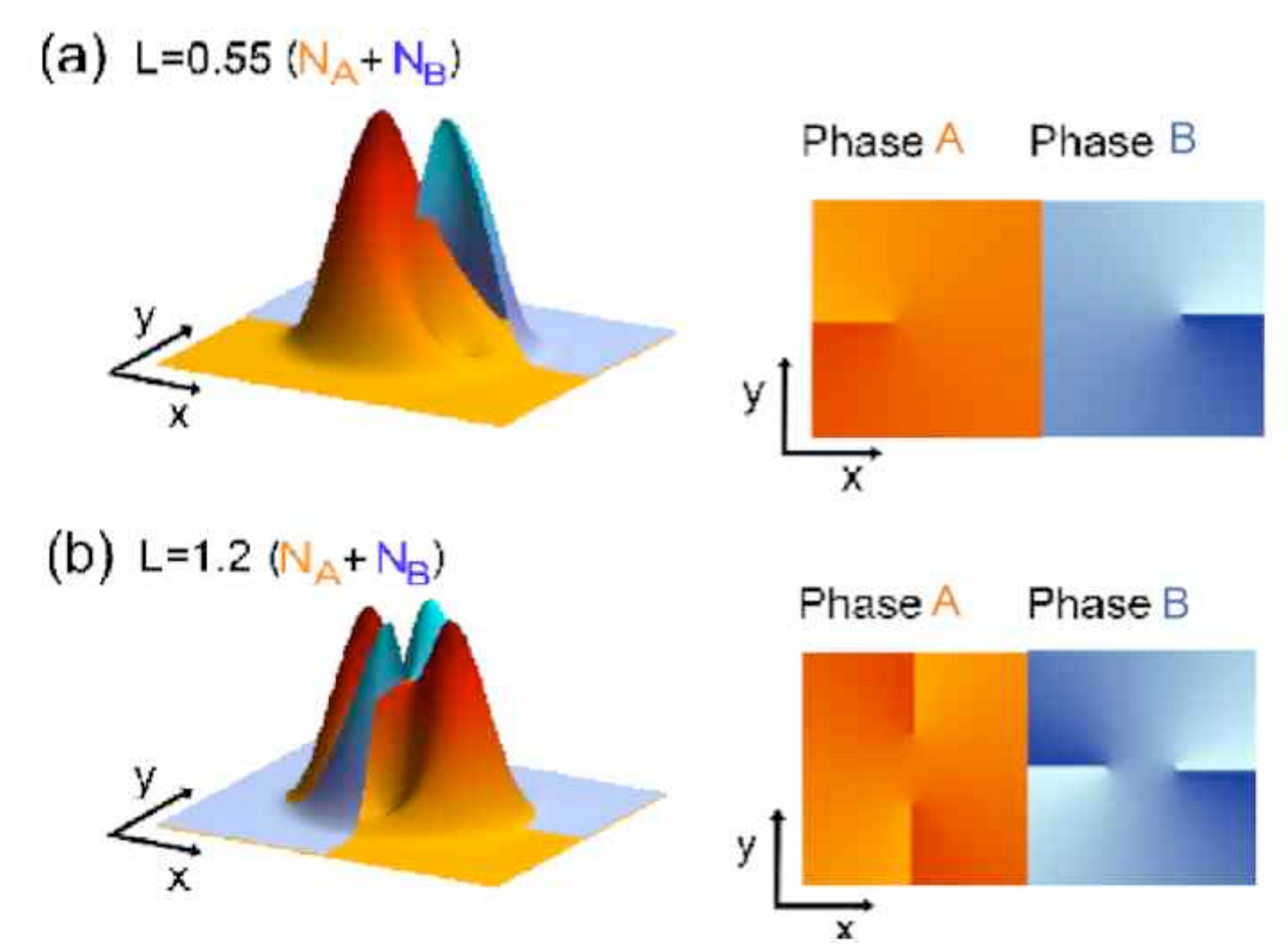}
\caption{({\it Color online}) Mean-field order parameters {\it (left)} and phases {\it (right)} 
of a symmetric condensate with $N_A=N_B$, 
at $L=1.2(N_A+N_B)$, for equal coupling strengths 
$g_{AA}=g_{AB}=g_{BB}=50$, showing a) one and b) 
two interlaced coreless vortices in the two components.
Component $B$ is only shown in a half-plane to make the vortex in the
component $A$ visible. After data from~\textcite{bargi2008}.}
\label{coreless_interlaced}
\end{figure}

When the cloud has equal populations of the two components, 
{\it i.e.}, $N_A=N_B$, a different scenario emerges: 
a vortex enters each of the components from ``opposite'' sides,  
reaching a minimum distance 
of one oscillator length from the center of the trap
when $L=N_A=N_B$~\cite{christensson2008a}. 
An example is given by the Gross-Pitaevskii solution
shown in the upper panel of Fig.~\ref{coreless_interlaced}.
Similarly to the one-component case, increasing rotation adds more vortices
to the cloud. For two equal components, the vortices become interlaced, 
with density maxima in one component located at the vortices in the other, 
minimizing the interaction energy between the different components
(lower panel of Fig.~\ref{coreless_interlaced}).
 In the limit of large $N$ and $L$ 
a lattice of coreless vortices is formed~\cite{kasamatsu2005b}. 

\begin{figure}
\includegraphics[width=.5\textwidth]{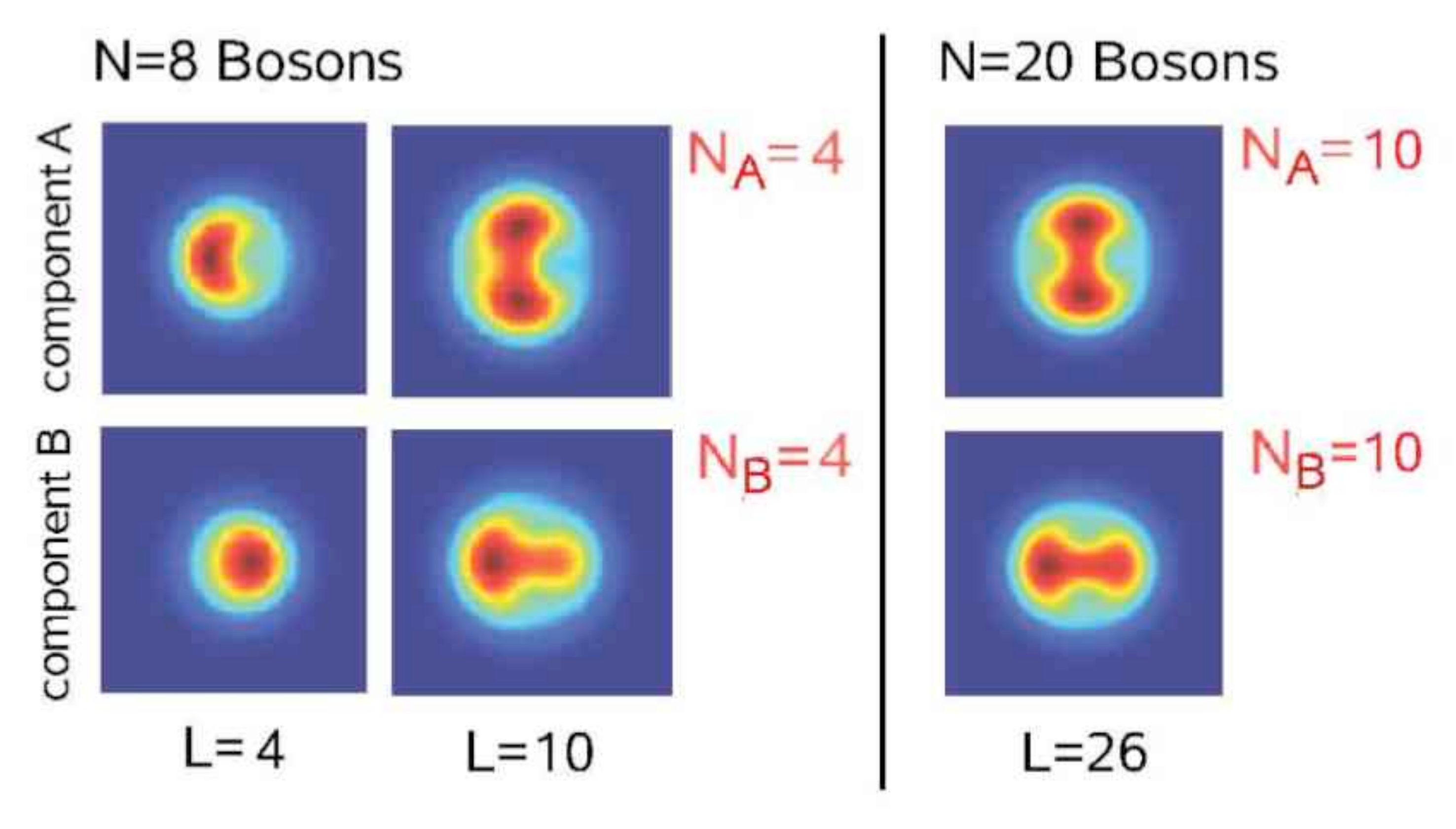}
\caption{({\it Color online}) As in Fig.~\ref{bosonspairNA2NB6}, but for equal components,
$N_A=N_B=4$ {\it (left panel)}. 
The density minima in one component coincide 
with the density maxima in the other component. 
This suggests that these states are finite-size precursors of 
interlaced vortex lattices that occur in the limit of large $N$. 
The picture becomes much more clear for 
larger particle numbers, as shown in the right panel 
for $N=20$ and $L=26$. After \textcite{bargi2009}.}   
\label{bosonspairNA4NB4N20}
\end{figure}

These Gross-Pitaevskii results are in good correspondence with
exact diagonalization results of few-particle systems.
The left panel of Figure~\ref{bosonspairNA4NB4N20} shows
conditional probability densities of a symmetric
configuration $N_A=N_B=4$.
When $L$ equals $N_A=N_B=4$, the clouds
separate, with a vortex hole emerging at the maximum density location in the
other component. These solutions correspond to a
Mermin-Ho vortex (or a meron pair, where each meron accounts for
half of the spin texture of the coreless vortex)
as obtained in Gross-Pitaevskii theory~\cite{kasamatsu2005b}.
For higher angular momenta, as here for $L=10$, the correlation functions
indicate interlaced vortices as in Fig.~\ref{coreless_interlaced} 
above, with density maxima in one component localizing at the 
minima (vortex cores) in the other component. 
The interlaced pattern of density minima and maxima becomes more
apparent with higher particle number as shown in the right panel of 
Fig.~\ref{bosonspairNA4NB4N20} for $N=20$ bosons, where $N_A=N_B = 10$, 
at angular momentum $L =26$. 

\begin{figure}
\includegraphics[width=.5\textwidth]{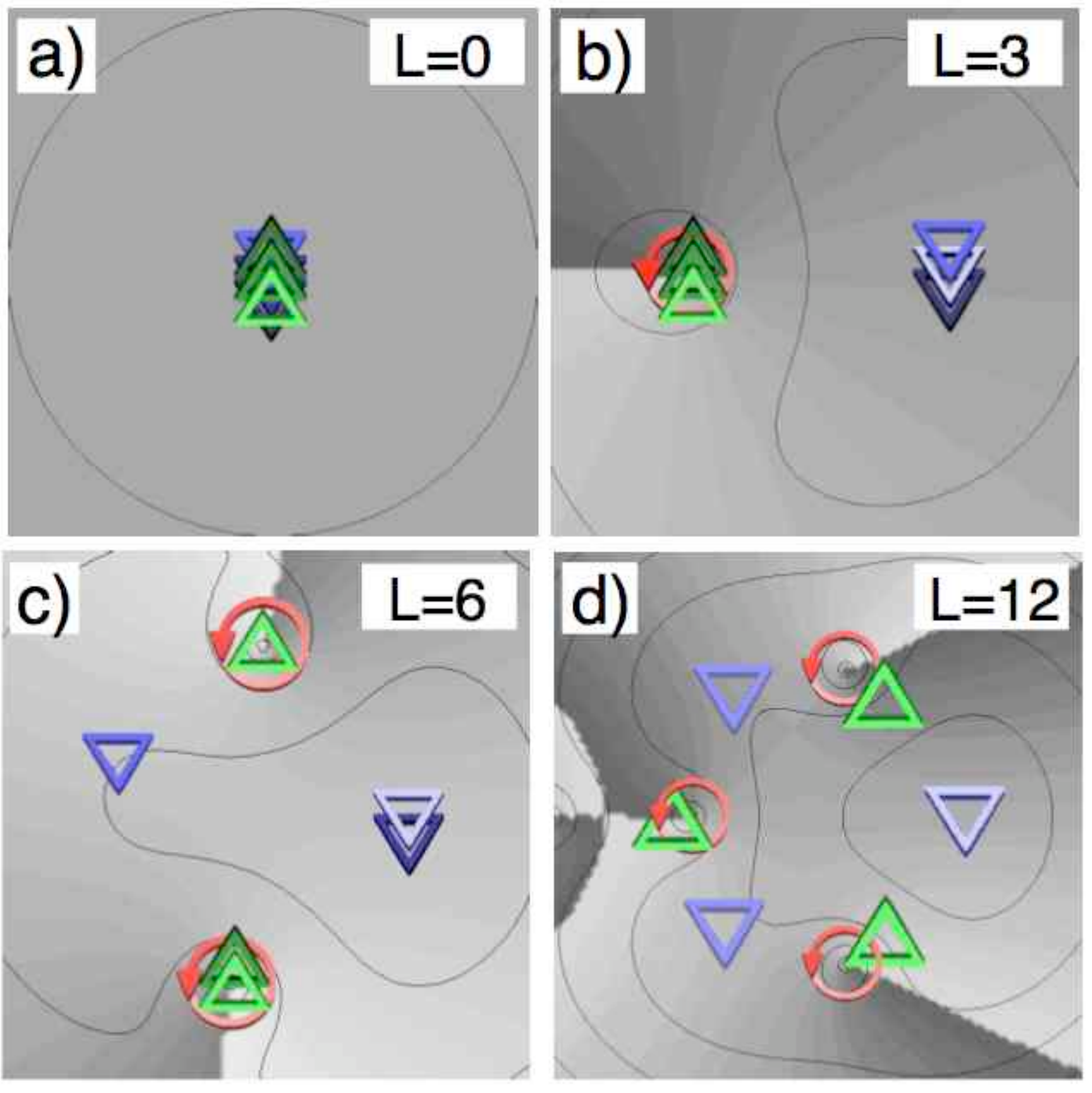}
\caption{({\it Color online}) Reduced wave functions in a symmetric system of $N_A=N_B=3$ bosons,
showing the correlations
between phase singularities (marked with circles) 
with the most probable positions
of the particles of opposite species (marked with triangles). 
This is an indication
for the formation of coreless vortices one-by-one in the
system as the angular momentum increases. The figure shows
a) the non-rotating state, b) a state with one coreless
vortex per particle species, c) two coreless vortices, and d)
three coreless vortices.
Note that for identical components $A$ and $B$, 
the reduced wave functions
for the two species are necessarily symmetric, and only one component is shown
here. After~\textcite{saarikoski2009}.
}    
\label{fig:boson2-cond-wave}
\end{figure}

The conditional probability densities average out the effect of phase 
singularities as signatures of vortices. However, the nodal 
structure of the many-body state may straightforwardly be 
probed by 
reduced wave functions~\cite{saarikoski2009} (see Sec.~\ref{sec:cond-wave}), 
as shown in Fig.~\ref{fig:boson2-cond-wave}
for a system with $N_A=N_B=3$ bosons.
Coreless vortices form one-by-one as the angular momentum
increases: in the example shown here 
for $L=6$ and $L=12$, the phase singularities in one component 
occur at the most probable positions of the particles of the other 
component, indicating formation of two and three coreless vortices, 
respectively, in each particle component (Fig.~\ref{fig:boson2-cond-wave}).

\subsubsection{Vortex lattices and vortex sheets}

Vortex lattices in two-component bosonic condensates may show a variety of 
different structures, depending on the strength and sign of the interspecies
interaction~\cite{mueller2002}.
In the antiferromagnetic case ($c_2> 0$), for weak interactions 
square lattices form, whereas for strong interactions the
vortices are arranged into triangular Abrikosov lattices.
In the former case the square
lattice is energetically favoured because the antiferromagnetic
interaction between adjacent vortex holes makes
a triangular lattice frustrated~\cite{kasamatsu2003}.
At $c_2=0$ the system has metastable states such as a stripe phase.
In the regime of ferromagnetic interspecies coupling ($c_2<0$),
spin domains spontaneously form. These vortex sheets form 
``serpentine-like'' structures that are nested into each 
other~\cite{kasamatsu2003,kasamatsu2009}.
A number of metastable lattice structures that were energetically
almost degenerate have also been found in 
an antiferromagnetic spin-1 BEC~\cite{kita2002}.

\subsection{Two-component fermion droplets}

Recent electronic structure studies of quantum dots with spin degrees of
freedom predicted the formation of coreless vortices
in fermion droplets analogously to the bosonic
case~\cite{petkovic2007,dai2007,koskinen2007,saarikoski2009}.
This comes as no suprise since 
analogies in the structure between fermion and boson states
(Sec.~\ref{sec:fermionboson})
are not limited to single-component systems, but an 
approximate mapping between two-component fermion and boson
states can be constructed as well.
In the following we discuss coreless vortices in fermion droplets
and some of the consequences of the fermion-boson analogy
with few-electron droplets as a particular example.

\subsubsection{Coreless vortices with electrons}

\begin{figure}
\includegraphics[width=.5\textwidth]{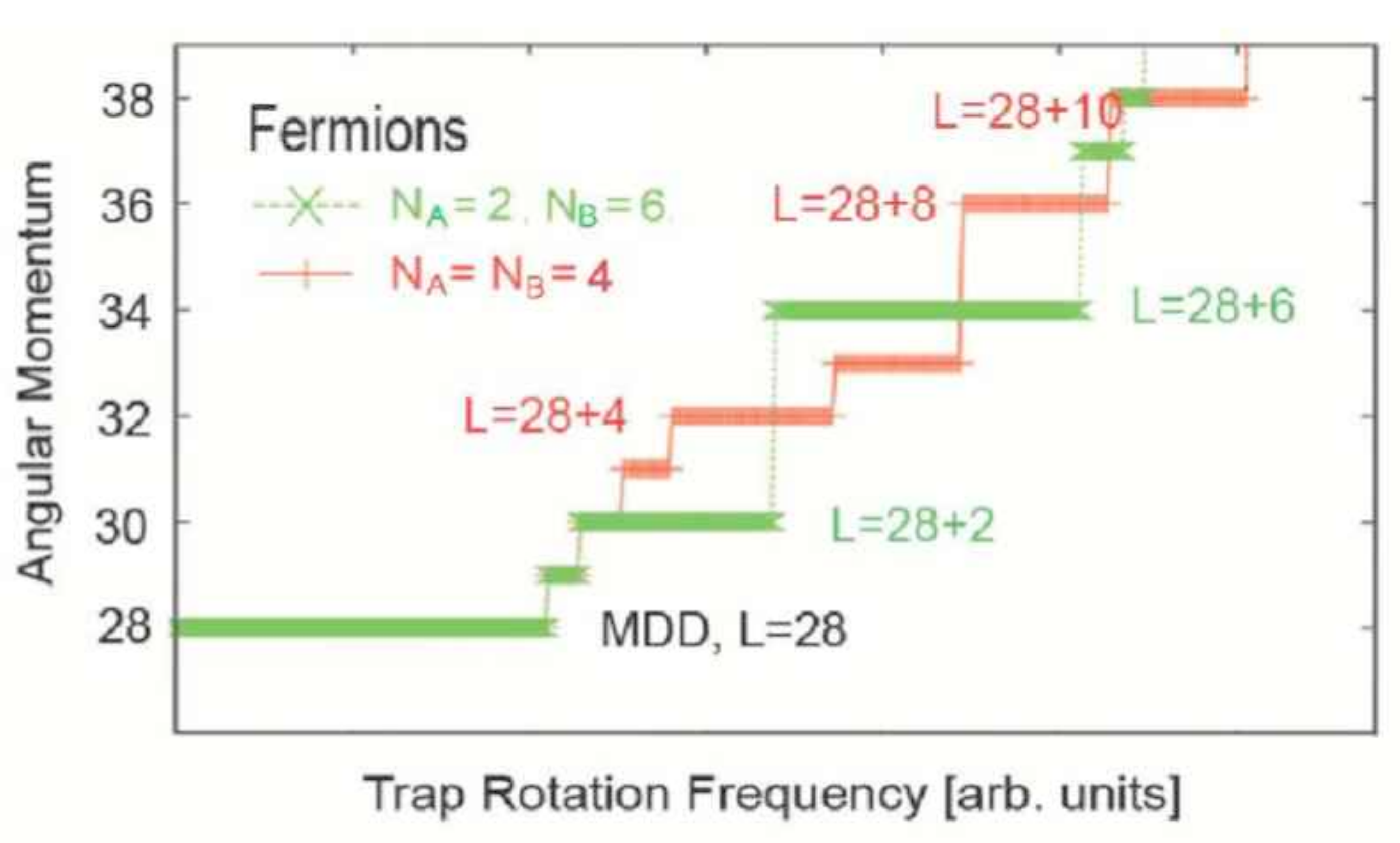}
\caption{Angular momentum as a function of the trap rotation frequency (in
arbitrary units) for $N=8$ fermions with symmetric (red) and asymmetric
(green) component occupations,  in Fig.~\ref{lomega_bosons2comp}
above. From \textcite{bargi2009}.}
\label{lomega_fermions2comp}
\end{figure}

The angular momentum for a system with eight fermions 
with both balanced ($N_A=N_B=4$) and unbalanced 
($N_A=2$, $N_B=6$) component sizes, is shown as a function 
of the trap rotation frequency in Figure~\ref{lomega_fermions2comp}. 
The staircase shape is strikingly similar
to the bosonic counterpart (Fig.~\ref{lomega_bosons2comp}) with
$L_{\rm boson}=L_{\rm fermion}-L_{\rm MDD}=L_{\rm fermion}-28$.
In the fermion case with asymmetric components $N_A=2$ and $N_B=6$,
the first pronounced plateaus appear at $L=L_{\rm MDD}+N_A=28+2$
and $L=L_{\rm MDD}+N_B=28+6$, which correspond to a coreless vortex in the
$A$ and $B$ component, respectively.
In the case of symmetric component occupations $N_A=N_B=4$
the first major plateau moves to $L=L_{\rm MDD}+4$ and
the coreless vortex configuration is analogous to
a meron pair~\cite{petkovic2007} in bosonic systems.
The lengths of these plateaus indicate that
coreless vortex states are very stable also in fermion systems.

\begin{figure}
\includegraphics[width=.45\textwidth]{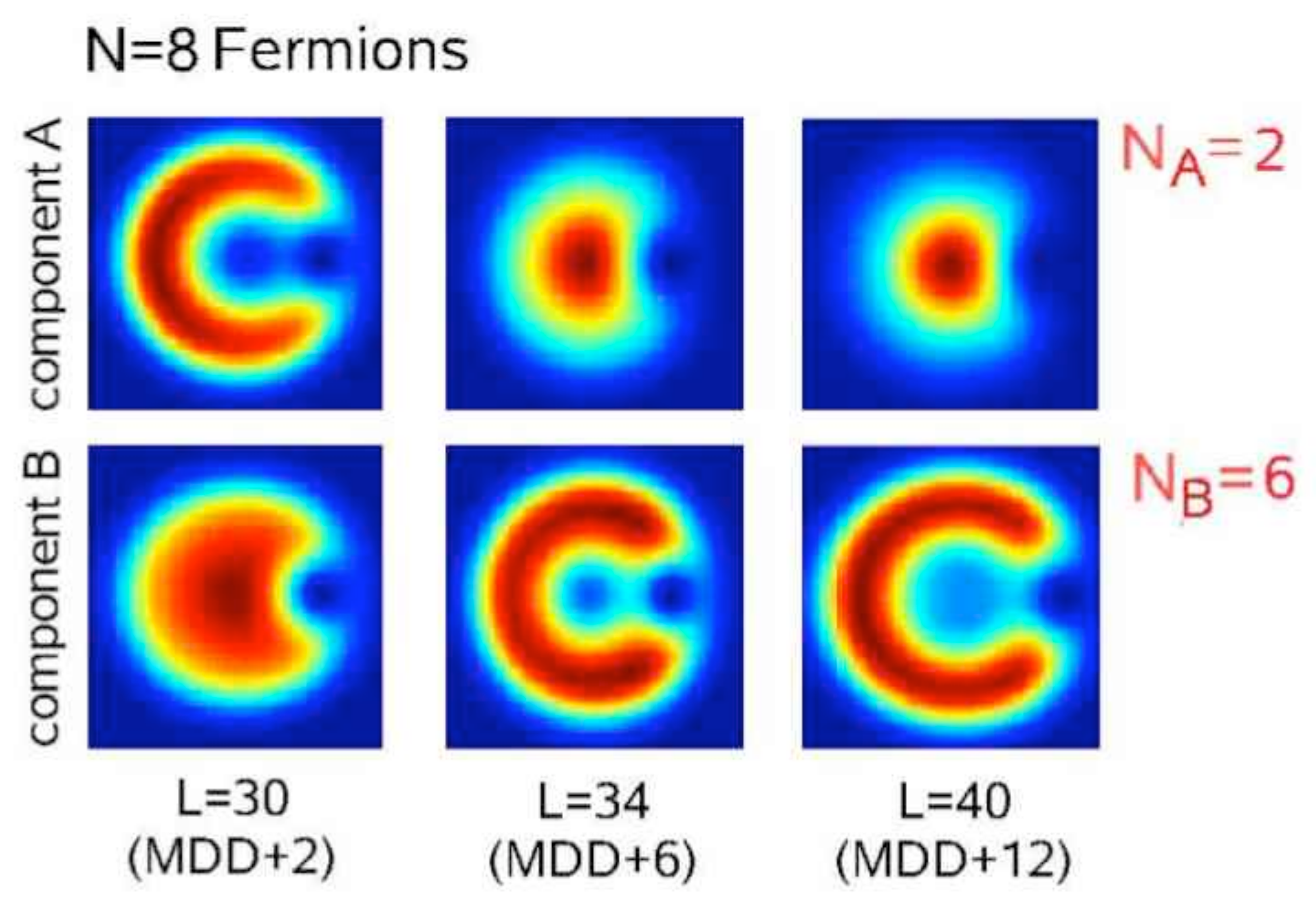}
\caption{({\it Color online}) Pair-correlated densities for fermions, 
as in Fig.~\ref{bosonspairNA2NB6}, but here for {\it fermions} with 
Coulomb interactions ($N_A=2$ and $N_B=6$).
Shown are angular momenta corresponding to the pronounced plateaus in 
Fig.~\ref{lomega_fermions2comp}. 
Compared to the bosonic case the densities show analogous structures 
except for an additional exchange hole at the reference point in component 
B which is reflected also in the component A due to Coulomb repulsion.
From~\textcite{bargi2009}.}
\label{fermionspairNA2NB6}
\end{figure}

\begin{figure}
\includegraphics[width=.45\textwidth]{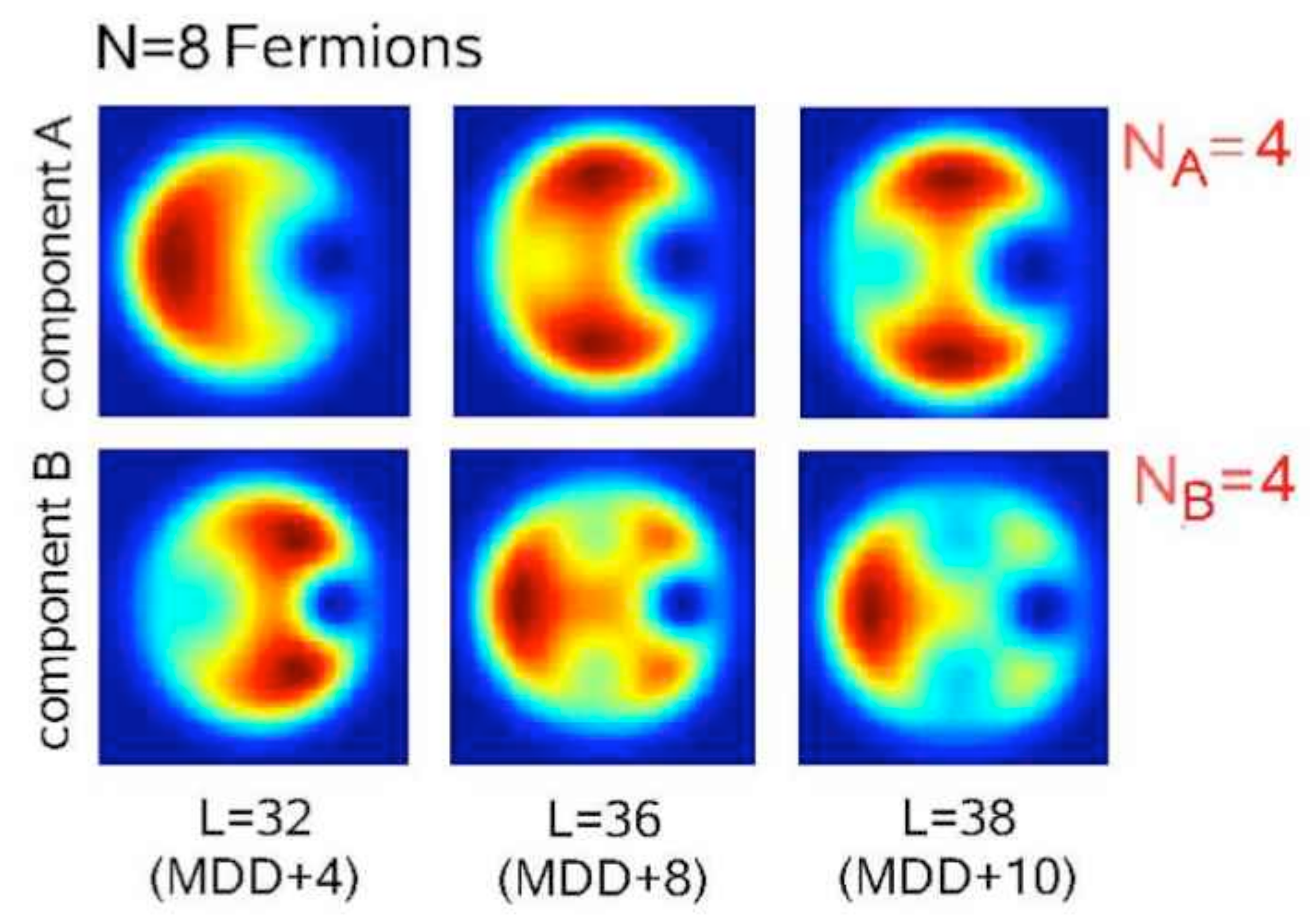}
\caption{({\it Color online}) Pair-correlated densities for fermions, as above, but for $S_z=0$, 
{\it i.e.}, equal components ($N_A=4$ and $N_B=4$). From~\textcite{bargi2009}.
}
\label{fermionspairNA4NB4}
\end{figure}

The fermionic ``quantum-dot'' analog to the unbalanced 
few-boson system (with $N_A=2$ and $N_B=6$)  discussed above would be a system 
with $N=8$ electrons and fixed $S_z =2$, which demands 
two spins antiparallel to the external magnetic field (component A) 
and six spins parallel to the field (component B).
Both components form compact maximum density droplets independently 
at $L_{\rm MDD} = 28$, that corresponds
to the $L=0$ non-rotating condensate in the bosonic case. 
When the angular momentum exceeds that of the MDD by two units of $\hbar $,
a hole forms at the center of the smaller component which is associated with a 
vortex state, while the larger one
remains a MDD. This can be clearly seen from the pair-correlated density 
shown in Fig.~\ref{fermionspairNA2NB6}. Note that in the case of fermions, 
there is a clearly visible exchange-correlation 
hole around the reference point in the pair-correlation. This is
due to the Pauli principle which is naturally absent in the
bosonic case. Due to the strong repulsion between the fermions, this 
hole is mirrored in the other component.
For larger angular momentum, multiply quantized vortices are found in the
larger component, in direct analogy to the bosonic case discussed above. 
This happens in our example for $L_{\rm fermion}=L_{\rm MDD}+6$ 
and $L_{\rm fermion} = L_{\rm MDD} + 12$ (shown in Fig.~\ref{fermionspairNA2NB6}). 
The case of equal components corresponds to fixed $S_z=0$. 
For $L=N_A=N_B$, just as in the bosonic case, a vortex appears at some distance
from the trap center, with a density maximum on the other side, and 
vice-versa. These textures are again similar to the ``meron'' pairs in the bosonic
two-component system discussed above. 
For higher angular momenta, the interlaced vortex lattice is seen for fermions 
at $L=L_{\rm MDD} +8$ and $L=L_{\rm MDD}+10$ (see Fig.~\ref{fermionspairNA4NB4}). 
(Note again the occurrence of the exchange hole, that should not be confused
with the holes of off-electron vortices.) 

\begin{figure}
\includegraphics[width=.5\textwidth]{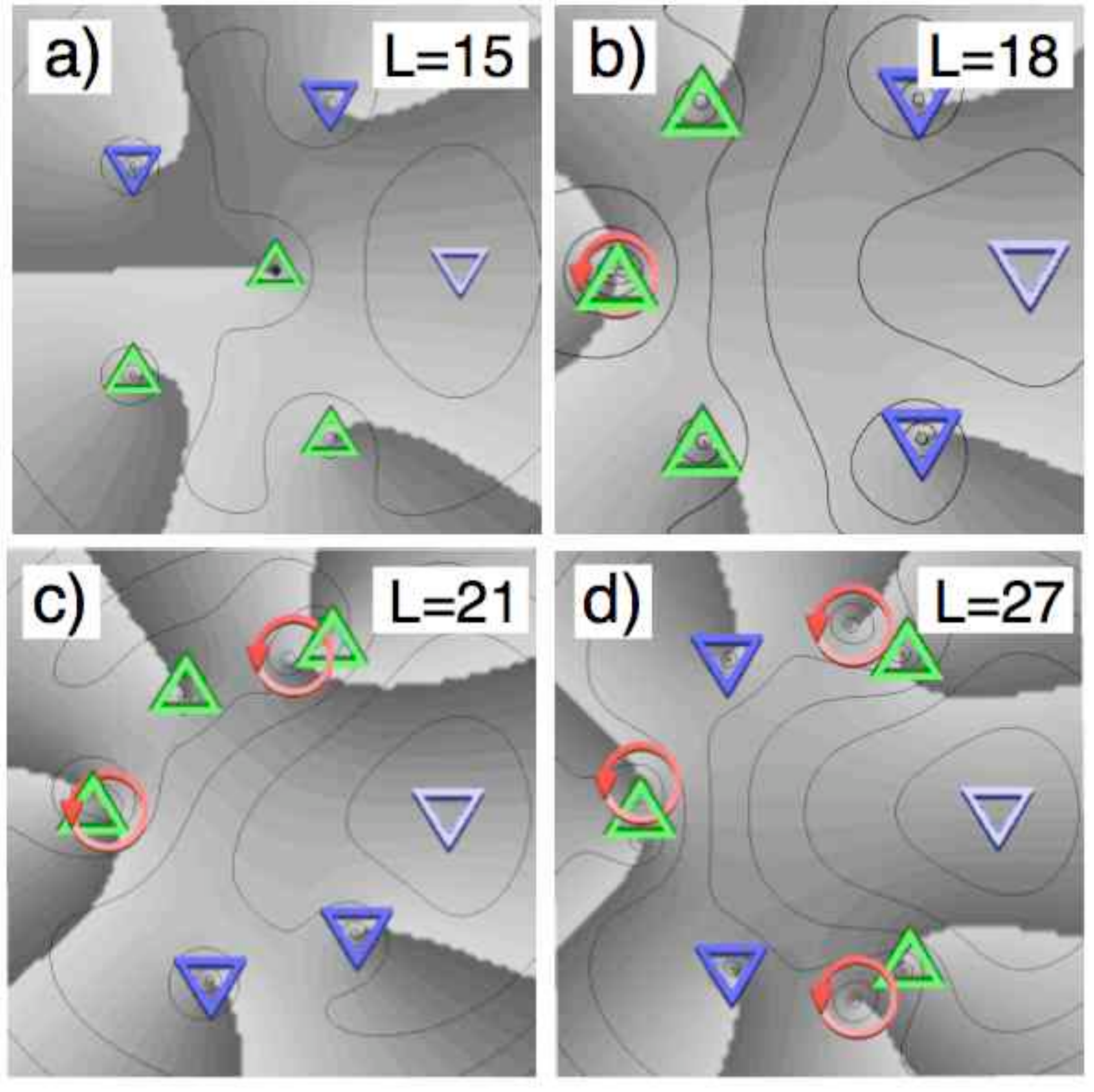}
\caption{({\it Color online}) 
Reduced wave functions in a two-component system.  
In a two-component fermion droplet with symmetric occupations
$N_A=N_B=3$ the reduced wave function in the lowest Landau level
reveals coreless vortices as correlations between phase singularities
(circles) with the most probable positions of the particles of opposite
spin (triangles). The figure shows a) the MDD state with total
spin $S=3$ and $S_z=0$ b) a state with one coreless
vortex per particle species c) two coreless vortices, and d)
three coreless vortices. This sequence of states is analogous to that
of a bosonic system in Fig.~\ref{fig:boson2-cond-wave}.
Note that vortices of the MDD state are not shown 
in order to ease the comparison to the bosonic case.
From~\textcite{saarikoski2009}.
}
\label{fig:fermion2-cond-wave}
\end{figure}

Figure~\ref{fig:fermion2-cond-wave} shows the reduced wave functions,
see Eq.~(\ref{condwave}),  for a two-component fermion droplet 
with Coulomb interactions
and $N=6$ particles, with symmetric component occupations $N_A=N_B=3$.
The sequence of states in this figure shows the formation of coreless
vortices one-by-one inside the fermion droplet, in analogy to the 
bosonic case
in Fig.~\ref{fig:boson2-cond-wave} above, 
with the angular momenta for boson and fermion systems shifted by 
$L_{\rm fermion}=L_{\rm boson}+L_{\rm MDD}$.
In comparison to the bosonic case, for fermions the 
Pauli vortices keep the particles further apart.

\subsubsection{Quantum dots with weak Zeeman coupling}

\begin{figure}
\includegraphics[width=.4\textwidth]{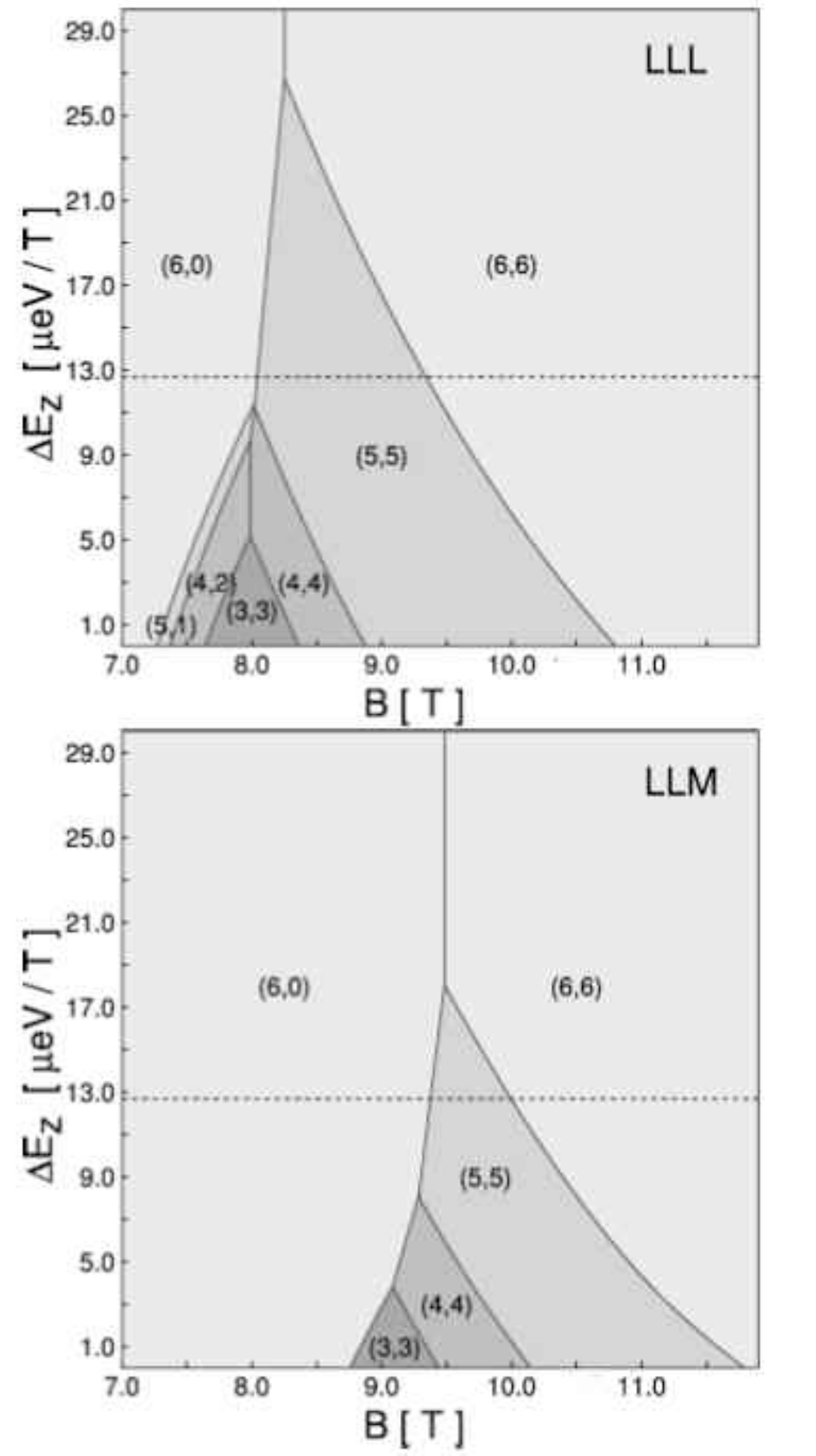}
\caption{Partially polarized states beyond the maximum density droplet
  reconstruction, obtained from a variational Monte Carlo study 
by~\textcite{siljamaki2002}. 
The diagrams show the different states of a 6-electron quantum dot
as a function of the magnetic field and the strength of the Zeeman coupling
per spin in the lowest Landau level approximation ({\it upper panel}, LLL)
and including Landau level mixing ({\it lower panel}, LLM).
The states are labeled as
($N_\uparrow,\Delta L$) where $N_\uparrow$ is the number of electrons
with spins parallel to the magnetic field and
$\Delta L=L-L_{\rm MDD}$ is the additional angular momentum with respect to the
MDD. The Zeeman coupling strength for GaAs is marked by dashed lines.
The confinement strength is $\hbar \omega = 5$~meV and the material parameters are for GaAs,
$m^*/m_e=0.067$ and $\epsilon_r=12.4$.
}
\label{fig:sami}
\end{figure}

The formation of coreless vortices, as discussed above, can be 
observed also in quantum dots where Zeeman coupling is weak.
Then, the first reconstruction of the MDD may not be directly into
the completely polarized states with one additional vortex, but into an
excitation which is reminiscent of the vortex state,  with one
spin flipped anti-parallel to the magnetic field.
This transition would be followed by a second one, 
involving a spin flip into the completely polarized state~\cite{oaknin1996}. 
\textcite{siljamaki2002} studied the effect of Landau level mixing
in the MDD reconstruction, using the variational quantum Monte Carlo method. 
They found significant
changes in the ground states for systems consisting of up to
7 electrons. Figure~\ref{fig:sami} shows 
the different states of a 6-electron quantum dot in the vicinity of the MDD.
The partially polarized state after the MDD has a leading 
determinant of the form $|01111100\ldots\rangle$ for the majority
spin component and $|100\ldots\rangle$ for the minority spin component:
the vortex hole at the center of the dot in the majority spin
component is  filled by a particle with opposite spin polarization.
Consequently, the state shows formation of a coreless vortex
and is completely analogous to the case of asymmetric particle
populations in a two-component bosonic systems, as discussed 
in Sec.~\ref{sec:asymmetry} above.
The minority spin component has a MDD-like structure,
which corresponds to the non-rotating component in the bosonic case,
and the majority spin component shows a single vortex
core localized at the center. 

\subsubsection{Non-polarized quantum Hall states}

\begin{figure}
\includegraphics[width=0.33\textwidth]{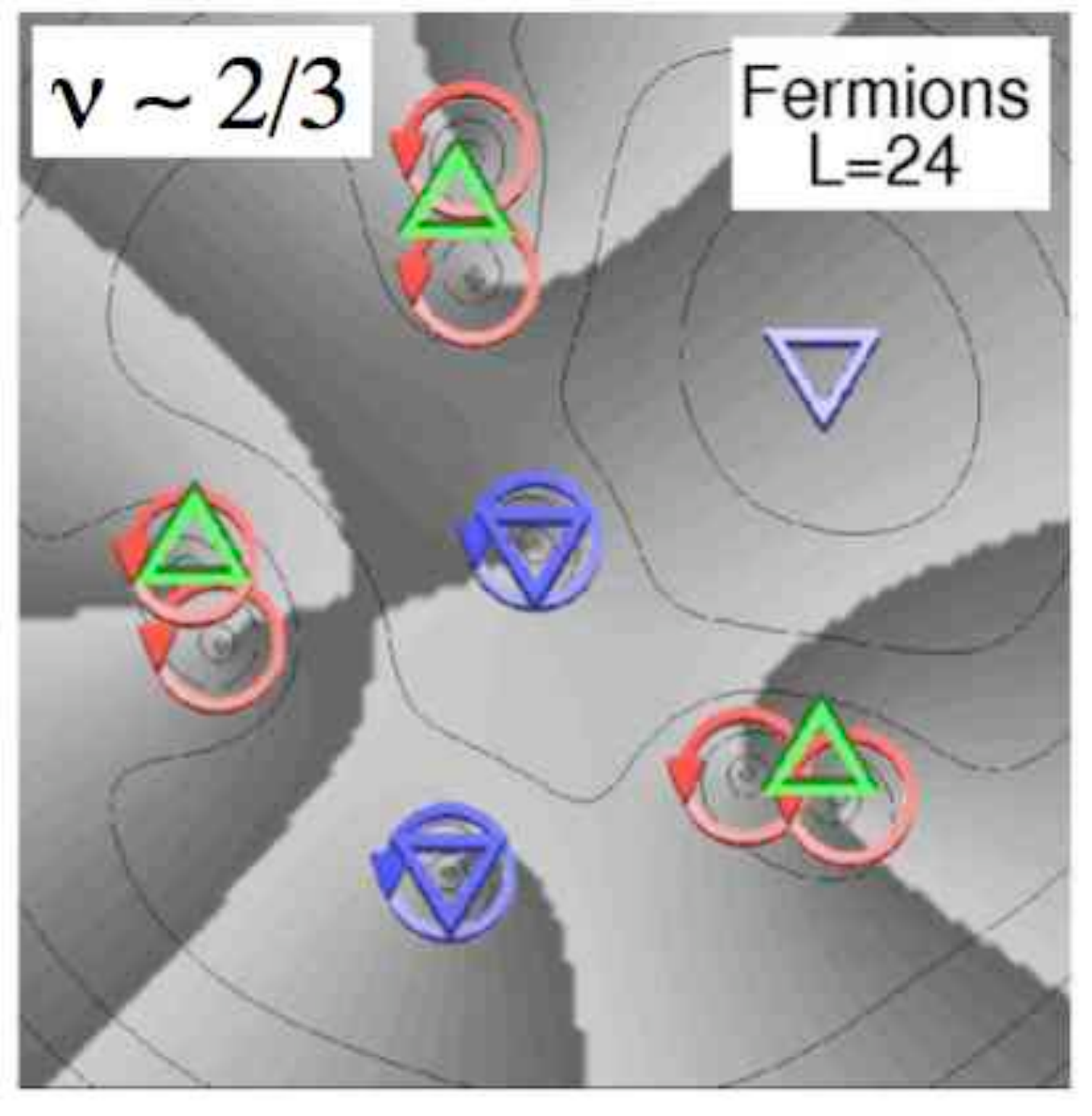}
\caption{({\it Color online}) 
Reduced wave function of the $L=24$ fermion state with
symmetric occupations $N_A=N_B=3$. The 
nodal structure closely corresponds to that of the $q=2, p=1$
Halperin-wave function
with one phase singularity in the component of the probing particle
and two phase singularities in the opposite component.
The approximate Landau level filling for the above finite size system
is  $\nu \approx 2/3$, just as for the 
Halperin state proposed
to describe the $\nu=2/3$ quantum Hall plateau.
The symbols in the figure were explained in Fig.~\ref{sepitys} above. 
From~\textcite{saarikoski2009}.
}
\label{halperinwave}
\end{figure}

In the regime of rapid rotation vortices are expected to attach to particles
also in two-component quantum droplets.
One of the studied model wave functions for two-component states
was introduced to explain the quantum Hall plateau at
$\nu=2/3$~\cite{halperin1983}
\begin{equation}
\psi =
\Pi_{i<j}^{N/2} (z_i-z_j)^q
\Pi_{k<l}^{N/2} (\tilde z_k-\tilde z_l)^q
\Pi_{m,n}^{N/2} (z_m-\tilde z_n)^p,
\label{halperinmod}
\end{equation}
where $q$ is an odd integer (due to fermion antisymmetry),
$p$ is a positive integer and the Gaussians have been omitted.
The last product in Eq. (\ref{halperinmod})
attaches $p$ vortices to each electron with opposite spin 
and these can be 
interpreted as coreless vortices.  The corresponding nodal
structure can also be found in spin-compensated few-electron systems
near the $\nu=2/3$ filling.
Figure~\ref{halperinwave} shows the reduced wave function
of the $N_A=N_B=3, L=24$ electron state where
one (Pauli) vortex is attached to each particle of the same
spin and two (coreless) vortices are attached to particles of the
opposite spin, in good agreement with the Halperin model
with $q=1$ and $p=2$~\cite{saarikoski2009}.
However, despite the correspondence in the nodal structures
the overlap of this state with the Halperin wave function has been found to
be small for large particle numbers, due to a mixing of spin
states in the Halperin model~\cite{koskinen2007}.

\subsection{Bose gases with higher spins} 

\begin{figure}
\includegraphics[width=.5\textwidth]{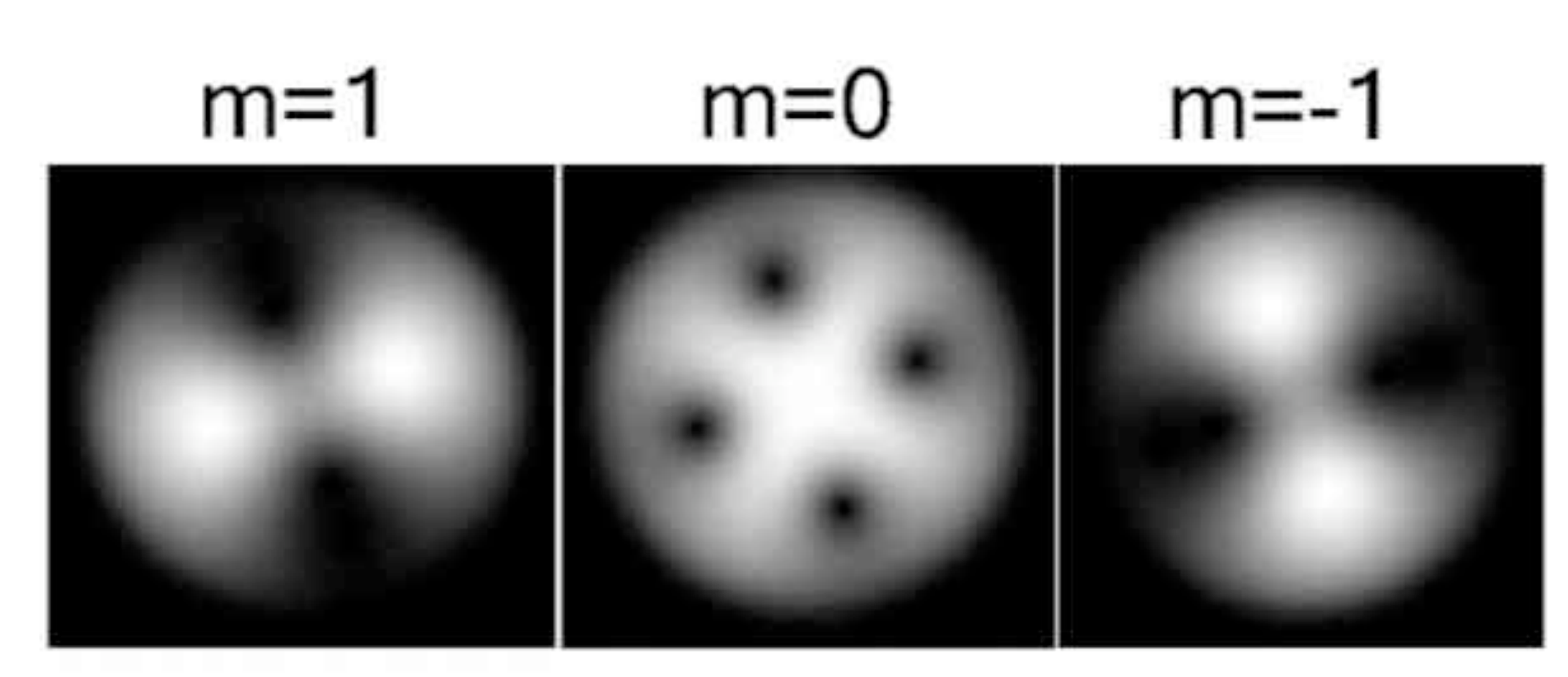}
\caption{Density plots of the Gross-Pitaevskii 
order parameters of the three components ($m=-1,0,1$) 
for a $T=1$ ferromagnetic condensate (see text). 
The calculation was performed for $1.7\times 10^4$ bosonic atoms of $^{87}$Rb.
Length units in the figure are in oscillator lengths. 
The total angular momentum per particle for the state shown was $L/N=1.85$,
and the rotation
frequency in units of the trap frequency was $\Omega  = 0.17$. 
 White color indicates maximum density. After \textcite{martikainen2002}.}
\label{skyrmion_martikainen}
\end{figure}

Experimentally, the investigations with two-component quantum gases have
been extended to higher pseudospins ($T=1$) \cite{leanhardt2003}.
For a rotating trap in the LLL approximation, 
the phase diagram of pseudospin $T=1$ bosons 
was studied by \textcite{reijnders2004}, both using mean-field approaches and
numerical diagonalization. 
The stability of the Mermin-Ho and Anderson-Toulouse vortices has been
demonstrated for rotating ferromagnetic condensates with pseudospin
$T=1$~\cite{mizushima2002c, mizushima2002b}. 
At small rotation the ground state is a coreless vortex. 
As an example, 
Fig.~\ref{skyrmion_martikainen} shows the ground state structure of a 
ferromagnetic $T=1$ spinor 
condensate for the three different components of
the order parameter~\cite{martikainen2002}.
The 3D trap was chosen with strong confinement in the 
$z$-direction of a harmonic trap, such that the system was effectively 
two-dimensional. 
The density distributions (where light color corresponds to the
maximum density) in the $x-y$-plane are shown for $m=1,0,$ and $-1$. 
The $m=\pm 1$ components show two coreless vortices in much
similarity to the two-component case discussed above. 
The third component, $m=0$, shows a regular array of four
vortices that occur 
at the same positions of the coreless vortices. 

\section{Summary and Outlook}

\label{sec:summary}
In finite systems with only a small number of particles,
vortex formation can be studied by a numerical diagonalization of 
the many-body Hamiltonian. Often, a reasonable 
approximation is to assume the confinement to be a
two-dimensional harmonic oscillator and to restrict the
single-particle basis to the lowest Landau level.
This is in particular the case in the limit of weak interactions. 
The close relationship of the many-body problem to the quantum
Hall liquid then helps to explain the vortex localization and the
similarity of vortex formation in boson and fermion systems.
The many-body energy spectrum, although experimentally yet inaccessible,
provides a wealth of information on the localization of vortices and their
mutual interactions. The energy spectrum should also allow an approximation
of the partition function and thus evaluation of temperature effects
in future studies~\cite{dean2002}. 

The exact diagonalization is limited to systems with only a few 
particles. Mean-field and density-functional methods
are necessary for capturing basic features
of vortices in larger systems.
In general, the density-functional methods describe the vortex structures
in excellent qualitative agreement with the exact diagonalization results.
In most density-functional 
approaches, the particles move in an effective field which
allows internal symmetry breaking, making the observation of vortices more
transparent than in the exact diagonalization method. However,
the present state-of-the-art density-functional 
approaches fail to describe properly the highly-correlated regime
at small filling fractions  
where vortices start to attach to particles, forming composites.

%\subsubsection{Prospects of future experiments}

Experimentally, clear signatures of vortices in
small electron droplets are still waiting to be observed.
Imaging methods of electron densities
in quantum dots may provide direct evidence of
vortex formation in the future~\cite{fallahi2005,dial2007,pioda2004}.
The predicted localization of vortices in asymmetric confinements and in the 
presence of pinning impurities open a possible way to 
direct detection of vortices by means of measurements of the 
charge density of the electron droplet.
Scanning probe imaging techniques have been developed
to visualize the subsurface charge accumulation~\cite{tessmer1998},
localized electron states~\cite{zhitenev2000} and
charge flow~\cite{topinka2003} of a quantum Hall liquid.
Similar methods could also turn out to be useful in probing
electron density of two-dimensional electron droplets in quantum dots.

In rotating traps the present observation techniques are based on
releasing the atoms from the trap and are limited to large atom numbers.
Naturally, the experimental goal has  been the study of large
condensates. Optical lattices, with a small number of atoms in each lattice
site, could in the future provide information of vortex formation in
the few-body limit.

Despite experimental and theoretical advances in studies of
rotating finite-size systems this review can provide only
glimpses of this rich field of physics where vorticity plays a central
role. Many important theoretical results presented here
remain unverified in experiments. 
Theoretical challenges remain as well, especially in
the regime of rapid rotation~\cite{baym2005} 
where strong correlations may lead to
emergence of exotic states. Vortex localization and ordering in
the transition regime to a quantum Hall liquid, as well as
the breakdown of this liquid state
into a crystalline one, are still lively discussed themes in the field.

%+++++++++++++++++++++++++++++++++++++++++++++++++++++++++++++++++++++

\section*{Acknowledgements}

S.M.R. and M.M. thank Georgios Kavoulakis and Ben Mottelson 
for many helpful discussions and advice, and for their collaboration on 
part of the subjects presented here. 

The authors are also indebted to 
S. Bargi, M. Borgh, K. Capelle, J. Christensson Cremon,
M. Koskinen, K. K\"arkk\"ainen, R. Nieminen, 
E. R\"as\"anen, S. Viefers, and others, for their 
collaboration. We thank 
J. Jain, C. Pethick, D. Pfannkuche, and V. Zelevinsky, as well as many others, 
for discussions. We also thank  
F. Malet for a careful reading of the manuscript.

Our work was financially supported by 
the Swedish Research Council, the Swedish Foundation for Strategic
Research, the Academy of Finland, and
the Magnus Ehrnrooth foundation. This article is the result of a collaboration 
within the NordForsk Nordic Network 
on ``{\it Coherent Quantum Gases - From Cold Atoms to Condensed Matter}''.

%\bibliographystyle{apsrmp}

%\bibliography{review10manninen.bib}

\end{document}